\documentclass[12pt,letter]{report}  
\usepackage{ifthen}
\usepackage{amssymb}
\usepackage{amsmath}
\usepackage{epsfig}
\usepackage{array}
\usepackage{verbatim}
\usepackage{udm_these_2e}
\maitrise
\titre{Heavy and Excited Leptons in the OPAL Detector?}
\nom{Erik~Elfgren}
\mois{Avril}
\annee{2002}
\president{Paul Taras}
\directeur{Georges Azuelos}
\jury{David London}

\newcommand{\gloss}[1]{{\sl #1}}
\newcommand{\ie}{i.~e. }
\newcommand{\eg}{e.~g. }
\newcommand{\etal}{et~al.}

\begin{document}
\setcounter{page}{0}
\pagetitre                     
\setcounter{page}{0}
\pagejury                      
\setcounter{page}{1}
\renewcommand{\thepage}{\roman{page}}

\begin{abstract}
Ce m\'emoire de Ma\^\i trise a pour object une recherche de leptons exotiques. 
Cette recherche a \'et\'e effectu\'ee \`a l'aide des donn\'ees enregistr\'ees par
le det\'ecteur OPAL utilisant les collisions $e^+e^-$ produites au sein de
l'acc\'el\'erateur LEP ``Large Electron Positron collider'' au CERN.
L'\'energie dans le centre de masse etait dans l'intervalle 183-209 GeV et la 
luminosit\'e integr\'ee \'etait de 663 pb$^{-1}$.
Le processus \'etudi\'e
est la production de leptons lourds par la voie du courant charg\'e suivi de la d\'esint\'egration du
$W$ en deux quarks. En particulier, on s'est pench\'e sur la production simple de leptons lourds
dans la r\'egion de masse 100-170 GeV. Aucune
\'evidence de l'existence d'une nouvelle particule n'a \'et\'e trouv\'ee.
Toutefois, les r\'esultats de cette analyse
imposent des limites sup\'erieures sur le m\'elange entre un lepton lourd et
le lepton ordinaire pour diff\'erentes masses du lepton lourd. Les limites sur
ces angles de m\'elange sont g\'en\'eralement meilleures que la valeur
nominale $\zeta^2\sim 0.005$.

{\bf Mots cl\'es} : physique, particules,  OPAL, leptons exotiques, leptons lourds, leptons
excit\'es, detection.

\end{abstract}

\chapter*{Abstract}
This M.Sc. thesis describes a search for exotic leptons. The search has been
performed 
using data from the OPAL
detector at the Large Electron Positron collider at CERN. The total integrated
luminosity was 663 pb$^{-1}$ with center of mass energies in the range of 183-209 GeV.
The work has been concentrated on the study of production of heavy leptons via
the charged current channel
and disintegration of the $W$ into two quarks. In particular, single production
of heavy leptons in the mass region 100-170 GeV has been extensively studied.
No evidence for any new particles has been found. The results translate into
upper limits on the mixing between the heavy and the ordinary lepton for
different heavy lepton masses. The limits on the mixing angles are generally
improved in comparison with the nominal value $\zeta^2\sim 0.005$.

{\bf Key words} : physics, particles, OPAL, exotic leptons, heavy leptons, excited leptons.

\chapter*{Preface}

This Master of Science project has been carried out at Universit\'e de
Montr\'eal under the supervision of Georges Azuelos. The research has been done
within the framework of the OPAL collaboration.

Regarding the notations used, a capital letter in fermion names
signifies a heavy fermion (e. g. $L$=heavy lepton, $E$=heavy electron) while a
star signifies an excited fermion (e. g. $u^*$=excited up-quark,
$\mu^*$=excited muon).
A fermion is either a lepton or a quark, and an
exotic particle is either heavy or excited.
A $\pm$ denotes a charged fermion (e. g. $L^\pm$=charged heavy
lepton) and a neutral lepton is synonymous to a neutrino.
A lepton $L$ can be either charged or neutral.
A bar denotes an anti-particle. Unless otherwise
specified, $c=\hbar=1$. For further details on symbols, constants, and
abbreviations, see Appendix \ref{App:Symb}. In the text, footnotes are used to
explain something in more detail, and they are not
needed for the basic comprehension. Words which appear {\sl slanted} are
explained in the Glossary in Appendix \ref{App:Glossary}.

I would also like to thank all the people who have helped me. First of all, my
supervisor Georges Azuelos who has always taken the time to answer my questions
properly and has given me a hand whenever I have had difficulties. He has given
me a lot of liberty, but also a great support. I would also like to thank
Robert McPherson who has been of great help in the simulations and
has shared his general reflections on the problem.
Finally, thanks to David London, Richard Teuscher and Alain
Bellerive for valuable discussions, and to R\'eda Tafirout for his preliminary
work on the subject.
\vspace{0.5cm}
\noindent Montr\'eal, May 2002.\\
\vspace{1.5cm}
\noindent Erik Elfgren

        \tableofcontents

	\pagenumbering{arabic}

\chapter{Introduction}
\label{Ch:1}

\section{Background}
Ever since the discovery of the three families/generations of light leptons and
quarks striking similarites have intrigued physicists:

\begin{itemize}
\item	Similarity between families.
\item	Similar group structure of leptons and quarks
\item	Mass hierarchy
\item	Equal charge
\end{itemize}

The Standard Model of particle physics \cite{SM} postulates three families of
leptons and quarks.
However, there is no underlying theory predicting these families to be three or 
predicting the number to be the same for leptons and for quarks.\footnote{
	In fact, in order for the Standard Model to be anomaly-free,
	the sum of the charge of all fermions must be zero, though this
	does not necessarily imply an equal number of leptons and quarks.
}
So far, it is only an experimental fact. Thus, a priori, there is no
reason why this number should be restricted to three only. In fact, the
situation is rather the opposite: there is theoretical justification
for {\it more} than three generations of leptons/quarks (cf. section
\ref{sec:IntroHL}).

\begin{figure}[here!]
\begin{center}
\includegraphics[width=6in]{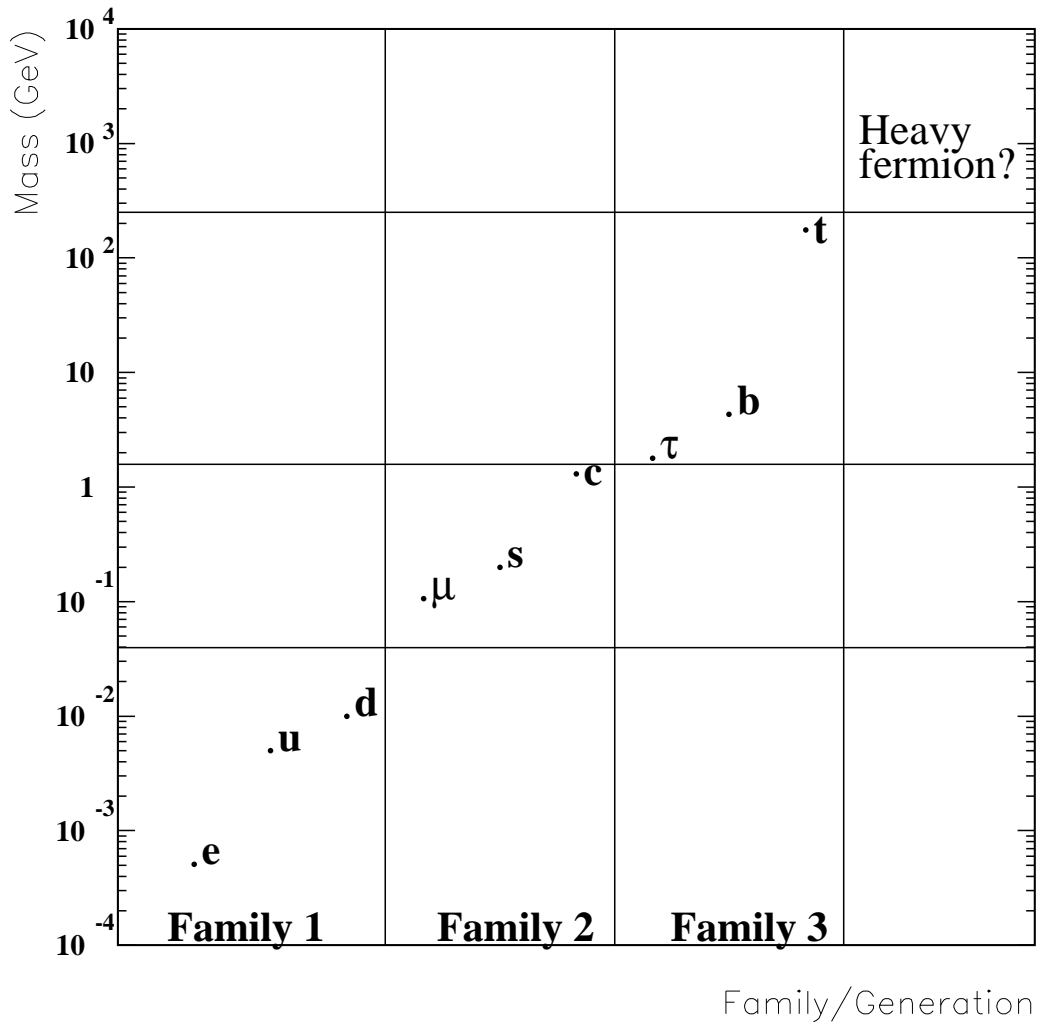}
\caption{The mass hierarchy of the three fermion generations}
\label{fig:Hierarchy}
\end{center}
\end{figure}

Within the lepton family there is nothing to distinguish an electron from a 
muon or a tau, except its mass (and its lepton number). Furthermore, the 
symmetry of the group structure of leptons and quarks (cf. table \ref{tab:LepQ})
is a remarkable coincidence. 
\begin{table}[here!]
\begin{center}
\begin{tabular}{|l|l|l|l|}
\hline
Leptons:     &  $\left(\begin{array}{cc}\nu_e\\e\\\end{array}\right)_L, \;
		e_R$ &
                $\left(\begin{array}{cc}\nu_\mu\\\mu\\\end{array}\right)_L, \;
		\mu_R$ &
                $\left(\begin{array}{cc}\nu_\tau\\\tau\\\end{array}\right)_L, \;
		\tau_R$ \\
Quarks:     &   $\left(\begin{array}{cc}u\\d'\\\end{array}\right)_L, \;
		u_R, d_R$ &
                $\left(\begin{array}{cc}c\\s'\\\end{array}\right)_L, \;
		c_R, s_R$ &
                $\left(\begin{array}{cc}t\\b'\\\end{array}\right)_L, \;
		t_R, b_R$ \\
\hline
\end{tabular}
\end{center}
\caption{
	The three families/generations of leptons
	(electrons, e, and electron neutrinos, $\nu_e$;
	muons, $\mu$, and muon neutrinos, $\nu_\mu$;
	taus, $\tau$, and tau neutrinos, $\nu_\tau$
	)
	and quarks
	(up, u; down, d; charm, c; strange, s; top, t; and bottom, b)
	. The quarks
	$d',s',b'$ are the weakly interacting mixed states of the mass
	eigenstates $d,s,b$. The indices L and R refer to left and right handed
	particles (for left-handed particles the spin is antiparallel with the
	direction of the particle, the spin points "backwards").}

\label{tab:LepQ}
\end{table}
In the past, symmetries of this kind have been recognized as a sign of
substructure. Also the remarkable mass distribution in figure
\ref{fig:Hierarchy} is an argument for either a substructure of fermions or for
a classification as a representation of a larger group. If a substructure
exists, then excited states ought to exist. If there is a larger group it might
very well contain heavy leptons. A priori, there is no relation between heavy
and excited leptons.

The hierarchy of the masses suggests the possibility of a fourth, even heavier, 
generation of leptons. Its mass would be of the order of 100 GeV, which could be
within the reach of the LEP accelerator where the sensitivity reaches up to
$M\sim 200$ GeV in single production and up to $M\sim 100$ GeV in double
production. The corresponding neutrino mass would, in principle, also have to
be large ($\gtrsim M_Z/2$) in order to be in accord with the measured invisible
width of the $Z$ (see section \ref{In:Constraints}).

Besides the arguments in favor of heavy and excited fermions there
are also several problems with the Standard Model. 
\begin{enumerate}
\item
	The Standard Model has more than 20 {\bf free parameters} which have
	to be determined experimentally (three lepton masses, six quark masses,
	$W^+, W^-, Z^0$ and Higgs boson masses, three coupling constants, four
	quark mixing angles, in addition to the masses and mixing angles of the
	neutrinos).
\item
	The form of the {\bf Higgs} potential is somewhat arbitrary.
	The {\sl Higgs particle} has not yet been found and the stability of
	its mass against radiative corrections is fine-tuned.
\item
	The nature and origin of the {\bf CP violation} is introduced ad
	hoc as a complex phase in the CKM matrix.
\end{enumerate}

\section{Outline of Thesis}
This thesis is divided into five chapters.
The introduction offers some reasons for the interest in the subject of exotic
leptons, as well as it presents a theoretical introduction to the
subject, and a summary of the results of previous research. A brief description
of the OPAL detector is also given. After that the
phenomenology of excited and heavy leptons is presented in chapter 2.
The word phenomenology refers to the characteristics of the exotic leptons, which can be
used to distinguish them from the background noise of ordinary particles. The
following chapter describes the simulations, both of the signal and of the
background. In this chapter, the process of event selection is also discussed.
This process is concentrated on the single production of heavy leptons, which
is the main issue of this thesis. Chapter four describes the data analysis 
itself with the results in the form of confidence limits on the masses and \gloss{mixing
parameter} involved in the theory. The conclusion in chapter 5 offers a summary
of the results, an outlook of what remains to be done, and possibilities of
improvement.

Below are presented some theoretical features of excited and heavy leptons.
However, this thesis will not dwell long on the theoretical properties of
different models, but rather will investigate their phenomenology and the
parameters of importance to detection. This will be the subject of Chapter
\ref{Ch:2}. So far, these theories are not (yet) supported by any experimental
evidence.

\section{Excited Fermions}
If a particle has an excited state, this means that it also has a substructure.
The excitation is nothing else than a rearrangement of the internal structure,
just like an atom or a molecule rearranges its constituents when 
(de)exciting. Excitation means stored potential energy, just like that in a
waterfall.

In the literature \cite{Preons}, the constituents of excited fermions are called
\gloss{preons}. Different models have been proposed; the number of preons is
usually, rather arbitrarily, set to two or three, where the three preon case can have two of them
bound together in a bosonic state (see \eg \cite{Sverker}).

Here are a few arguments in favour of the existence of excited leptons:

\begin{itemize}
\item	
	There are unstable and mixed leptons and quarks, and neutrinos are
	oscillating \cite{nuOsc}. From a philosophical viewpoint, these are not
	desirable properties for a fundamental particle. In the past, these
	effects have also been shown to be signs of substructure (\eg atomic
	and hadronic decays, Kaon oscillation etc.).

\item	
	In some preon models \cite{Sverker} the masses of the $Z^0$ and
	$W^\pm$ can be explained without the introduction of a Higgs boson. The
	$Z^0$ and $W^\pm$ are considered to be preon-antipreon states in
	analogy with the strong force that can be said to be mediated by
	quark-antiquark pairs (mesons).
	This would certainly be of interest if the Higgs particle is not found
	in the next generation of accelerators.

\end{itemize}

The catch with the preon models is that they must be able to reproduce {\it all}
experimental data with the new configuration and properties of the preons. This 
is already the case for \eg the Supersymmetry and the Grand Unified Theory models,
since they are extensions of the Standard Model. However, {\it new}
particles have to be shown to reproduce current data. Even though there
remain several theoretical problems within current preon models they still have
many attractive features as mentioned in the previous paragraph.

In this thesis, the detailed description of a specific model is not important,
as we study the phenomenological properties of excited leptons from a general
point of view. 

\section{Heavy Fermions}
\label{sec:IntroHL}
Several theories predict the existence of new, heavy fermions.
These extensions of the Standard Model generally have a mass scale at which
a certain symmetry is restored. If we can obtain accelerator energies reaching
this mass scale, we can directly probe the validity of the theory. However,
these energies are generally far too high ($M \sim 10^{12}$ TeV) to be reached 
for our present accelerators ($E \sim 1$ TeV). 

The most ambitious of the theories is the Superstring Theory ($M \sim 10^{16}$
TeV), aiming at unifying all the known forces: electromagnetic. weak, strong
and gravitational. It is currently the only viable theory to do so. The Grand
Unified Theories (unifications at $M \sim 10^{15}$ TeV) attempt all the above, except that they
do not include gravity. Supersymmetry Theory, which basically is a way of
explaining the mass hierarchy discussed above, is an integral part of both the
Superstring Theory and most of the Grand Unified Theories.

A characteristic property of these extended models is that they predict a zoo
of new particles, which have to be detected in order to be validated.
The experimental
advantage is that they offer some concrete predictions to search for.

The Grand Unified groups have fermion representations which contain the
Standard Model quarks and leptons but often also additional fermions.
The wide range of masses of the
fermions suggests that some new leptons could have masses of about 100 GeV,
which is within reach of LEP.

There are three popular unifying groups $SU(5)$, $SO(10)$, and $E_6$, the
latter two contain new fermions. The group $SO(10)$ \cite{SO10} contains a
right-handed Majorana neutrino and it is one of the simplest groups in which the
Standard Model could be conveniently embedded ($SU(5)$ is the simplest). 
The exceptional group E6 \cite{E6} 
contains several singlet
neutrinos as well as new charged leptons and quarks
and it is an acceptable four-dimensional
field theoretical limit of Super String Theory,

There are four phenomenologically different types of heavy leptons
(compare with table \ref{tab:LepQ}):\\
\begin{itemize}
\item{\bf Sequential}\\
{\indent        A fourth generation of fermions with the same basic properties
	as the other three.}\\
\begin{tabular}{l}
        $\left(\begin{array}{cc}N\\L^\pm\end{array}\right)_L; \; L^\pm_R$ \\
        $\left(\begin{array}{cc}U\\D\end{array}\right)_L; \; U_R, D_R$ \\
\end{tabular}

\item{\bf Mirror}\\
\indent Doublets have right chirality and singlets have left (opposite of the
        Standard Model).\\
\begin{tabular}{l}
        $\left(\begin{array}{cc}N\\L^\pm\end{array}\right)_R; \; L^\pm_L$ \\
        $\left(\begin{array}{cc}U\\D\end{array}\right)_R; \; U_L, D_L$ \\
\end{tabular}\\
Mirror fermions occur in many extensions of the Standard Model trying to restore
the left-right symmetry \cite{LeftRight} at the scale of the electroweak symmetry breaking.

\item{\bf Vectorial}\\
\indent Both the right and left chirality are doublets.\\
\begin{tabular}{l}
        $\left(\begin{array}{cc}N\\L^\pm\end{array}\right)_L; \;
        \left(\begin{array}{cc}N\\L^\pm\end{array}\right)_R$ \\
        $\left(\begin{array}{cc}U\\D\end{array}\right)_L; \;
        \left(\begin{array}{cc}U\\D\end{array}\right)_R$ \\
\end{tabular}\\
Vector fermions occur \eg in the group $E_6$.

\item{\bf Singlet}\\
\indent Both the right and left chirality are singlets.\\
\begin{tabular}{l}
        $L^\pm_L, \; L^\pm_R$ ; $N_L, \; N_R$\\
        $U_L, \; D_R$ ; $U_R, \; D_L$\\
\end{tabular}\\
Singlet fermions are found both in $E_6$ and in $SO(10)$.
\end{itemize}


In this thesis the focus will be on sequential heavy leptons, also called
fourth generation leptons. These new leptons are of the same character as
the ones already known.

An experimental reason for searching for this type of leptons is the recent
evidence for the mass of neutrinos \cite{NeutrinoMasses}. The see-saw mechanism
\cite{SeeSaw} can be used to generate massive neutrinos. The mechanism
predicts new neutrinos with masses, which could be reachable at LEP.

\section{Present Contraints}
\label{In:Constraints}
In this section, the present experimental limits in the search for exotic
(\ie\ heavy and excited) fermions and in particular, for exotic leptons, are outlined.
In the past, several searches have been performed in order to find traces
of excited and heavy leptons.

Lower limits on the masses of heavy leptons were obtained
in ${\rm e^+e^-}$ collisions at centre-of-mass energies,
$\sqrt{s}$, around $ M_{\rm Z}$ \cite{ref:hllep1}, and
recent searches at $\sqrt{s}=$~130-140~GeV \cite{ref:hlOPAL15,ref:hllep15},
$\sqrt{s} =$~161~GeV \cite{ref:hlOPAL161,ref:hlL3172},
$\sqrt{s} =$~172~GeV
\cite{ref:hlL3172,ref:hlOPAL172},
$\sqrt{s} =$~130-183~GeV \cite{ref:stableOPAL183,ref:DELPHI183,ref:OPAL183},
and $\sqrt{s} =$~200-209~GeV \cite{Teuscher}
have improved these limits.
Excited leptons have been sought at
$\sqrt{s} \sim M_{\rm Z}$
\cite{ref:ellep1},
$\sqrt{s}=$~130-140~GeV
\cite{ref:elopal15,ref:ellep15},
$\sqrt{s}=$~161~GeV \cite{ref:elopal161,ref:ellep161},
$\sqrt{s}=$~172~GeV \cite{ref:hlOPAL172},
$\sqrt{s}=$~183~GeV \cite{ref:OPAL183,ref:DELPHI183},
$\sqrt{s}=$~189~GeV \cite{ref:L3189},
and at the HERA ep collider \cite{ref:herasearches}.
If direct production is kinematically forbidden,
the cross-sections of processes
such as ${\rm e^+e^-} \rightarrow \gamma\gamma$ and
${\rm e^+e^-} \rightarrow {\rm f \bar{f}}$  \cite{ref:opalgg,ref:2f}
are sensitive to new particles at higher masses through loop corrections.

For double production of heavy leptons, masses up to 103 GeV have been explored
\cite{Teuscher}.
For single production of heavy leptons, this thesis covers masses up to 170 GeV.
In both
cases it is the charged current channel that has been studied because it has higher
branching ratio than the neutral current channel.

For single production of excited leptons in all important decay modes, the most
recent results are found in \cite{OPAL_EL} where all LEP data are combined.



Most searches for new particles, predicted by models beyond the Standard Model,
assume that the new particles decay promptly at the primary interaction vertex
due to very short lifetimes. These searches would not be sensitive to
long-lived heavy particles, which do not decay in the detectors. For example,
heavy neutral singlet leptons with masses below 2 GeV will live long enough to
escape the LEP and the SLAC Linear Collider detectors before decaying
\cite{ExtLim}
and through their mixing they would contribute insignificantly to the width
of the $Z$ resonance and therefore would be they are disguised within the $N_\nu =3$
light-neutrino number measurements \cite{Nequal3}. These measurements also give
a mass limit $M_L\gtrsim M_Z/2$ GeV.
The disguise is not watertight from a theoretical
point of view but this is of no experimental importance.

The possibility of light, long-lived new leptons has recently been addressed in
\cite{SLAC} while \cite{ExtLim} gives some older limits in the light of
kaon decays, the absence of forward neutral decaying particles in neutrino
beams from conventional sources or beam dumps and high statistics neutrino
experiments. Furthermore, singlet neutrinos do not have full couplings to the
$Z$ boson and would not contribute to the decay width of the $Z$, thus escaping
detection.


Several models predict such long-lived particles.
One example is the Minimal SuperSymmetric Model \cite{MSSM}.
Some models beyond the Standard Model would also predict the existence of
particles with fractional electric charge. As an example,
leptoquarks~\cite{ref:lq} could be long-lived. There are also some long-lived
hadronic states with fractional charge predicted by some modified 
Quantum Chromo Dynamics models~\cite{ref:dec-quarks}.

\subsection{Mixing}
\label{sec:Mixing}
The \gloss{mixing} between a heavy fermion and an ordinary fermion (by Flavour Changing
Neutral Currents) is parametrized by a constant
$\zeta$. In order to be general, this would have to be a mixing matrix
containing the mixing between all the new fermions and the old ones
$\zeta_{Ff}$ where $F$ represents all the new fermions and $f$ all the old ones.
However, the intergenerational mixing is generally supposed to be very
weak (see section \ref{Ph_Gen}).
We will assume 
only one
effective mixing parameter $\zeta$ for every new lepton.
\begin{figure}[here!] \begin{center}
\includegraphics[width=2.0in]{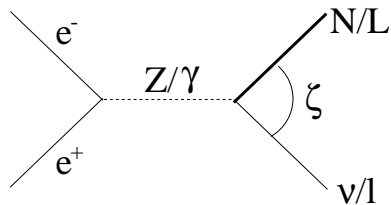}
\caption{Mixing,$\zeta^2$, between an ordinary and a heavy lepton.}
\end{center} \end{figure}
The mixing between fermions will alter their couplings to the electroweak
gauge bosons ($Z,W^\pm,\gamma$). If this mixing is large, the high
precision measurements\footnote{
	Measurements of total, partial and invisible decay widths as well
	as forward-backward and polarization asymmetries.
} of the $Z$ boson would be significantly altered, which is not the case. 
Thus, even the generational mixing must be rather small. The upper experimental
limit on the mixing parameter is of the order of $\zeta^2\lesssim 0.005$
\cite{MixLim,Mixing}. However, in some cases, the mixing could be higher
\cite{LeptMix}. If the left and
the right-handed mixing angles are equal for leptons, their mixing can be even
further constrained, to the order of $\zeta^2\lesssim 0.0001$ \cite{LeptMix}. This
is due to the contributions from lepton loops to the anomalous magnetic moment
($g-2$)$_{e\mu}$, which would be too large unless the lepton mass is very
large.

The decay length of the heavy neutrino is inversely proportional to the square of
the mixing. If the mixing parameter is smaller than ${\cal O}(10^{-12})$
\cite{ref:hlOPAL172}, the heavy neutrino is too long-lived to be discovered in the
detector.

\section{The Detector}
\label{sec:Det}
\subsection{The LEP Accelerator}
The accelerator and storage ring had a circumference of about 27 km.
Four experiments, called ALEPH, DELPHI, L3 and OPAL were placed
equidistant along the ring\cite{LEP}.
The peak luminosity of the machine reached about $10^{32}$ cm$^{-2}$s$^{-1}$.

\subsection{The OPAL Detector}

\begin{figure}\begin{center}
\includegraphics[width=14cm]{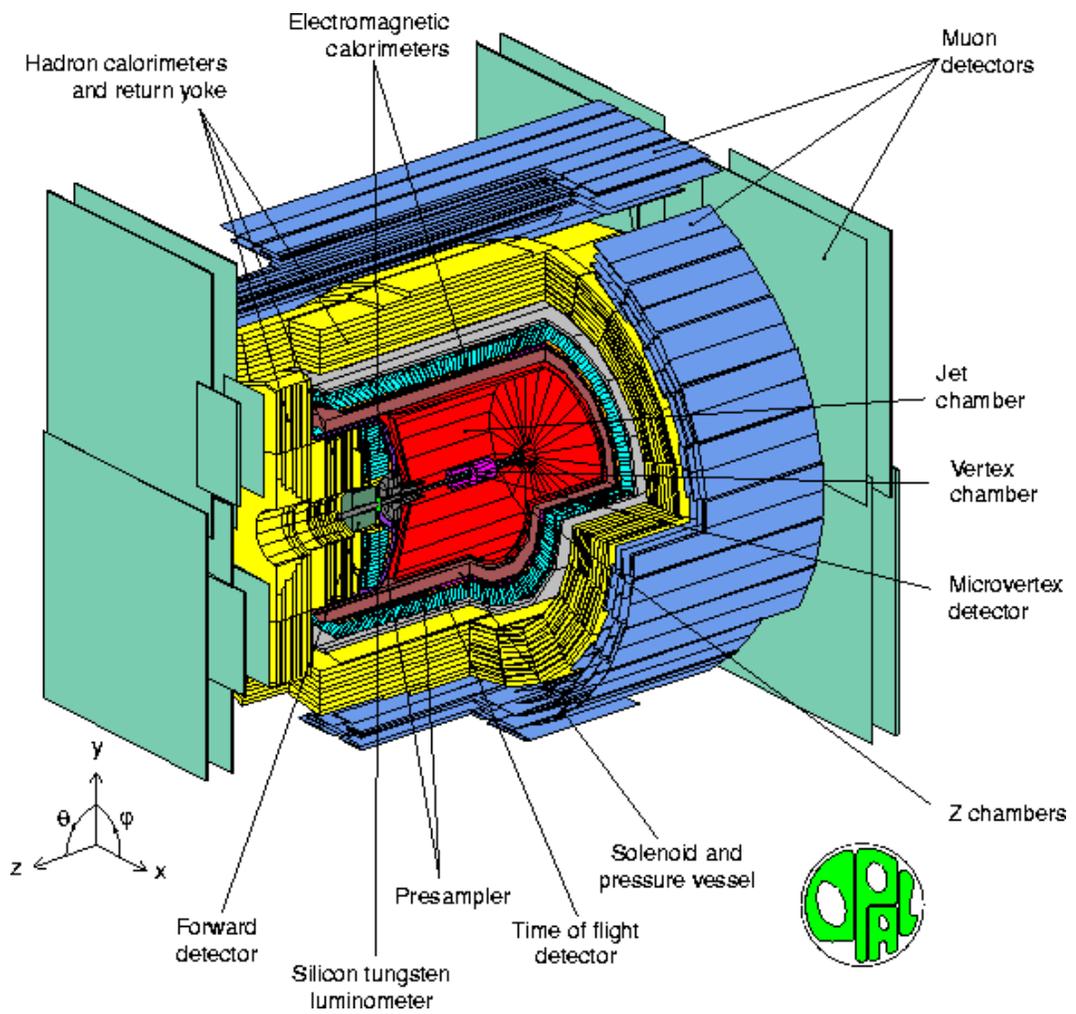}
\caption{The OPAL detector}
\label{fig:OPALdet}
\end{center}\end{figure}

This is a summary of the OPAL detector (see figure \ref{fig:OPALdet}), the
details can be found in \cite{OPAL}.

A trigger determines when a collision
has occurred and causes the event to be recorded.
It is set off when (i) an "essential" detector (such as the
electromagnetic or hadronic calorimeters) detects a signal above a
predetermined threshold or (ii) when a signal is observed in the same
$\eta-\phi$ region (out of 6 $\eta$ and 24 $\phi$ divisions) in a number of
subdetectors.

The OPAL right-handed coordinate system is defined such that the $z$ axis is in 
the direction of the electron beam, the $x$ axis is horizontal and points 
towards the centre of the LEP ring, and $\theta$ and $\phi$ are the polar and 
azimuthal angles, respectively. 


The OPAL detector consists of seven different parts:
\begin{itemize}
\item	Central Detector
\item	Electromagnetic Calorimeter
\item	Hadron Calorimeter
\item	Muon Detector
\item	Forward Detector
\item	Silicon Tungsten Detector
\end{itemize}

\subsubsection{Central Detector}
The central detector consists of a system of tracking chambers, providing 
charged particle reconstruction over 96\% of the full solid angle inside a 
0.435 T uniform magnetic field. It consists of a two-layer
silicon microstrip vertex detector, a high-precision drift chamber,
a large-volume jet chamber and a set of $Z$-chambers measuring
the track coordinates along the beam direction.

Closest to the interaction point is the microvertex detector to pin-point
the earliest vertices. It is placed about 7 cm from the center of the
beam pipe.

Then comes the vertex detector which is a jet-drift chamber providing track
reconstruction with high precision. It has a radius of about 20 cm.

The jet chamber and the Z-chamber also do track reconstruction with the
only difference that the Z-chamber only measures the z-coordinate of
the tracks.

\subsubsection{Electromagnetic Calorimeter}
The electromagnetic calorimeter covers almost 99 \% of the solid angle and
can be used to identify electrons and photons.

The first part of this calorimeter is the time of flight chamber which permits
low energy (0.6-2.5 GeV) charged particle identification.

Then comes presampling devices assuring the measurement of premature
electromagnetic showers. After that the barrel lead glass calorimeter
produces Cerenkov light used to identify the higher energy particles.

Finally there is an endcap presampler and calorimeter.

\subsubsection{Hadron Calorimeter}
The hadron calorimeter detects and measures the energy of jets, showers
of particles concentrated in a specific direction. The jets are often
the remnants of a quark.

\subsubsection{Muon Detector}
There are four muon detectors placed at the outer perimeter of the detector.
A muon leaves an electromagnetic trace from the inner of the detector all the
way out to the muon
chamber. The muon is the only particle to do this since the other particles
disintegrate before the muon chamber.

\subsubsection{Silicon Tungsten Detector}
The Silicon Tungsten Detector is an electromagnetic calorimeter 
used to measure the luminosity.

\subsubsection{Forward Detector}
Electromagnetic calorimeters close to the beam axis complete the geometrical 
acceptance down to 24 mrad on both sides of the interaction point.


\chapter{Phenomenology}
\label{Ch:2}

\section{Introduction}
Basically there are two mutually supporting approaches relating physical
principles, theory, and experiment. In the first approach, we start with some
physical principles and from those we can deduce a theory.
In our case we are concerned about the
origin of the heavy leptons and the constitution of excited leptons. This part
of the theory is briefly treated in chapter \ref{Ch:1}. In the second approach
we investigate how our theoretical predictions would manifest themselves in
experiments and we study the experimental parameters of the predictions. 
The latter approach will be presented in this chapter.
Most of the properties are common to \gloss{exotic} (heavy or excited)
leptons and quarks, even though the present thesis is concentrated on exotic
leptons.

When the results are general, this will be made explicit by the use of the word
fermion instead of lepton.


\section{General}
\label{Ph_Gen}
\begin{figure} \begin{center}
	\includegraphics[width=2.5in]{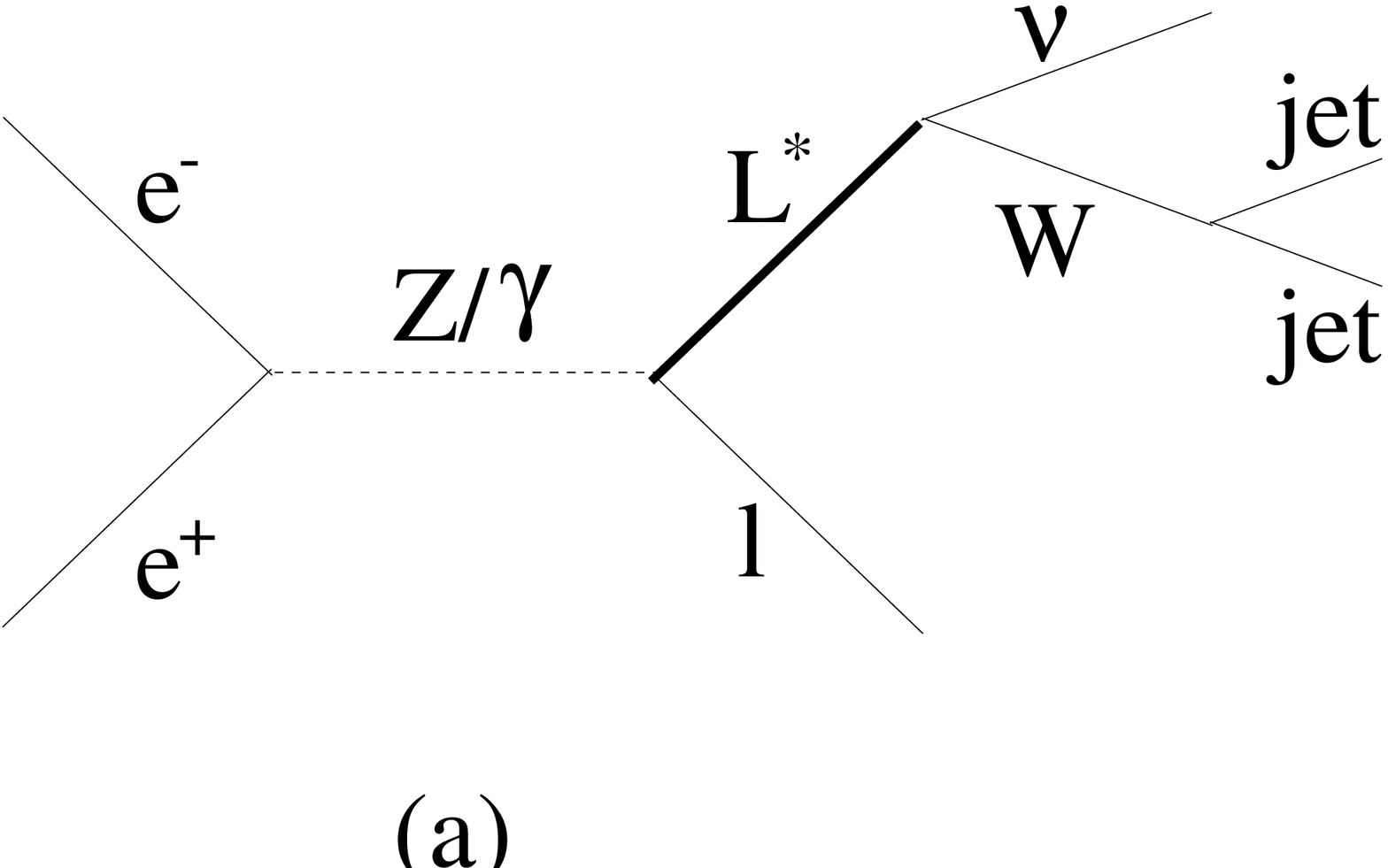}\hspace{1cm}
	\includegraphics[width=2.5in]{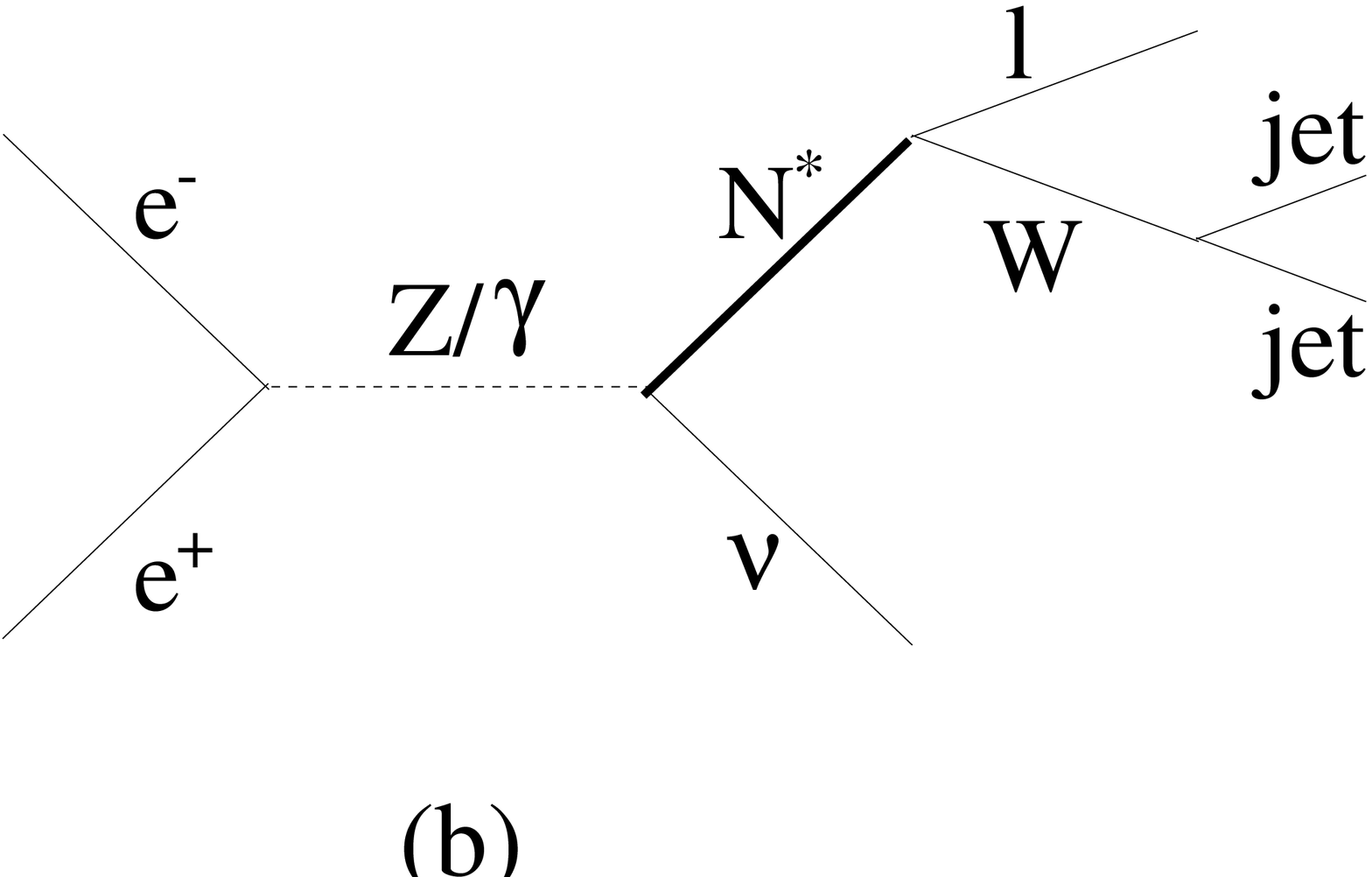}
\caption{$s$-channel single production processes for heavy/excited fermions.
	(a)~Charged exotic (heavy or excited) lepton, $L^*$.
	(b)~Neutral exotic lepton, $N^*$.}
\label{fig:Single}
\end{center} \end{figure}

In LEP there are basically two ways of producing an exotic lepton. The first
way is by single production, in which two electrons interact via a vector boson,
which in turn decays into an exotic lepton and an ordinary lepton (see figure
\ref{fig:Single}). The second one is double production, in which the vector
boson decays into a pair of exotic leptons (see figure \ref{fig:Double}).

If the mass difference between the exotic fermion and the ordinary fermion to
which it decays, is less than the mass of the intermediary vector boson, the
latter is virtual, leading to a three body final state. Otherwise, the boson is
real, giving a two-body phase space for the decay.

The general formulae for heavy fermions are described in \cite{DjProdDec}.
These formulae account for single production of these heavy fermions as well as
for double production of them including all possible decay channels for the
different types of fermions (as described in \ref{sec:IntroHL}). The exotic
fermions have electromagnetic and weak couplings, except for the singlet
neutrinos which only couple through the mixing. The Lagrangian of the
interactions of the new fermions is:

\begin{equation}
	L = \sum_{V=\gamma,Z,W^\pm,...?}g_VJ_V^\mu V_\mu
\end{equation}
where implicit summation over $\mu$ is assumed, $V_\mu$ is the four vector
of the vector boson field $V$, $J_V^\mu$
is the fermion current coupling to $V$ and complex conjugation
of the charged current is understood. $V$ allows for Standard Model as well as
new vector bosons.

$g_V$ are the coupling constants:

\begin{equation}
	g_\gamma = e, \quad
	g_Z = \frac{e}{\sin\theta _W\cos\theta _W}, \quad
	g_W = \frac{\sqrt{2}e}{\sin\theta _W}
\end{equation}
for the Standard Model vector bosons. In this thesis, it will be supposed that
any new gauge boson will be heavy enough not to affect the result. Furthermore,
it is also supposed that the mixing between the new vector bosons and
the $Z$ is too weak to interfere with the current research.

A coupling $\zeta$ allows 
the \gloss{mixing} between the exotic fermions and the ordinary
fermions. In this thesis, the mixing is supposed not to be intergenerational.
In other words, a heavy fermion of electron-type would not mix with the regular
muon, tauon, their neutrinos or the quarks. This is motivated by the fact that
large intergenerational mixing would introduce flavour changing neutral currents
at tree level. However, these currents are severely limited by experimental data
\cite{E6}, thus putting a tight constraint on intergenerational mixing.

\subsection{Comments About Cross Section Generation}
In order to generate the cross sections in the sections below, a number
of (rather general) conditions have been taken into account. Spin-correlations
have been included as well as initial state radiation. Final state radiation
has not been used, due to limitations in the EXOTIC \cite{EXOTIC} software.
The effect has, however, been shown to be negligible \cite{SingleTopPN}.
The fermions that interest us are leptonic, and they are all supposed to
belong to the first generation because their cross sections are much higher
than in the second and third generationi, \cite{DjProdDec}.

For heavy leptons, the mixing angle has been taken as $\zeta^2=0.005$,
and for excited leptons the gauge coupling parameters $f=f'=1$ (see
section \ref{Ph_ExLep}).

\section{Double Production of Exotic Leptons}

\begin{figure} \begin{center}
	\includegraphics[width=3.5in]{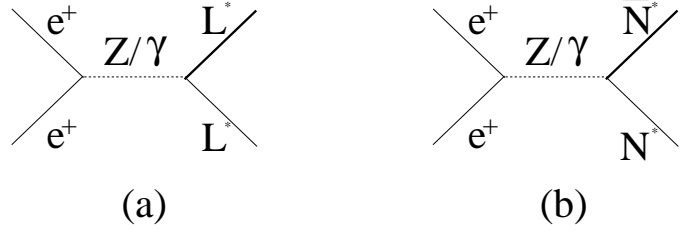}
\caption{$s$-channel double production processes for heavy/excited fermions.
	(a)~Charged exotic leptons, $L^*$.
	(b)~Neutral exotic leptons, $N^*$.}
\label{fig:Double}
\end{center} \end{figure}

Pair production of heavy leptons is mainly achieved through the $s$-channel
(see figure \ref{fig:Double}) via a virtual photon or $Z^0$. There is also a
contribution from the $t$-channel but it is quadratically suppressed
\cite{DjProdDec} by the mixing angle and is very small. Masses up to
$\sqrt{\hat s}/2$ can be probed in double production.

The principal production channels are:
\[
	e^+e^- \rightarrow E^+E^- \rightarrow e^+e^-ZZ \rightarrow e^+e^- +
	4\,{\rm jets}
\]
\[
	e^+e^- \rightarrow N\bar N \rightarrow e^+e^-WW \rightarrow e^+e^- +
	4\,{\rm jets}.
\]
$W/Z$ leptonic decay has a comparatively low branching ratio, and the neutrino
identification can be problematic.

The unpolarized differential cross section for pair production of heavy leptons
is:
\begin{equation}
	\frac{d\sigma}{d\cos\theta} =
	\frac{3}{8}\sigma_0 N_c \beta_F\left[(1+\beta_F^2\cos^2\theta)Q_1 +
	(1-\beta_F^2)Q_2 + 2\beta_F\cos\theta Q_3\right],
\end{equation}

where $\sigma_0 = 4\pi\alpha^2/3s$, $s = (\ell + \bar\ell)^2$, $\ell$ 
is the four vector of the ordinary lepton,
$\beta_F = \sqrt(1-4m^2/s)$ (velocity of the final state fermion),
 $m$ is the mass of the fermion, and $N_c$ is
the number of colours, \ie three for quarks and one for leptons.
The generalized charges $Q_i$ are rather comlicated and can be found in
equation (3.4) in \cite{DjProdDec}.

\begin{figure}[here!] \begin{center}
\includegraphics[width=4.0in]{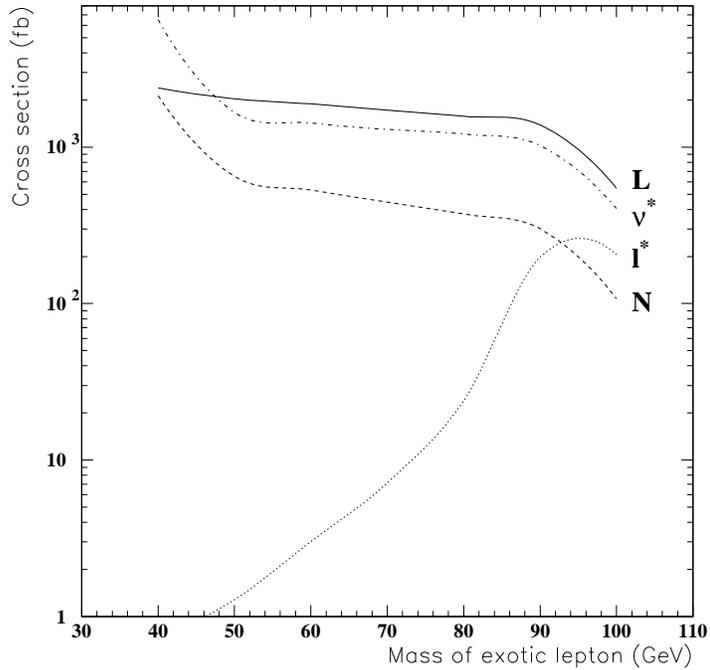}
\caption{ Cross sections for double production of different types of exotic
	leptons at $E_{CMS}=207$ GeV. All the types of heavy charged leptons, $L$, have the
	same cross section $\sigma_L$ and all the types of heavy neutral
	leptons, $N$, have the same cross section $\sigma_N$. 
	The different types of heavy leptons are discribed in section \ref{sec:IntroHL}.
	The heavy lepton
	case assumes a mixing angle $\zeta^2=0.005$. The excited lepton case
	assumes a mass scale $\Lambda^2=1$ TeV and $f=f'=1$ (see section \ref{Ph_ExLep}).
	problems. }

	\label{fig:Xsec2M}
\end{center} \end{figure}

The cross section is plotted in figure \ref{fig:Xsec2M}.

In this thesis there is only a general discussion on double production
of exotic leptons, as the data has already been analysed in detail in
\cite{Teuscher}.

\section{Single Production of Exotic Leptons}
\label{1Prod}

\begin{figure}[here!] \begin{center} 
        \includegraphics[width=5.0in]{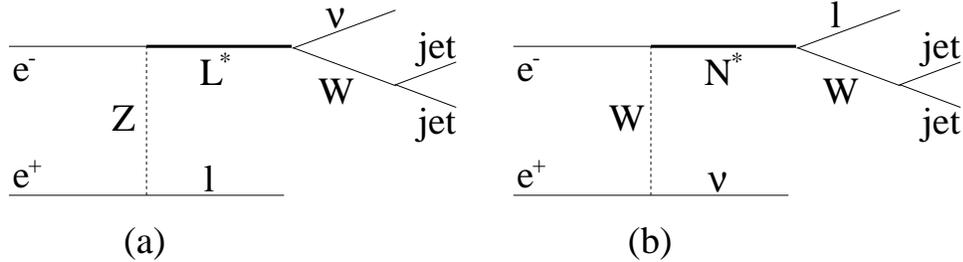}
\caption{$t$-channel single production processes for heavy/excited fermions.
        (a)~Charged exotic lepton, $L^*$.
        (b)~Neutral exotic lepton, $N^*$.}
\label{fig:tSingle} 
\end{center} \end{figure}

For exotic quarks and second and third generation leptons, single production
proceeds only through the $s$-channel \cite{AzDjSigBg}. This means that a photon or a $Z^0$
splits into an exotic fermion and an ordinary one. However, for the
electron type, additional $t$-channel production is allowed (see figure
\ref{fig:tSingle}). This means $W$ exchange for neutral leptons and $Z$
exchange for charged leptons. The $t$-channel is obviously not available
for muon- and tau-type heavy leptons because LEP is an $e^+e^-$ collider.
The $t$-channel increases the cross section by
several orders of magnitude making a search for electron-type heavy leptons
favourable.

For single production of heavy fermions to be possible, there has to be a
mixing between the ordinary and the exotic fermions. Otherwise, flavour
conservation would make the production impossible.

The heavy fermions decay into their light counterparts and a vector
boson, again through mixing. Thus, the final state is rather complicated,
containing four different particles. However, the final state particles are
not distributed randomly. This is a fact that can be used to distinguish them
from the background noise.
Their distribution pattern and properties of the decay
particles can also help us to identify the type of the heavy lepton (see
section \ref{sec:IntroHL}), which is crucial to the understanding of the origin
of heavy leptons. In particular, the angular distributions are useful for this
purpose \cite{AzDjSigBg} as the different types of heavy leptons have the same
type of interaction, differing only in their angular distributions.

\begin{figure}[here!] \begin{center}
\includegraphics[width=4.0in]{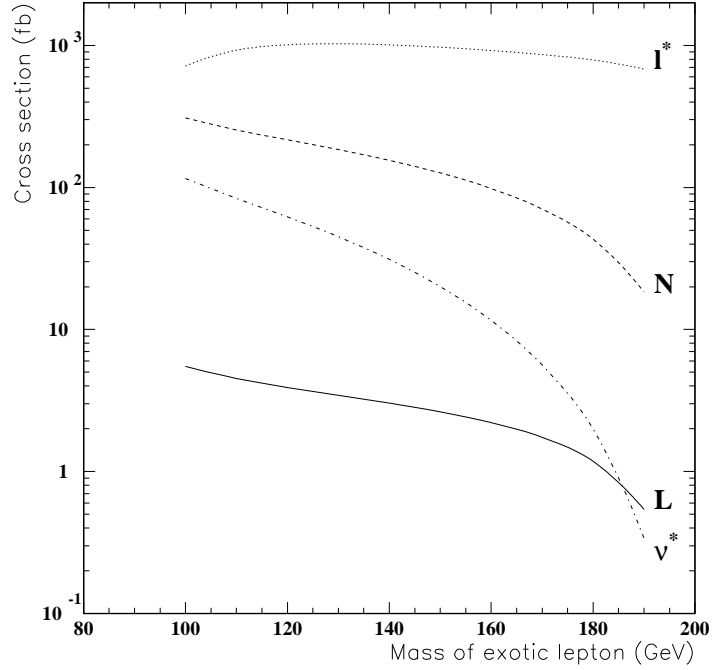}
\caption{Cross sections for single production of different types of exotic
leptons at $E_{CMS}=200$ GeV. All the different types of heavy leptons (as described in section \ref{sec:IntroHL}) have the same cross section.
The heavy lepton case supposes a mixing angle $\zeta^2=0.005$.}
\label{fig:XsecM}
\end{center} \end{figure}
\begin{figure}[here!] \begin{center}
\includegraphics[width=7cm]{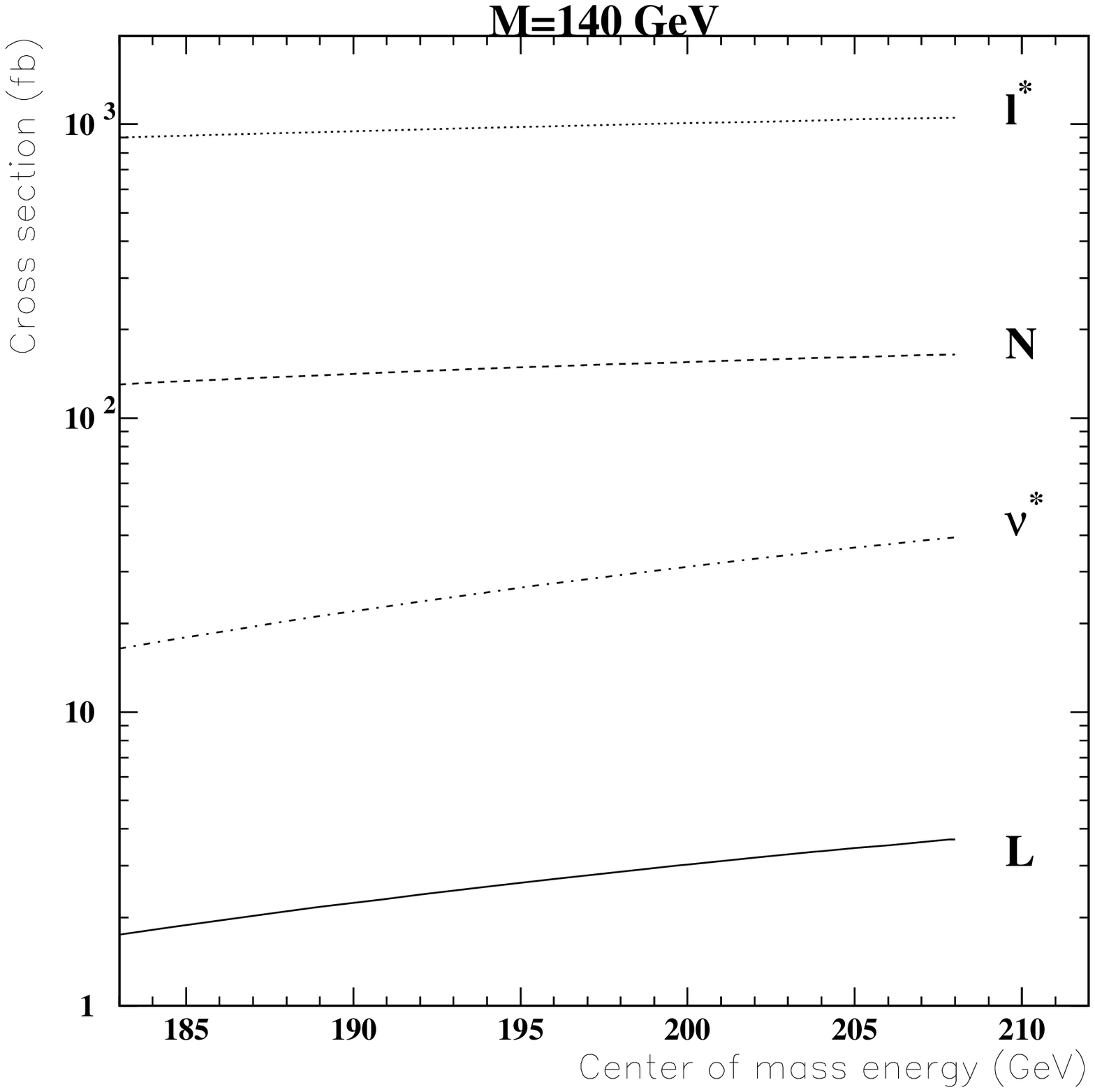}
\includegraphics[width=7cm]{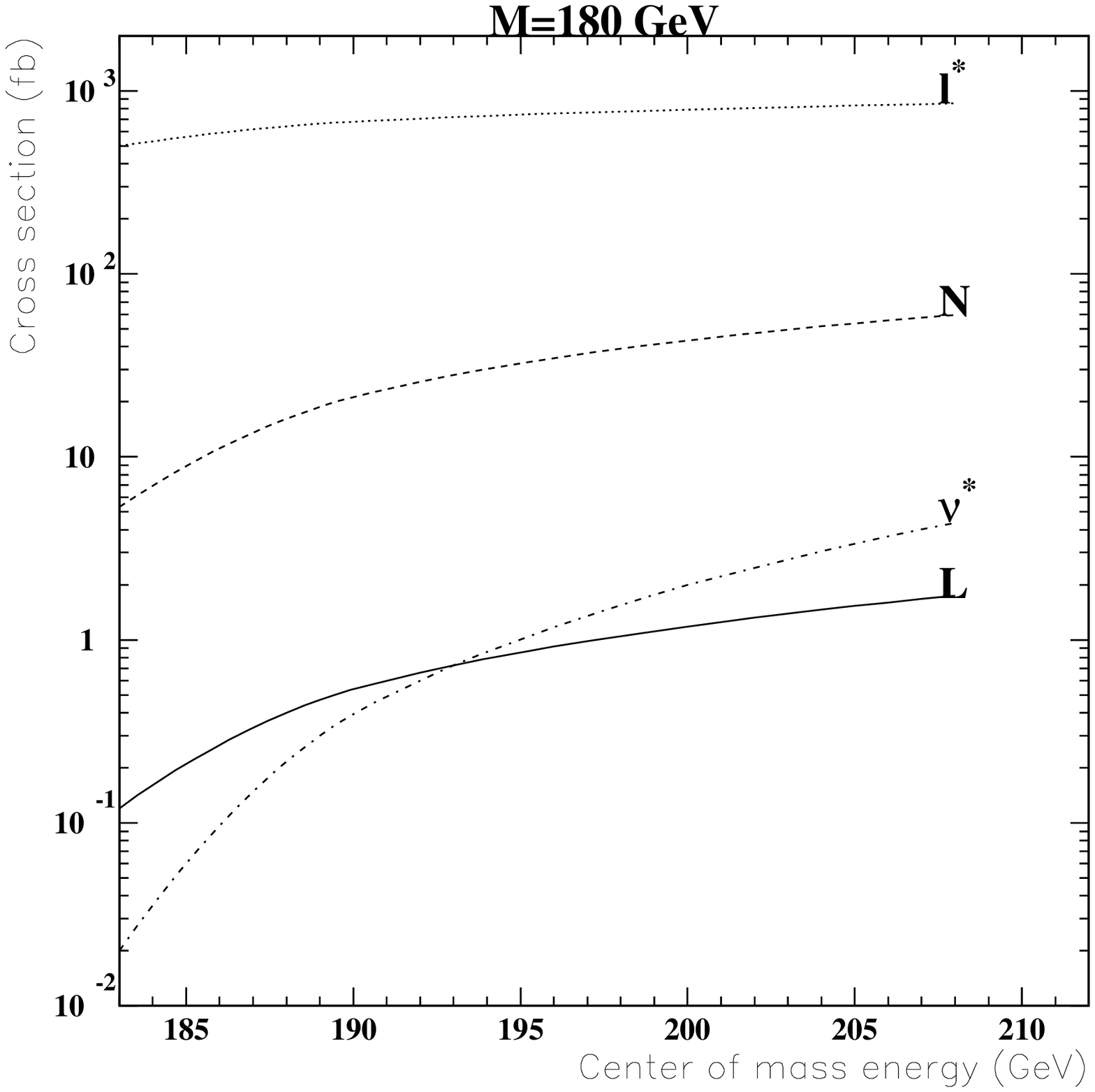}
\caption{Cross sections for single production of different types of exotic
	leptons as a function of energy. All the types of heavy leptons
	have the same cross section. The heavy lepton case supposes a mixing
	angle $\zeta^2=0.005$.}
\label{fig:XsecE}
\end{center} \end{figure}

The calculated cross sections for different signals are given in
figures \ref{fig:XsecM} and \ref{fig:XsecE}, presenting the cross section as a
function of mass and energy, respectively. The cross section for masses close to
the $Z^0$ mass, is not well treated by the inital state radiation method
(PHOTOS \cite{PHOTOS} used by EXOTIC \cite{EXOTIC}) for single production of exotic leptons.


A rough estimate for the number of expected events for a total integrated
luminosity of ~660 pb$^{-1}$ would be:\\
\begin{tabular}{llll}
$\ell^*:$	&	660	&	$L:$	&	2	\\
 $\nu^*:$	&	15	&	 $N:$	&	100	\\
\end{tabular}
\\
supposing the mass of the exotic fermion to be 140 GeV and the center of mass
energy to be 200 GeV. However, at such a low mass, the cross section will be
roughly the same for the different center of mass energies (see table \ref{tab:Xsec}).


This cross section as a function of energy is plotted in figure \ref{fig:XsecE}.

\subsection{Branching Ratio $\Gamma_{CC/NC}$}
The branching ratio for heavy fermion decay through the charged current (decay
through $W^\pm$) versus the neutral current (decay through $Z^0$) is independent of
the center of mass energy, as well as of type and charge of the exotic lepton.
This result has been obtained with EXOTIC \cite{EXOTIC} which is based on the
formulae in \cite{AzDjSigBg}.

The expressions for the partial decay widths for on-shell vector bosons are:
\begin{eqnarray}
	\Gamma(F_{L,R} \rightarrow Zf) &=& \frac{\alpha}
	{32\sin^2\theta_W\cos^2\theta_W}
	(\zeta_{L,R}^{fF})^2\frac{m_F^3}{M_Z^2}(1-v_Z)^2(1+2v_Z)\\
	\Gamma(F_{L,R} \rightarrow Wf) &=& \frac{\alpha}
	{16\sin^2\theta_W\cos^2\theta_W}
	(\zeta_{L,R}^{fF})^2\frac{m_F^3}{M_W^2}(1-v_W)^2(1+2v_W)
\end{eqnarray}
where $v_{W,Z} = (M_{W,Z}/m_F)^2$ and $\Gamma_{tot} = \Gamma_{CC}+\Gamma_{NC}$.
This means that for low fermion masses $m_F$, the charged current channel will be
strongly dominating, and even when $m_F>>m_{W,Z}$ the charged current channel will
have twice as high a cross section.


\begin{figure}[here!] \begin{center}
\includegraphics[width=4.0in]{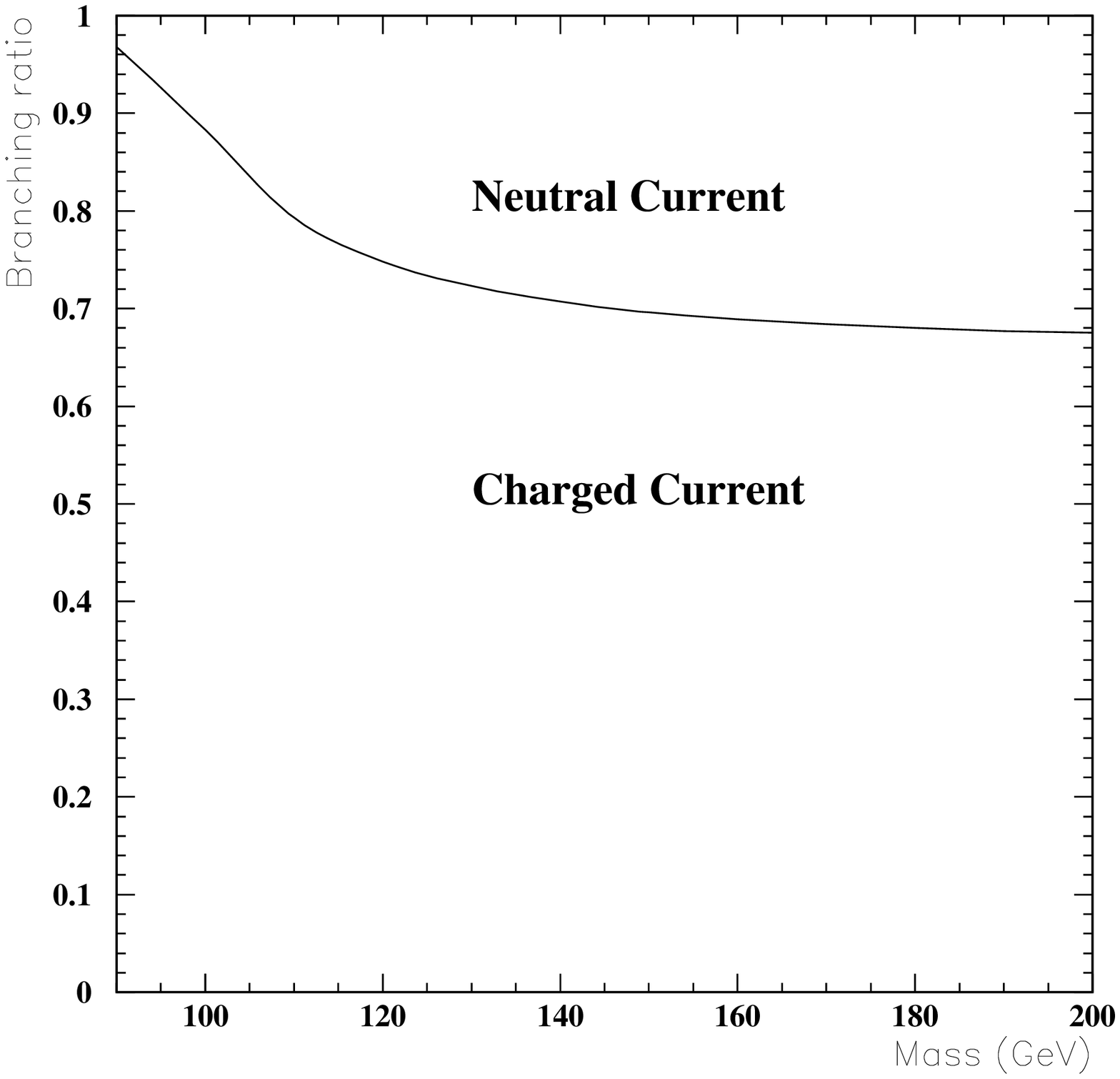}
\caption{
	Branching ratio $L\rightarrow W/Z + \ell$ for single production of
	heavy leptons as a function of energy. All the
	types of heavy leptons, charged as well as neutral, have the same branching ratio.}
\label{fig:Current}
\end{center} \end{figure}

\subsection{Heavy Leptons}
\label{sec:Ph_HL}
The polarized cross section of a heavy lepton is given by
\begin{equation}
	\frac{d\sigma^{\textrm{unpol}}}{d\cos \theta} = \frac{3}{8}\sigma_0
		N_c\beta_F\left[(1+\beta_F^2\cos^2\theta)Q_1 + (1-\beta_F^2)Q_2 + 
		2\beta_F\cos\theta Q_3\right]
\label{eq:xsec}
\end{equation}
where $\sigma_0 = 4\pi\alpha^2/3s$, $s = (\ell + \bar\ell)^2$ and 
$\beta_F = \sqrt{1-4m^2/s}$ and $N_c$ is the number of colours, \ie three
for quarks and one for leptons.
The values of the cross section are presented in table \ref{tab:Xsec}.

\begin{table}[here!]
\begin{tabular}{lllllllll}
\hline
                       Mass:  &     100  &     110  &     120  &     130
                              &     140  &     150  &     160  &     170  \\
\hline
$\sigma(N_e)_{\sqrt{\hat s}=183}$  &  305.42  &  245.44  &  202.14  &  165.09  
                              &  130.24  &   96.34  &   63.42  &   32.15  \\
$\sigma(N_e)_{\sqrt{\hat s}=189}$  &  306.62  &  248.80  &  207.63  &  172.70  
                              &  139.96  &  108.09  &   76.90  &   46.66  \\
$\sigma(N_e)_{\sqrt{\hat s}=192}$  &  307.25  &  250.41  &  210.17  &  176.21  
                              &  144.42  &  113.48  &   83.14  &   53.50  \\
$\sigma(N_e)_{\sqrt{\hat s}=196}$  &  308.12  &  252.46  &  213.37  &  180.60  
                              &  150.01  &  120.24  &   90.98  &   62.20  \\
$\sigma(N_e)_{\sqrt{\hat s}=200}$  &  309.05  &  254.45  &  216.41  &  184.73  
                              &  155.24  &  126.56  &   98.35  &   70.41  \\
$\sigma(N_e)_{\sqrt{\hat s}=202}$  &  309.51  &  255.41  &  217.88  &  186.70  
                              &  157.74  &  129.58  &  101.86  &   74.36  \\
$\sigma(N_e)_{\sqrt{\hat s}=204}$  &  310.01  &  256.37  &  219.31  &  188.63  
                              &  160.17  &  132.51  &  105.27  &   78.19  \\
$\sigma(N_e)_{\sqrt{\hat s}=205}$  &  310.24  &  256.84  &  220.02  &  189.56  
                              &  161.35  &  133.93  &  106.94  &   80.08  \\
$\sigma(N_e)_{\sqrt{\hat s}=206}$  &  310.49  &  257.30  &  220.72  &  190.49  
                              &  162.52  &  135.35  &  108.58  &   81.93  \\
$\sigma(N_e)_{\sqrt{\hat s}=207}$  &  310.74  &  257.79  &  221.40  &  191.42  
                              &  163.66  &  136.73  &  110.20  &   83.76  \\
$\sigma(N_e)_{\sqrt{\hat s}=208}$  &  311.00  &  258.24  &  222.09  &  192.32  
                              &  164.80  &  138.10  &  111.80  &   85.56  \\
\hline
\end{tabular}
\caption{Cross sections, $\sigma_j$ (fb) for single heavy neutrino production for different
	masses, $M_N$ (GeV) and center of mass energies, $\sqrt{\hat s}$ (GeV).}
\label{tab:Xsec}
\end{table}

\subsubsection{Decay}

The possible decay channels for a heavy lepton are:
\[
  {\rm N \rightarrow \ell^\pm W^\mp\quad\quad,\quad\quad
       N \rightarrow L^\pm W^\mp\quad\quad,\quad\quad
       N \rightarrow \nu_\ell Z},
\]
\[
  {\rm L^\pm \rightarrow \nu_\ell W^\pm \quad\quad ,\quad\quad
       L^\pm \rightarrow N W^\pm \quad\quad,\quad\quad
       L^\pm \rightarrow \ell^\pm Z}.
\]
The decay of a heavy fermion into another (lighter) heavy fermion is
possible, though not favourable, because of their smaller mass difference
compared to that between a heavy and an ordinary lepton. However, if the mixing
angle $\zeta^2$ is very small, decay into another heavy lepton could still be a
possibility. From the fact that the charged leptons in general are much heavier
than their neutral counterparts, it is plausible that the same would be the
case for heavy leptons. In this case, the channel $N\rightarrow L^\pm W^\mp$
would be kinematically closed.

\subsection{Excited Leptons}
\label{Ph_ExLep}
Excited leptons are supposed to have the same electroweak $SU(2)\times U(1)$
couplings ($g$ and $g'$) as ordinary leptons. However, unlike ordinary leptons,
they are supposed to be grouped into both left-handed and right-handed
iso-doublets with vector couplings to the gauge bosons ($Vf^*f^*$). The
$Vf^*f^*$ coupling is of magnetic type. If the excited leptons were not
iso-doublets, the light leptons would
radiatively acquire a large anomalous magnetic moment, due to the $\ell^*\ell V$
($V=Z, W, \gamma$) interaction \cite{Boudjema}.

The effective Lagrangian of excited leptons is \cite{Boudjema}:
\begin{equation}
  {\cal L}_{\ell\ell^*} =
  \frac{1}{2 \Lambda} \bar{\ell}^*\sigma^{\mu\nu}
  \left[g f \frac{ \mbox{\boldmath $\tau$} }{2}
    \mbox{\boldmath {\rm\bf W}}_{\mu\nu} +
  g^\prime f^\prime \frac{Y}{2} B_{\mu\nu} \right] \ell_{\rm L} +
  {\rm hermitian~conjugate},
  \label{eqLll}
\end{equation}
which describes the generalized magnetic de-excitation of
the excited states through the decay of the excited lepton.
The matrix
$\sigma^{\mu\nu}$ is the covariant bilinear tensor,
\mbox{\boldmath $\tau$} are the Pauli matrices,
${\rm\bf W}_{\mu\nu}$
and $B_{\mu\nu}$ represent the fully gauge-invariant
field tensors,
and $Y$ is the weak hypercharge.
The parameter $\Lambda$ has units of energy and can be regarded as
the compositeness scale, while $f$ and $f'$ are the weights
associated with the different gauge groups. The relative values of $f$ and $f'$ also affect the size of the
single-production cross-sections and their detection efficiencies.
Either the photonic decay,
the CC decay, or the NC decay will have the largest
branching fraction, depending on their respective couplings~\cite{Boudjema}:
\[
 {f_{\gamma} = e_f f' + I_{3L}(f-f')         \quad , \quad
 f_W        = \frac{f}{\sqrt{2} s_w}          \quad , \quad
 f_Z        = \frac{4I_{3L}(c^2_w f + s^2_w f') - 4e_fs^2_w f'}{4 s_w c_w}
}
\]
where $e_f$ is the excited fermion charge, $I_{3L}$ is the weak isospin,
and $s_w(c_w)$ are the sine (cosine) of the Weinberg angle $\theta _w$.

The integrated cross section for excited leptons is plotted as a function of
mass and energy in figures \ref{fig:XsecM} and \ref{fig:XsecE} respectively.


The possible decay modes for an excited lepton are:
\[
  {\rm \nu_\ell^*  \rightarrow \nu_\ell\gamma     \quad\quad , \quad\quad
       \nu_\ell^*  \rightarrow \ell^\pm W^\mp     \quad\quad , \quad\quad
       \nu_\ell^*  \rightarrow \nu_\ell Z,
  }
\]
\[
  {\rm \ell^{*\pm} \rightarrow \ell^\pm \gamma    \quad\quad , \quad\quad
       \ell^{*\pm} \rightarrow \nu_\ell  W^\pm     \quad\quad , \quad\quad
       \ell^{*\pm} \rightarrow \ell^\pm  Z.
  }
\]

\chapter{Simulation and Selection}
\label{Ch:MC}

\section{Introduction}
This chapter describes the Monte Carlo generated signals of exotic leptons and
their background. It also includes cut-optimization, which is applied on the
Monte Carlo samples in order not to be biased by the data. The Monte Carlo
samples have been generated for several different energies and normalized with
respect to the corresponding integrated luminosity at LEP (see table
\ref{tab:Lum}). Furthermore, they have been grouped into three regions, 183-189
GeV, 192-200 GeV and 202-209 GeV. There are two reasons for this grouping.
The first reason is that the cross sections of the sought signals is so small
that statistically, they would not show at one single energy. The second is
that the grouping simpifies the calculations. The total integrated luminosity is
663 pb$^{-1}$ after detector cuts. Detector cuts means basically that all
detectors are required to have registered a clean signal.

\begin{table}[here!]
\begin{tabular}{|l|r|r|r|r|r|r|r|r|r|r|r|r|}\hline
$E_{CMS}$ [GeV] & 183 & 189 & 192    &    196    &    200    &    202    &    204    &    205    &    206    &    207    &    208  \\
$\int{\cal L}dt$ [pb$^{-1}$] & 57.3 & 187 & 28.7&    71.1&    74.3&    38.1&         6.5 &      69.6&       16.6&    106&    7.7\\\hline
$\Sigma\int {\cal L}dt$&  \multicolumn{2}{|c|}{244.5}  & \multicolumn{3}{|c|}{174.0}  & \multicolumn{6}{|c|}{244.9} \\ \hline
\end{tabular}
\caption{ Integrated luminosities at LEP for different center of mass energies.
	The total integrated luminosity, $\Sigma\int{\cal L}dt = 663.5$ pb$^{-1}$.	The energies in the center of mass system are approximate within 1-2 GeV.}

\label{tab:Lum}
\end{table}

After having been generated, all Monte Carlo events (both signals and
backgrounds as described later) are processed through the GOPAL \cite{GOPAL} system, which is a
detector simulator based on GEANT 3. GEANT provides tools to allow the user
to define the geometrical parameters of his detector, using standard shapes. 
GEANT also deals with tracking of particles through this detector, including
the necessary physics processes (scattering, decays, interactions). All Monte
Carlo results (and later on also the data) are then processed through the same
system of pre-selection.\footnote{
	The pre-selection scheme is named the Tokyo Multihadronic Event
	Selection (TKMH) \cite{TKMH} and basically means that only good
	multihadronic events are kept.
} and then through selection of events with desired
characteristics. The detector itself is described briefly in section
\ref{sec:Det}.

\section{Signal Generation}
\label{Sec:Assumptions}

The simulation of the signal have been done with EXOTIC \cite{EXOTIC},
a Monte Carlo generator for single and pair production of heavy and
excited fermions in e$^+$e$^-$ colliders. All spin correlations in the
production and in the decay processes are included, as well as the transition betweeen
a virtual and a real intermediate vector boson at $M_N\sim M_Z,M_W$.
The hadronization of quarks is
done with the JETSET \cite{JETSET} package. The formulae for the generation of
exotic fermions are based on \cite{DjProdDec,zerwas}.

EXOTIC was chosen because it is the most versatile generator. For heavy
leptons, TIPTOP \cite{TIPTOP} is another possible generator, but it is more
limited. It only generates pair-produced sequential leptons within the
framework of the Standard Model. PYTHIA \cite{PYTHIA} could also have been used, but it is
restrained to generate fourth generation sequential fermions and it does not
include spin-correlations.

For the generation of excited fermions, existing Monte Carlo generators do not
take into account the polarization of the $\ell^*$ in their decay process
which is important for the angular distribution of $\ell^*$.
Besides, decays via $Z^0$ and $W$ have not been considered. For further
discussion on the limits of alternative generators, see the EXOTIC write-up
\cite{EXOTIC}.

Within the OPAL collaboration, several topics have already been covered, like
heavy lepton double production \cite{Teuscher} and excited leptons with photon
decay \cite{Vachon}. For further discussions see chapter \ref{Ch:1}.

The assumptions made for the signal generation are summarized and motivated
in the following sections.

\subsection{General}
\begin{itemize}
\item	Leptons, not fermions\\
	The fact that the accelerator collides $e^+e^-$ favours
	the production of leptons. Exotic quarks might
	also be produced, but we know that a fourth quark would have to be
	heavier than the top quark and therefore it would not be observable
	(see however \cite{SingleTop}).
\item	Single production for $M>100$ GeV\\
	This thesis is primarily focused on single production, because
	the final state ($jj\ell\nu$) is clean and easy to identify. 
	Furthermore, double production of heavy leptons is already
	covered by \cite{Teuscher} up to heavy lepton mass 103 GeV.
\item	Charged Current Channel\\
	The charged current channel is selected in preference to the neutral
	current channel, due to its higher branching ratio. Even for high
	masses, the branching ratio into a $W$ is roughly double that into
	a $Z^0$. For further details, see figure \ref{fig:Current}.
\item	$W\rightarrow jj$\\
	The branching ratio of $W\rightarrow jj$ is about 70 \%,
	while the branching ratio of $W\rightarrow \ell\nu$ is about 20 \% (not counting
	the $\tau$ neutrino, which is not easily identified).
	Further, the final state signature $\ell\nu jj$ is rather
	easy to identify and to reconstruct. Using the $W\rightarrow \ell\nu$
	channel would imply a combinatory problem between the
	two leptons, as well as it would lead to severe difficulties in
	distinguishing the neutrinos.
\item	Minimum cross section 15 fb\\
	If the number of signal events is required to be at least ten,
	then the smallest allowed cross section is $\sigma\gtrsim 15$ fb 
	for the total integrated luminosity of 663.5 pb$^{-1}$.
	If the number of signal events is significantly lower than ten
	they will most likely disappear during the cuts.
\item	No final state radiation in EXOTIC\\
	There is a bug in the OPAL version of the code of EXOTIC making it
	impossible to use final state radiation. The effect, however, has been
	found to be small, \cite{SingleTopPN}.\\
	For masses around 90 GeV, even the initial state radiation encounters
	some difficulties, thus producing unphysical results. This is due to the
	transition between a three body problem with a virtual vector boson, and
	a two body problem with a real vector boson at the $Z^0$ peak
	($M\sim90$ GeV). Further discussion can be found under the section
	ISRGEN in \cite{EXOTIC}.
\item	The use of spin correlations.\\
	This means that the angular distributions should be correctly generated.
\item	No new vector bosons light enough to affect the results.
\end{itemize}

\subsection{Heavy Leptons}
\label{sec:MCHL}
\begin{itemize}
\item	Heavy electron generation, neither heavy $\mu$, nor heavy $\tau$.\\
	As we have mentioned in chapter \ref{Ch:1}
	intergenerational mixing is assumed not to exist. Futhermore, only
	the first generation is studied, due to the non-existence of the
	$t$-channel in the other generations. The $t$-channel
	gives a much higher contribution to the cross section than the
	$s$-channel. For example,
	in the second generation, the number of observed events can be
	predicted to peak at about three, which would be hidden by the
	LEP background (see chapter \ref{Ch:2}).

\item	$\zeta_L^2=\zeta_R^2=0.005$\\
	The current upper limit on the \gloss{mixing} of heavy leptons and ordinary
	leptons (see section \ref{sec:Mixing}).
\item	Assuming $M_L>M_N$
	so that the channel $N\rightarrow L^\pm\ell^\mp$ is kinematically
	closed.
\item	Left-handedness\\
	There is no major difference between right-handed and left-handed	
	leptons, except in the $\theta$ distribution for heavy neutrinos.
	The left-handed state has been chosen in order to limit the
	calculations. 
\end{itemize}

\subsection{Excited Leptons}
\begin{itemize}
\item	Only generation of $e^*$, not $\mu^*$ or $\tau^*$.\\
	The electrons are the easiest to detect and identify.
\item	The gauge coupling parameters $f=f'=1$.\\
	As an extension to the current study, the case where $f=-f'=1$ could
	be studied.
\item	The compositeness scale is, rather aritrarily, set to $\Lambda = 1$ TeV.\\
	The cross section is inversely proportional to the square of the mass
	scale.
\end{itemize}


\section{Backgrounds}
\label{MC_Bg}
The backgrounds that are likely to cloud the signature of the signal
(electron, neutrino and two jets) are (ordered approximately by importance as
can be seen in tables \ref{tab:VarVar100} and \ref{tab:VarVar170}):
\begin{enumerate}
\item   llqq\\
        This is an irreducible background as it has exactly the same signature
        as the signal. The main production process is $WW,ZZ \rightarrow$ llqq.
        For technical reasons the eeqq background has been generated and treated
        separately.
\item   qq$(\gamma)$\\
        The cross section of this process is very high but on the other hand,
	the final state is quite different from the final state of the signal.
	in order to confuse the two of them,
        the lepton must come \eg from one of the jets (b-jet or c-jet) and
        the missing energy could come in the form of radiation or a particle
        disappearing along the beampipe. The production process is $Z/\gamma^*
        \rightarrow$ qq.
\item   qqqq\\
        Two showers are recognized as jets, while the other two give rise to
        individual particles. Some of these include leptons and neutrinos, which
        is the signature of the signal.
        The principal sources of this background are $WW,ZZ \rightarrow$ qqqq.
\item   $\gamma^*\gamma$qq\\
        If one of the electrons from $e^+e^- \rightarrow e^+e^-\gamma^*\gamma
        \rightarrow e^+e^- qq$ is undetected, then this final state correspond
        to that of the signal. $\gamma\gamma$qq has not been considered, due
        to the fact that the virtuality of the photons must be rather high.
\item   ee$\tau\tau$\\
        The two $\tau$ decay quickly to produce the jets and missing energy.
        $\gamma\gamma\tau\tau$ has not been included because of its rather
        far-fetched signature.
\end{enumerate}
The explanations above show how the backgrounds can be confounded with the
signals. Note, however, that they are only examples and that other confusions
also are possible.

The backgrounds have been generated by the OPAL collaboration with KK2f
\cite{kk2f}, Pythia \cite{PYTHIA} PHOJET \cite{PHOJET} and grc4f \cite{grc4f}
(cf. table \ref{tab:GenBg}).

Alternatively, 
a number of other generators could have been used, for instance
KORALW \cite{KORALW} for the llqq and qqqq backgrounds.

The $\gamma\gamma$qq (where $\gamma$ is a real photon) background has not been
considered due to the fact that the virtuality of the photons must be
rather high to be a real background to our process. Even when one of the
photons can be considered to be virtual, the background is hardly noticeable
(see table \ref{tab:PrelCut}).

The $\ell\ell\ell\ell$ background has not be treated either because of its
relatively low cross section ($\sim 3$ pb) in combination with the remoteness
of the final state particles compared to the signal.

\begin{table}[here!]
\begin{tabular}{llrrr}
\hline
{\bf Bg} &{\bf Generator}  &{\bf Simulated events/energy}\\
\hline
qq($\gamma$)        &  KK2f/PY6.125 &  200k ($\sim$2 fb$^{-1}$)\\
llqq                &  grc4f 2.1    &  5 fb$^{-1}$ \\
eeqq                &  grc4f 2.1    &  5 fb$^{-1}$ \\
qqqq                &  grc4f 2.1    &  5 fb$^{-1}$ \\
ee$\tau\tau$        &  grc4f 2.1    &  5 fb$^{-1}$ \\
$\gamma^*\gamma qq$ &  HERWIG       &  150k-350k  ($\sim$.5-1 fb$^{-1}$)\\
\hline
\end{tabular}
\caption{Background processes. The $\gamma^*\gamma qq$ background has
	150k events for E=183, 189, 196 and 200 GeV;
	350k events for E=192, 202, 204, 205, 207, 208 GeV and
	330k events for E=206 GeV.}
\label{tab:GenBg}
\end{table}

\begin{table}[here!]
\begin{tabular}{|l|r|r|r|r|r|r|r|r|r|r|r|r|}\hline    
$E_{CMS}$                           & 183   & 189   & 192   & 196    & 200    & 202    & 204    & 205    & 206    & 207    & 208   \\\hline
$\sigma_{qq\gamma}$ [pb]            & 109   & 99.4  & 94.8  & 90.1   & 85.6   & 83.4   & 82.1   & 81.3   & 79.5   & 79.4   & 78.4  \\
$\sigma_{llqq}$ [pb]                & 8.11  & 8.68  & 8.95  & 9.14   & 9.28   & 9.32   & 9.37   & 9.39   & 9.40   & 9.42   & 9.43  \\
$\sigma_{eeqq}$ [pb]                & 26.7  & 25.5  & 41.4  & 40.5   & 39.5   & 39.0   & 38.7   & 38.3   & 38.2   & 38.1   & 37.8  \\
$\sigma_{qqqq}$ [pb]                & 7.86  & 8.42  & 8.66  & 8.82   & 8.91   & 8.94   & 8.96   & 8.97   & 8.97   & 8.97   & 8.97  \\
$\sigma_{ee\tau\tau}$ [pb]          & 1.88  & 1.83  & 1.82  & 1.80   & 1.78   & 1.75   & 1.75   & 1.74   & 1.74   & 1.73   & 1.73  \\
$\sigma_{\gamma^*\gamma qq}$ [pb]   & 296   & 302   & 310   & 310    & 314    & 321    & 324    & 325    & 326    & 327    & 328   \\\hline
\end{tabular}
\caption{ Cross sections for background processes for different energies.
	Energy given in GeV.
	There are no simulations for $203, 209$ GeV, but their luminosities
	are so low that it does not matter.}
\end{table}

For specific energies, where the luminosity is high and the number of generated
events are low, the number of generated background events might be
insufficient. Notably, this is the case for the $\gamma^*\gamma qq$ background,
where the 189 GeV energy is particularly flagrant. The number of generated
events correspond to a luminosity of about 500 pb$^{-1}$ while the data
contains roughly 200 pb$^{-1}$. Unfortunately, at this moment, there are no
more of these background events available. However, this background is very
sensitive to the preliminary cuts (see section \ref{MC_PrelCut}) and is quickly
eliminated. This means that the lack of simulations is not really a problem.

The only other background where simulations might be insufficient is the
qq($\gamma$) final state. In the worst case (at $E=189$ GeV), the number
of generated events correspond to about 2 fb$^{-1}$, which is ten times the
data luminosity. This should be rather sufficient.

\section{Variables and Cuts}

\subsection{Description of Variables}
\label{sec:DescrVar}
The electron mentioned below refers to the most energetic electron detected,
while the neutrino is deduced as the missing energy and momentum. The missing
energy and momentum can either be a neutrino, which passes right through the
detector, or particles that pass so close to the beam pipe that they escape
detection. Often it is both, making the determination of the exact neutrino
energy and momentum difficult.
The $W$ is reconstructed from the two jets. The jets are identified with the
Durham jet-finding algorithm \cite{Durham} forcing reconstruction of only two
jets. The quality of this jet-finding is characterized by $y_{12}$ but this
variable has not proved to be of any use in the cuts.

\begin{itemize}
\item	$|\cos\theta_{\textrm{vis}}|$\\
	$\theta_{\textrm{vis}}$ is the angle between the missing\footnote{
		Conservation of momentum gives that 
		$\vec p_{\textrm{missing}} =-\vec p_{\textrm{visible}}$ and thus
		$\theta_{\textrm{vis}}=\pi - \theta_{\textrm{mis}}$.
	} momentum
	and the $z$-axis.
	\includegraphics[width=2.8cm]{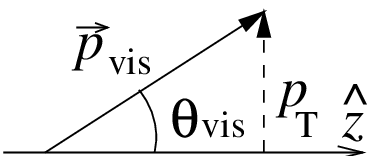}
\item	$M_{\ell\nu}$\\
	is the reconstructed invariant mass of the neutrino and
	the lepton with the highest energies. This variable will
	have a peak around $M_W=80$ GeV for all backgrounds containing $W$.
	This means that it is very useful to eliminate the llqq
	background.
\item	$E_\nu$\\
	is the energy of the neutrino, which is reconstructed supposing
	that there are only two jets and one charged lepton except for
	the neutrino. This variable is very useful to eliminate the qq($\gamma$)
	background.
\item	$E_\ell$\\
	is the energy of the most energetic lepton, which should be the lepton
	shown in figure \ref{fig:Single}. Unfortunately this is not always the
	case (cf. \ref{sec:HCL}). This variable is very useful to eliminate the
	qq($\gamma$) background.
\item	$M_N$\\
	is the mass of the reconstructed neutral lepton ($N$), i. e. the
	invariant mass of the two jets and the most energetic lepton.
\item	$M_L$\\
	is the mass of the reconstructed charged lepton ($L^\pm$), i. e. the
	invariant mass of the two jets and the most energetic neutrino (missing
	$E, \vec p$).
\item	$\theta_{W\ell}$\\
	is the angle between electron and the $W$ direction in the center
	of mass system of the reconstructed $W$.
	\includegraphics[width=2.8cm]{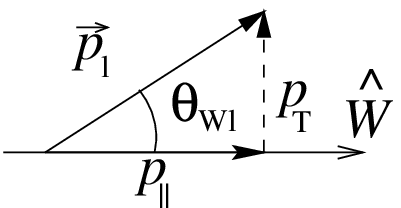}
\item	$\theta_{Wl\nu}(\hat z)$\\
	is the angle between the reconstruced $W$ and the invariant $l\nu$
	in the laboratory coordinate system.
	This variable could be used to discriminate the signal from the
	irreducible $WW$ background.
	\includegraphics[width=2.8cm]{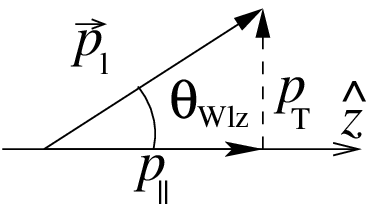}
\item	$\theta_{N\ell}$\\
	is the angle between the reconstructed $N$ and the $\ell$ in the
	coordinate system of the reconstructed $N$. This variable has not
	proved to be very useful to distinguish the signal from the background,
	but it might, for example, be used to distinguish the type of heavy
	lepton.
	\includegraphics[width=2.8cm]{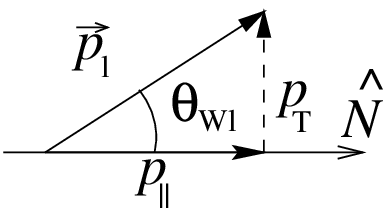}
\item	Lepton type\\
	is useful to distinguish heavy leptons, which should basically
	only give electron-type recoil leptons. 
\end{itemize}
Note that the angle variables make sense in the heavy neutral
lepton case. This is also the only case that has been studied thoroughly.

\subsection{Preliminary Cuts}
\label{MC_PrelCut}
First of all, a standard set of OPAL cuts is applied on the Monte Carlo
samples, the Tokyo Multihadronic Event Selection \cite{TKMH}. These cuts are
applied in order to have nice jets, which is advantageous in the present
analysis.

As seen in section \ref{MC_Bg}, the backgrounds that have only photons and/or
quarks in their final state have a rather high cross section. However, due to
the lack of leptons, these backgrounds can be severely reduced by
requiring a high energy of the lepton and of the neutrino. All the signals of exotic
neutrinos have rather high $E_\ell$ and $E_\nu$ because of their kinematics.
However, the heavier the exotic lepton is, the more energy is available for the
recoil lepton and the less for the neutrino. It would eventually be possible
to combine these two variables and cut only on the combination as
the two energies are kinematically interrelated. For the moment, cutting
each one individually still produces good results.

For all exotic neutrinos and their backgrounds, the following preiminary cuts
have been applied:

\[
E_\nu>10 \,{\rm GeV}, \quad\quad E_\ell>5 \,{\rm GeV}.
\]
These cuts reduce the signals with a factor of about 1.3-1.5 (depending on the
mass of the exotic lepton) while the quark backgrounds mentioned above are
reduced more than sevenfold. The other backgrounds are moderately reduced as
can be seen in table \ref{tab:PrelCut}.
\begin{table}[here!]
\begin{tabular}{lll}
\hline
Process             &  Reduction    &  \# of events after prel. cuts\\
\hline
qq($\gamma$)        &  7.57         &  9787        \\
llqq                &  1.88         &  23002       \\
eeqq                &  2.78         &  3571        \\
qqqq                &  7.51         &  5868        \\
ee$\tau\tau$        &  1.56         &  37          \\
$\gamma^*\gamma qq$ &  7.4          &  64          \\
\hline
\end{tabular}
\caption{The effect of the preliminary cuts. The reduction is 
	defined as the number of events before divided by the number
	of events after the preliminary cuts.
}
\label{tab:PrelCut}
\end{table}


\begin{figure} \begin{center}
\includegraphics[width=7.0cm]{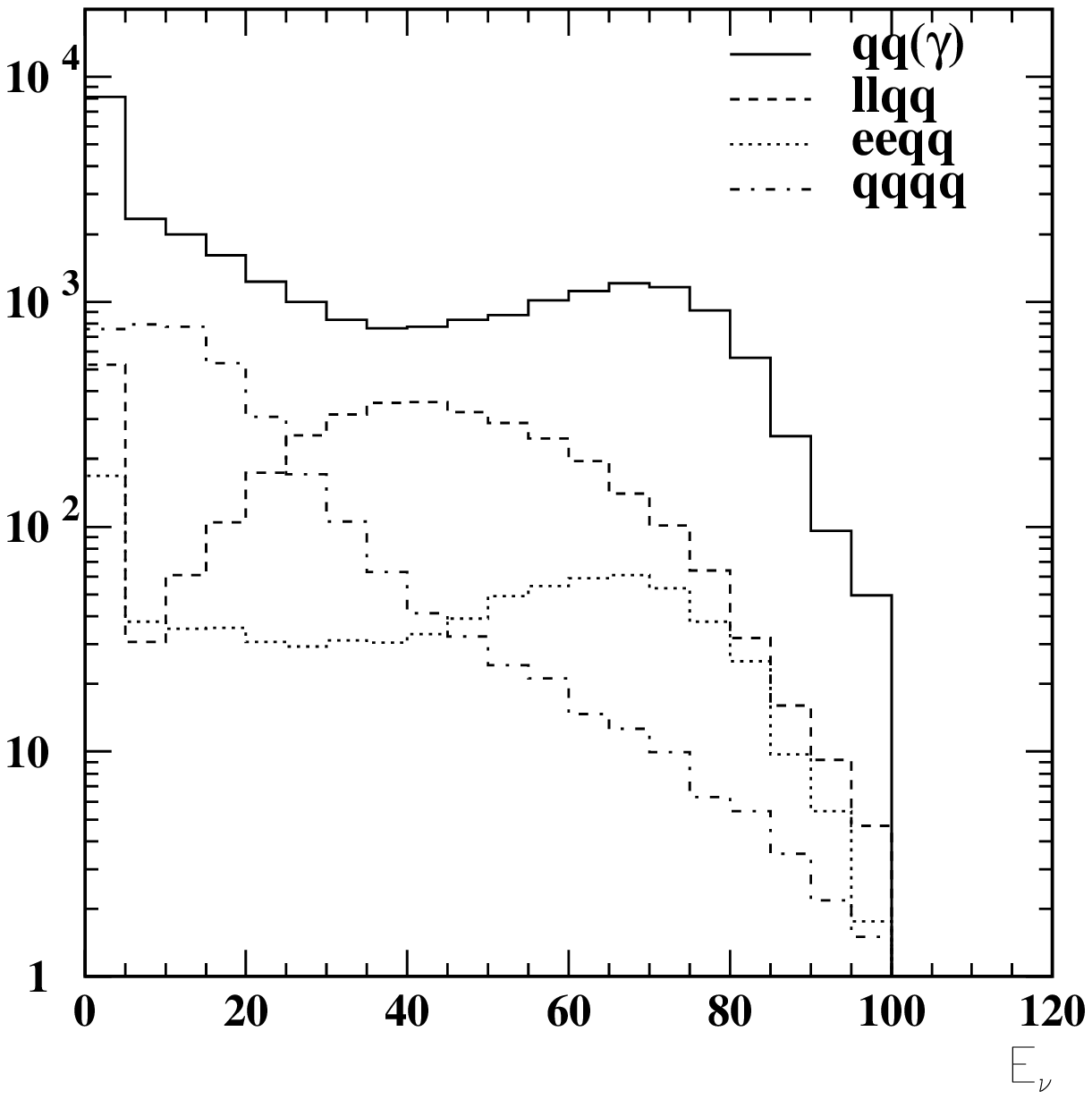}
\includegraphics[width=7.0cm]{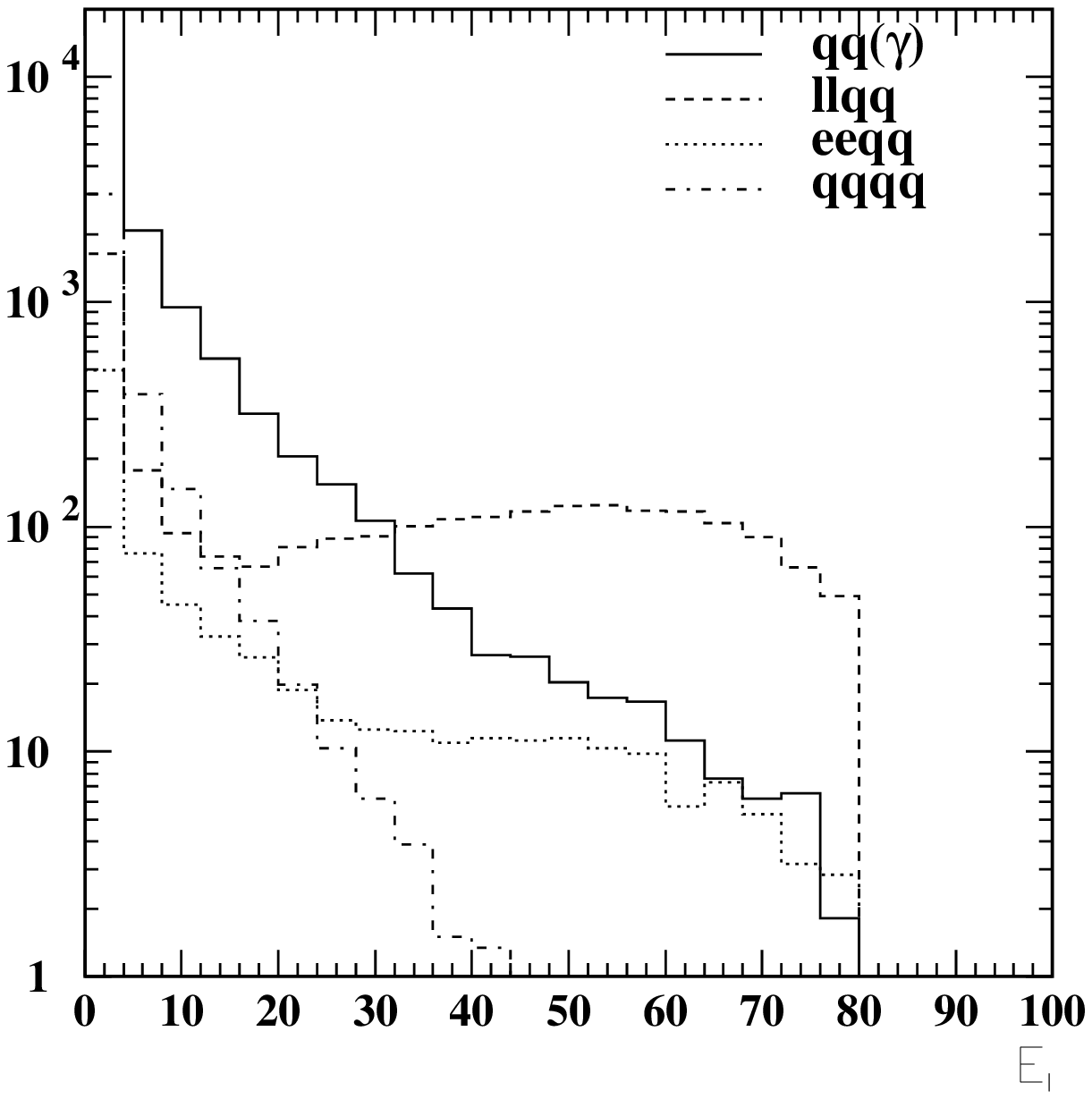}
\includegraphics[width=7.0cm]{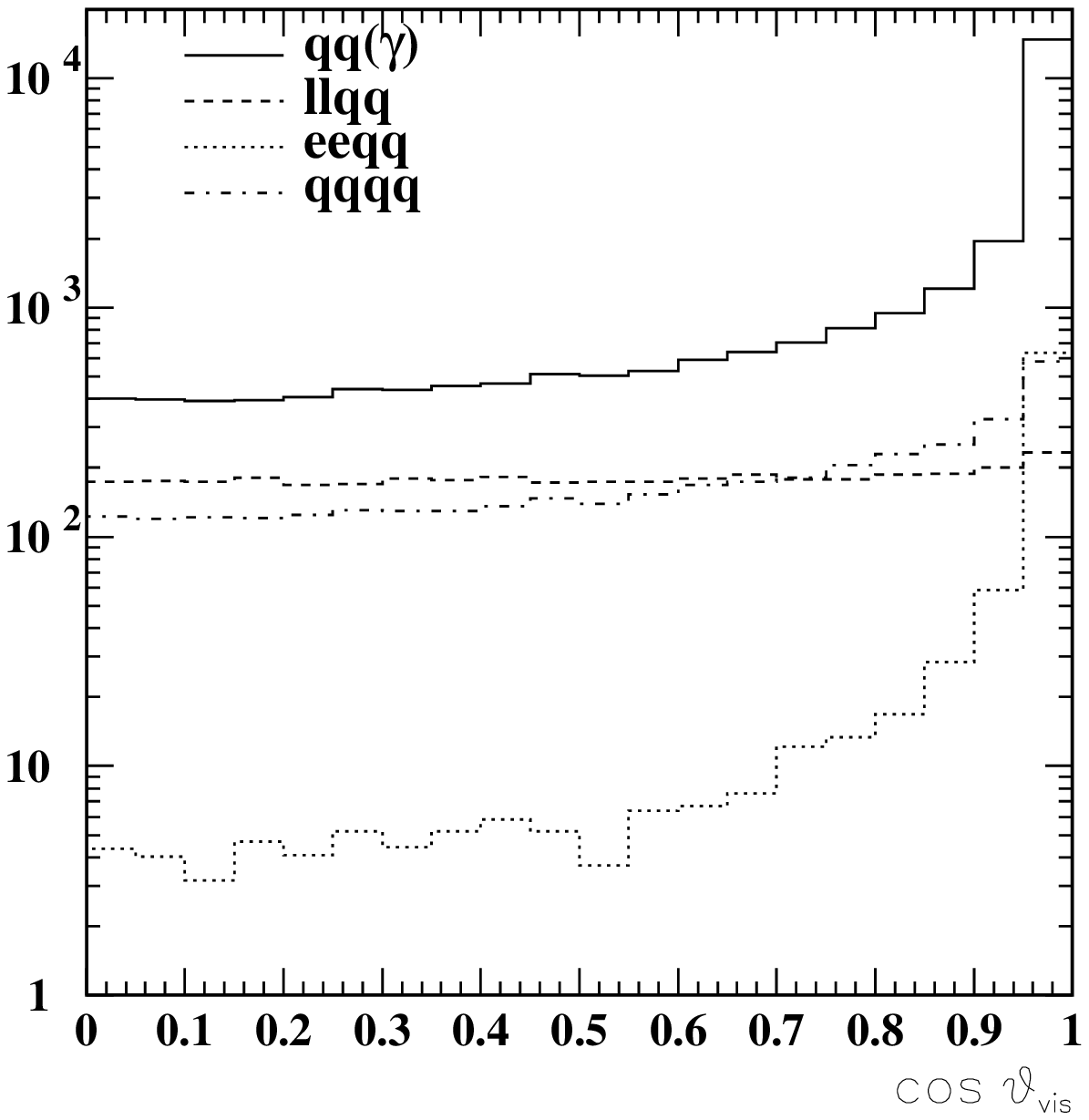}
\includegraphics[width=7.0cm]{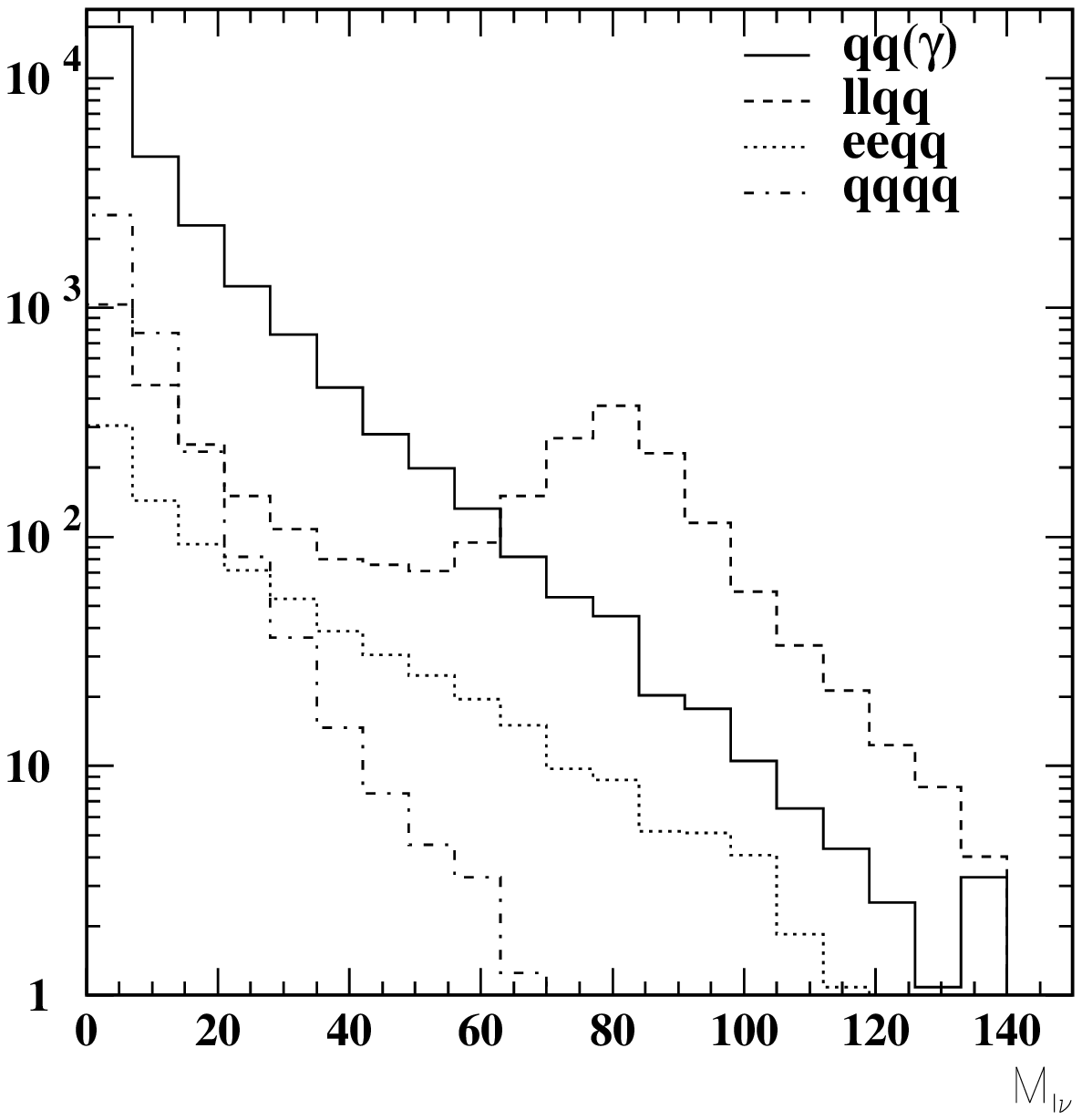}
\caption{Comparison of the principal variables for different backgrounds. 
	The integrated luminosity $\int{\cal L}dt = 79.3$ pb$^{-1}$
	and the center of mass energy $E_{CMS} = 200$ GeV.}
\label{fig:BgVar}
\end{center} \end{figure}

\begin{figure} \begin{center}
\includegraphics[width=7.0cm]{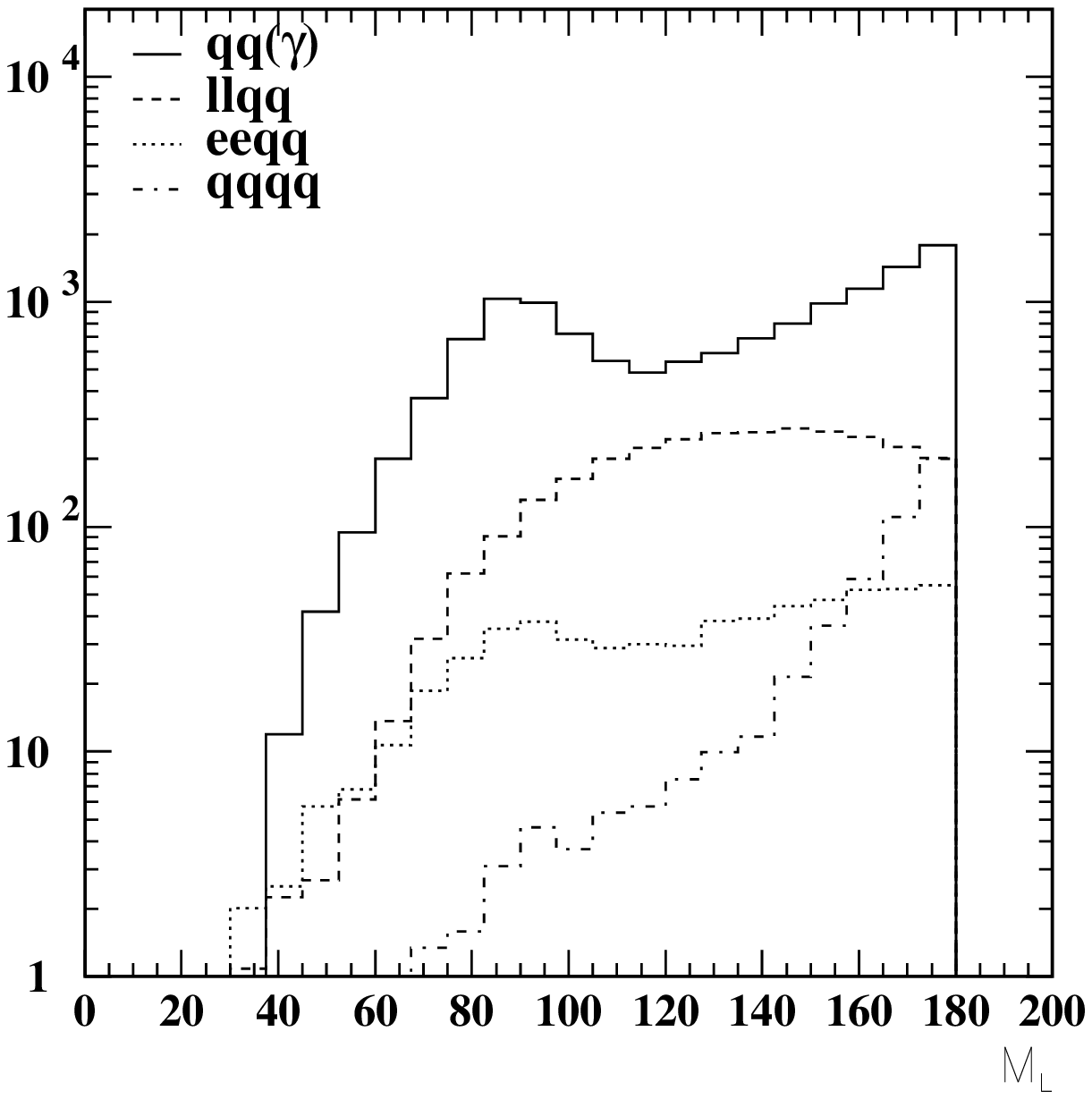}
\includegraphics[width=7.0cm]{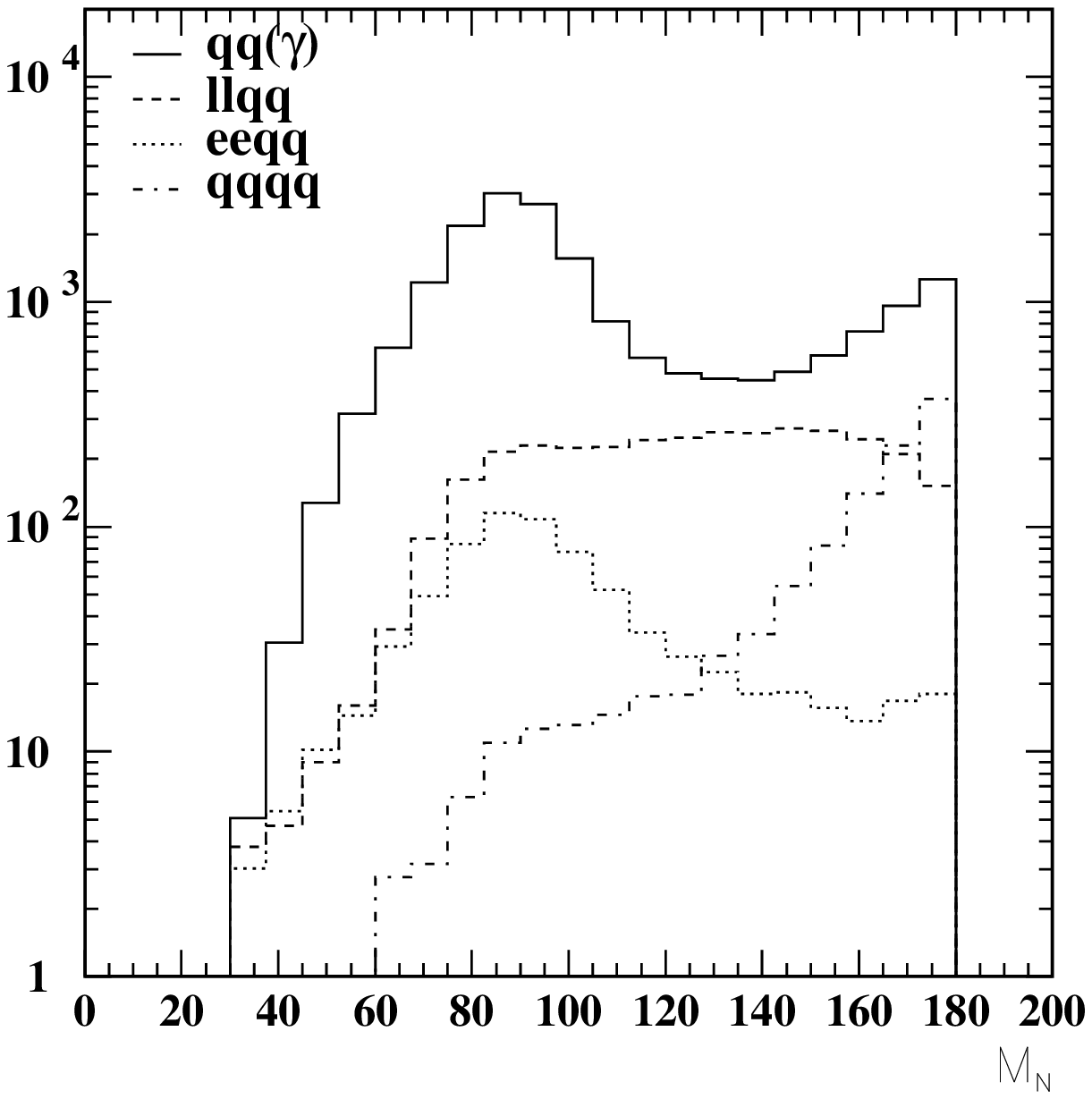}
\caption{Comparison of reconstructed masses $M_L$ and $M_N$
	 for different backgrounds}
\label{fig:BgMass}
\end{center} \end{figure}

\begin{figure} \begin{center}
\includegraphics[width=7.0cm]{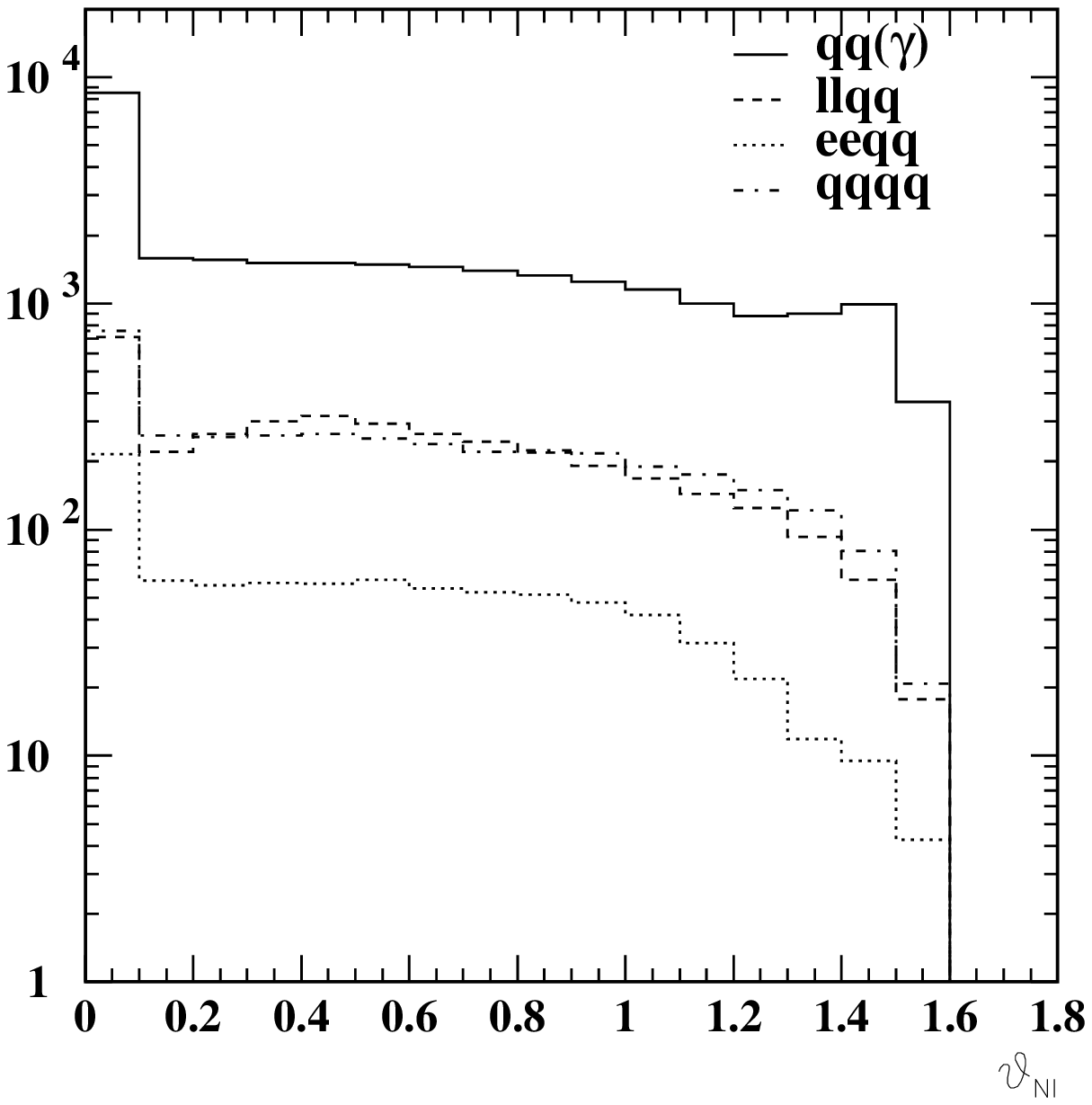}
\includegraphics[width=7.0cm]{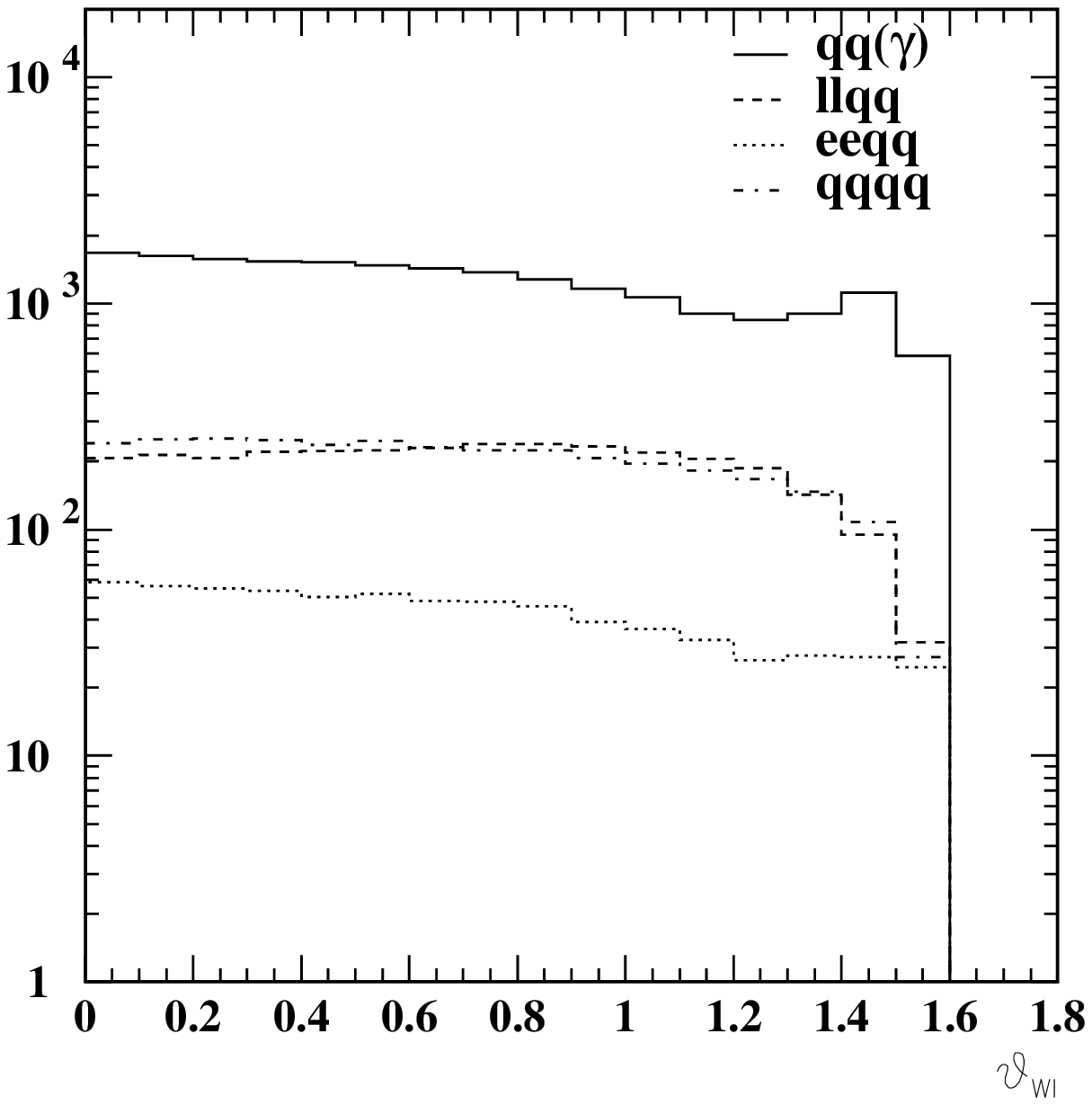}
\includegraphics[width=7.0cm]{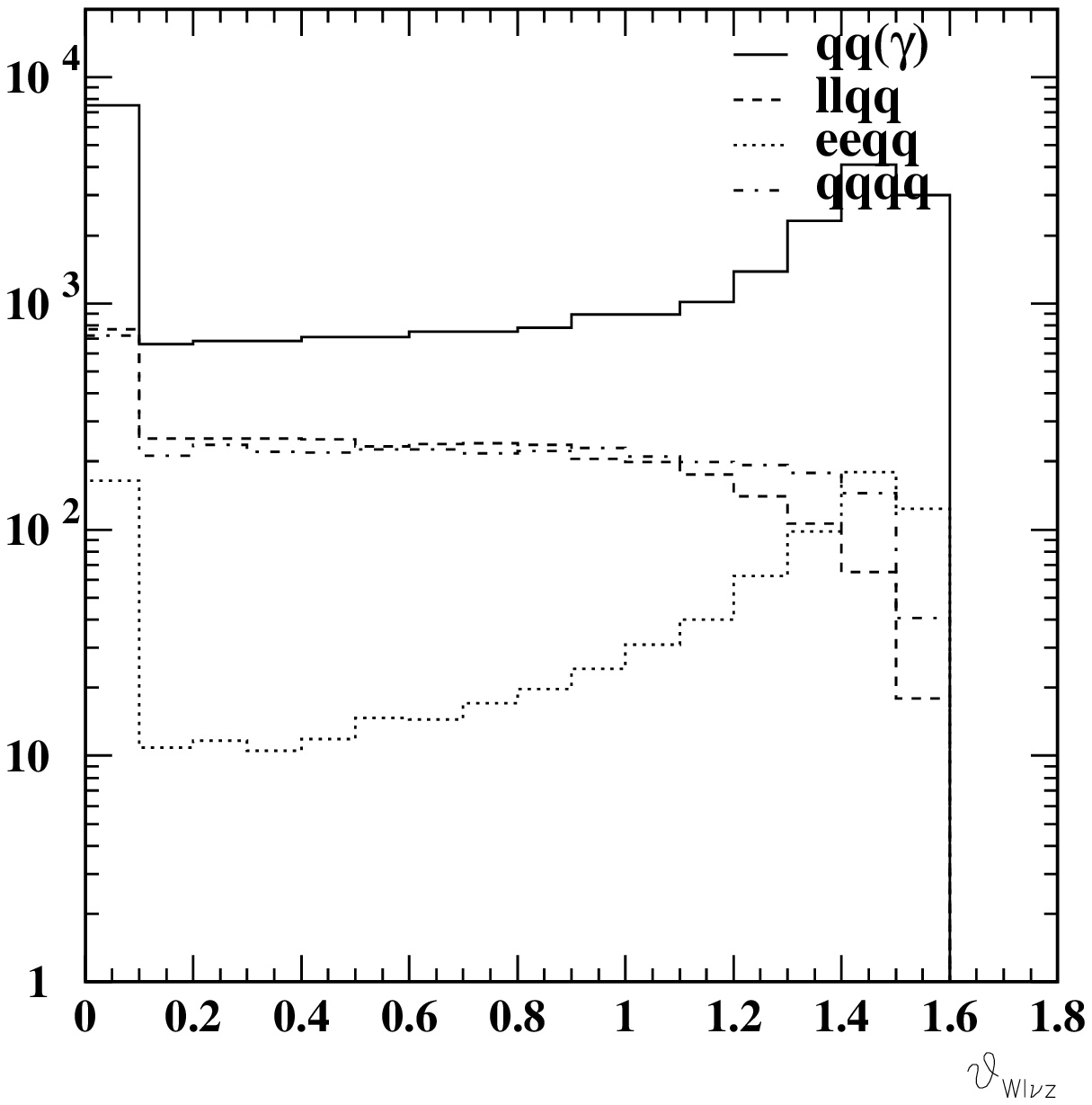}
\includegraphics[width=7.0cm]{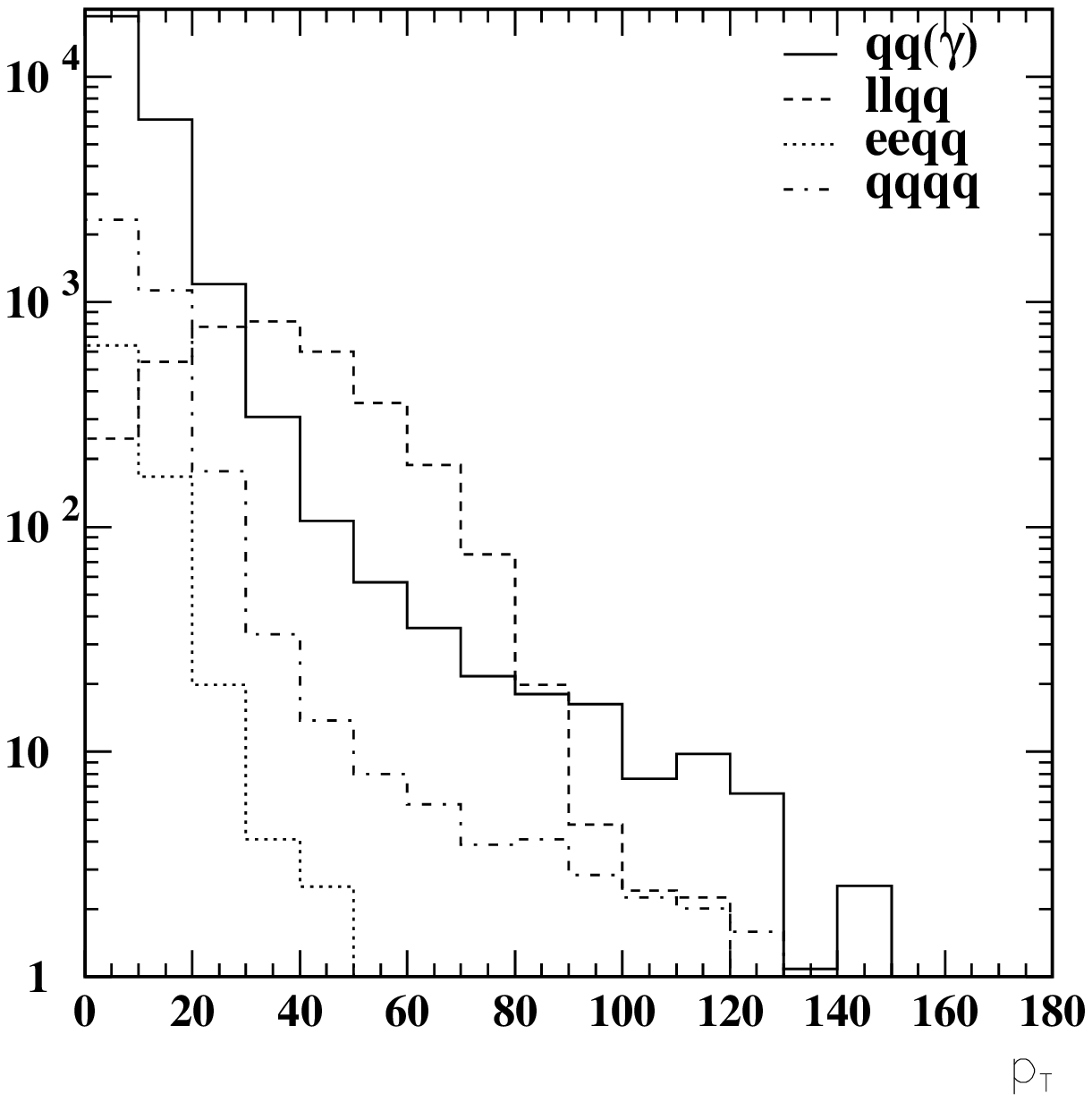}
\caption{Comparison of variables relevant for exotic neutral lepton
	 production for different backgrounds}
\label{fig:BgVarN}
\end{center} \end{figure}




\section{Complete Analysis of Heavy Neutrinos}
Heavy neutrinos have been extensively studied for $M_N =$ 100, 110, 120, 130,
140, 150, 160, 170 GeV. Higher masses have not been studied due to the
difficulties present already at 170 GeV (see table \ref{tab:Ev170}). It is noteworthy that the masses are
high enough to avoid the problems with initial state radiation, as has been
discussed in section \ref{Sec:Assumptions}. Each mass and energy region has been
individually treated and optimized.

The analysis has been divided into three energy regions in order to optimize
the different cuts more effectively. The first region is 180-190 GeV, containing
244.5 pb$^{-1}$, the second region is 191-200 GeV, 
containing 174.04 pb$^{-1}$, and the last region is 201-209 GeV. containing
244.92 pb$^{-1}$.

\subsection{Cuts}
\label{sec:MC_Cuts}
There are many different methods to optimize the cuts. It matters in what
order the cuts are applied. It also matters whether all optimizing
cuts are applied at the same time, or if they are applied one by one, with an
inspection of the results before moving on to the next cut. Different methods
give different results although the conclusions have proved to be basically the
same, independently of the method used. Apart from the method described below,
one could start optimizing from the cuts of a higher or lower mass or energy
region. However, this would induce a correlation between different masses and
energy regions, which is not necessarily real.

Due to the kinematics of the different masses it is not possible to make good
mass independent cuts. However, it is possible to make cuts that are 
functions of the mass (see tables \ref{tab:Cuts180}--\ref{tab:Cuts200}).
For an explanation of the different variables, cf. section \ref{sec:DescrVar}.



The optimization of cuts has been made in three steps, starting from
some very loose general cuts:
$5<E_\ell<100$,
$10<E_\nu<100$,
$|\cos\theta_{mis}|<0.95$,
$5<M_{\ell\nu}<70$.

The first step is a visual estimation of appropriate cuts for the variables
$M_N$, $E_\ell$, $E_\nu$, $|\cos\theta_{mis}|$, $\theta_{Wl}$,
$\theta_{Wl\nu}(\hat z)$, and $M_{\ell\nu}$.

At this point, every cut is slightly altered ($\pm 5$ GeV or $\pm$0.1 radians), and
those changes are kept, which improve $S/\sqrt{B}$, without too much decreasing
the signal. The alteration is then repeated, until no more significant
improvements are achieved. This step is done automatically, and because of this
a rather high improvement (in the order of several percent) is required to
continue. Note that in some cases this optimization implies a broadening of the
general cuts.

The last step is a manual fine-tuning of the cuts with the aim of having
a high $S/\sqrt{B}$, but still a with signal larger than one. In the evaluation,
$S^{3/2}/\sqrt{B}$ is also considered in order to promote high signal cuts.
Finally, we require that the decay lepton from the $N$ should be an electron,
which cuts none of the real signal (supposing that the heavy lepton is of
E-type without any mixing to the $\mu$ or $\tau$) but should cut more than half
of the background.

\subsubsection{Statistical Error}
The last step in the optimization procedure is somewhat arbitrary in the
choice between keeping a lot of signal events and obtaining a high
$S/\sqrt{B}$. This is remedied by evaluating the results as they change by
varying the cuts ($\pm 2.5$ GeV or $\pm$0.05 radians) and observe the changes.
More about this in the next chapter.

\subsubsection{Comments}
The $E_\ell$ is very low for low masses, while $E_\nu$ is high for low masses.
$\theta_{Wl}$ also increases with the mass. Generally, the widths of the
distributions grow with the mass.
In the end of the fine-tuning, the cut on the $\theta_{Nl}$ turns out to be
superfluous. 
The cuts are summarized in tables \ref{tab:Cuts180}--\ref{tab:Cuts200}.

The principal variables and their cuts are plotted below. Two cases have been
chosen, $M_N=100$ GeV and $M_N=170$ GeV, both for the center of mass energy
range of 191-200 GeV. Everything outside the region marked by the arrows is
cut away. As the cuts depend on order, each cut is shown with
the other cuts already applied.
Note that the binning sometimes is too small,
giving the illusion of an excess in the data. Furthermore, the Monte Carlo
simulation are often less than one while data obviously is integral. This
will also produce the impression that the data does not follow the simulated
background.

For each background, the number of remaining events after each consecutive cut
is presented in tables \ref{tab:VarVar100} and \ref{tab:VarVar170}. This is
useful to see how much a specific background is affected be a certain cut.
This could eventually be used to find a different way of optimizing, targeting
the most dominant background.

\begin{table}[here!]
\begin{tabular}{llllllll}
\hline
	$M_N$ &
	$|\cos\theta_{\textrm{vis}}|$ $\in$& 
	$M_{\ell\nu}$ $\in$&
	$E_\nu$ $\in$&
	$E_\ell$ $\in$&
	$M_N$ $\in$&
	$\theta_{W\ell}$ $\in$&
	$\theta_{W\ell\nu}(\hat z)$$\in$ \\

\hline
 100  & $      [.15,.9]$ &$       [15,73]$ &$   [60,\infty)$ &$        [8,95]$ &
        $      [85,113]$ &$       [0,.75]$ &$     [.05,1.4]$ \\
 110  & $     [.15,.95]$ &$        [8,60]$ &$       [55,80]$ &$       [15,30]$ &
        $     [103,125]$ &$     [.15,1.4]$ &$      [0,1.55]$ \\
 120  & $       [0,.95]$ &$        [5,60]$ &$      [48,100]$ &$       [20,30]$ &
        $     [108,138]$ &$     [.5,1.35]$ &                 \\
 130  & $       [0,.95]$ &$        [5,60]$ &$      [43,100]$ &$      [28,100]$ &
        $     [120,150]$ &                 &$       [0,1.5]$ \\
 140  & $       [0,.95]$ &$        [8,63]$ &$       [38,65]$ &$       [28,50]$ &
        $     [133,158]$ &$     [.55,1.5]$ &$      [0,1.35]$ \\
 150  & $       [.05,1]$ &$       [13,63]$ &$       [28,55]$ &$       [40,55]$ &
        $     [143,170]$ &$     [.95,1.4]$ &$      [0,1.35]$ \\
 160  & $     [.05,.95]$ &$       [10,55]$ &$       [20,45]$ &$       [50,70]$ &
        $     [150,183]$ &$     [1.1,1.5]$ &$       [0,1.2]$ \\
 170  & $         [0,1]$ &$       [18,65]$ &$       [10,30]$ &$       [55,78]$ &
        $     [158,188]$ &$    [1.2,1.45]$ &$       [0,1.3]$ \\
\hline

\end{tabular}
\caption{Summary of cuts for the different masses $M_N$ for the energy region
	180-190 GeV. Energies and masses
	are given in GeV, angles in radians. For explanations of variables,
	see section \ref{sec:DescrVar}.
	For readability, 2.5 and 7.5 are displayed as 3 and 8,
	respectively.}
\label{tab:Cuts180}
\end{table}

\begin{table}[here!]
\begin{tabular}{llllllll}
\hline
	$M_N$ &
	$|\cos\theta_{\textrm{vis}}|$ $\in$&
	$M_{\ell\nu}$ $\in$&
	$E_\nu$ $\in$&
	$E_\ell$ $\in$&
	$M_N$ $\in$&
	$\theta_{W\ell}$ $\in$&
	$\theta_{W\ell\nu}(\hat z)$$\in$ \\
\hline

 100  & $     [.15,.95]$ &$       [23,78]$ &$       [60,90]$ &$       [10,23]$ &
        $      [88,115]$ &$        [0,.8]$ &$       [0,1.6]$ \\
 110  & $     [.05,.95]$ &$        [8,70]$ &$       [60,85]$ &$       [13,28]$ &
        $     [103,128]$ &$      [0,1.35]$ &$      [0,1.45]$ \\
 120  & $       [0,.95]$ &$       [10,63]$ &$       [50,80]$ &$       [20,38]$ &
        $     [105,133]$ &$     [.2,1.35]$ &$     [.05,1.4]$ \\
 130  & $       [0,.95]$ &$       [13,73]$ &$       [50,78]$ &$       [25,40]$ &
        $     [120,150]$ &$    [.55,1.35]$ &$     [.05,1.2]$ \\
 140  & $         [0,1]$ &$       [10,68]$ &$       [38,70]$ &$       [30,48]$ &
        $     [118,155]$ &$      [.7,1.4]$ &$       [0,1.5]$ \\
 150  & $         [0,1]$ &$        [8,85]$ &$       [35,58]$ &$       [38,53]$ &
        $     [138,168]$ &$     [.85,1.5]$ &$       [0,1.4]$ \\
 160  & $       [0,.95]$ &$       [13,63]$ &$       [25,48]$ &$       [38,65]$ &
        $     [150,180]$ &$    [.95,1.45]$ &$      [0,1.05]$ \\
 170  & $         [0,1]$ &$       [23,73]$ &$       [15,40]$ &$       [53,78]$ &
        $     [160,193]$ &$    [1.05,1.4]$ &$       [0,.95]$ \\
\hline

\end{tabular}
\caption{Summary of cuts for the different masses $M_N$ for the energy region
	191-200 GeV. Energies and masses
	are given in GeV, angles in radians.}
\label{tab:Cuts190}
\end{table}

\begin{table}[here!]
\begin{tabular}{llllllll}
\hline
	$M_N$ &
	$|\cos\theta_{\textrm{vis}}|$ $\in$&
	$M_{\ell\nu}$ $\in$&
	$E_\nu$ $\in$&
	$E_\ell$ $\in$&
	$M_N$ $\in$&
	$\theta_{W\ell}$ $\in$&
	$\theta_{W\ell\nu}(\hat z)$ $\in$\\
\hline

 100  & $      [.1,.95]$ &$       [20,75]$ &$      [65,105]$ &$       [10,23]$ &
        $      [83,115]$ &$        [0,.9]$ &$      [.1,1.5]$ \\
 110  & $       [0,.95]$ &$       [10,73]$ &$       [63,95]$ &$       [13,28]$ &
        $     [100,128]$ &$      [0,1.15]$ &$    [.05,1.45]$ \\
 120  & $      [.1,.95]$ &$       [10,73]$ &$       [60,95]$ &$       [20,33]$ &
        $     [108,135]$ &$     [.2,1.35]$ &$    [.05,1.45]$ \\
 130  & $        [.1,1]$ &$       [20,73]$ &$       [55,75]$ &$       [28,40]$ &
        $     [115,145]$ &$     [.45,1.3]$ &$       [0,1.4]$ \\
 140  & $      [.1,.95]$ &$       [18,75]$ &$       [48,68]$ &$       [30,45]$ &
        $     [128,160]$ &$      [.7,1.3]$ &$       [0,1.5]$ \\
 150  & $       [.45,1]$ &$      [25,123]$ &$       [43,58]$ &$       [38,73]$ &
        $     [135,163]$ &$     [.55,1.3]$ &$         [0,1]$ \\
 160  & $         [0,1]$ &$       [15,93]$ &$       [35,58]$ &$       [33,65]$ &
        $     [150,180]$ &$     [.85,1.4]$ &$       [0,1.1]$ \\
 170  & $       [.05,1]$ &$       [20,98]$ &$       [20,45]$ &$       [48,70]$ &
        $     [150,188]$ &$       [1,1.4]$ &$       [0,.95]$ \\
\hline

\end{tabular}
\caption{Summary of cuts for the different masses $M_N$ for the energy region
	201-210 GeV. Energies and masses
	are given in GeV, angles in radians.}
\label{tab:Cuts200}
\end{table}

\begin{figure}[here!] \begin{center}
\includegraphics[width=7.0cm]{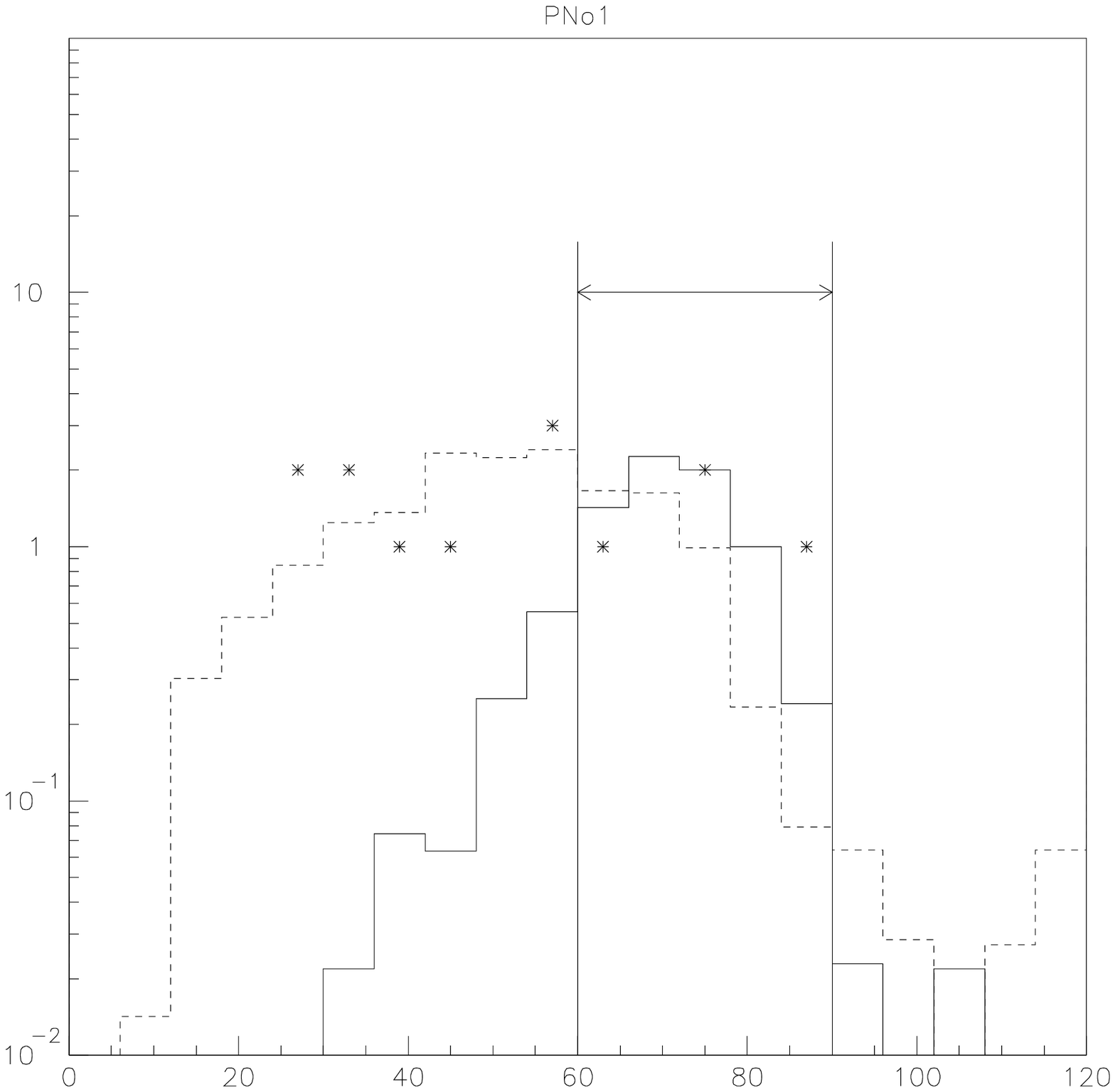}
\includegraphics[width=7.0cm]{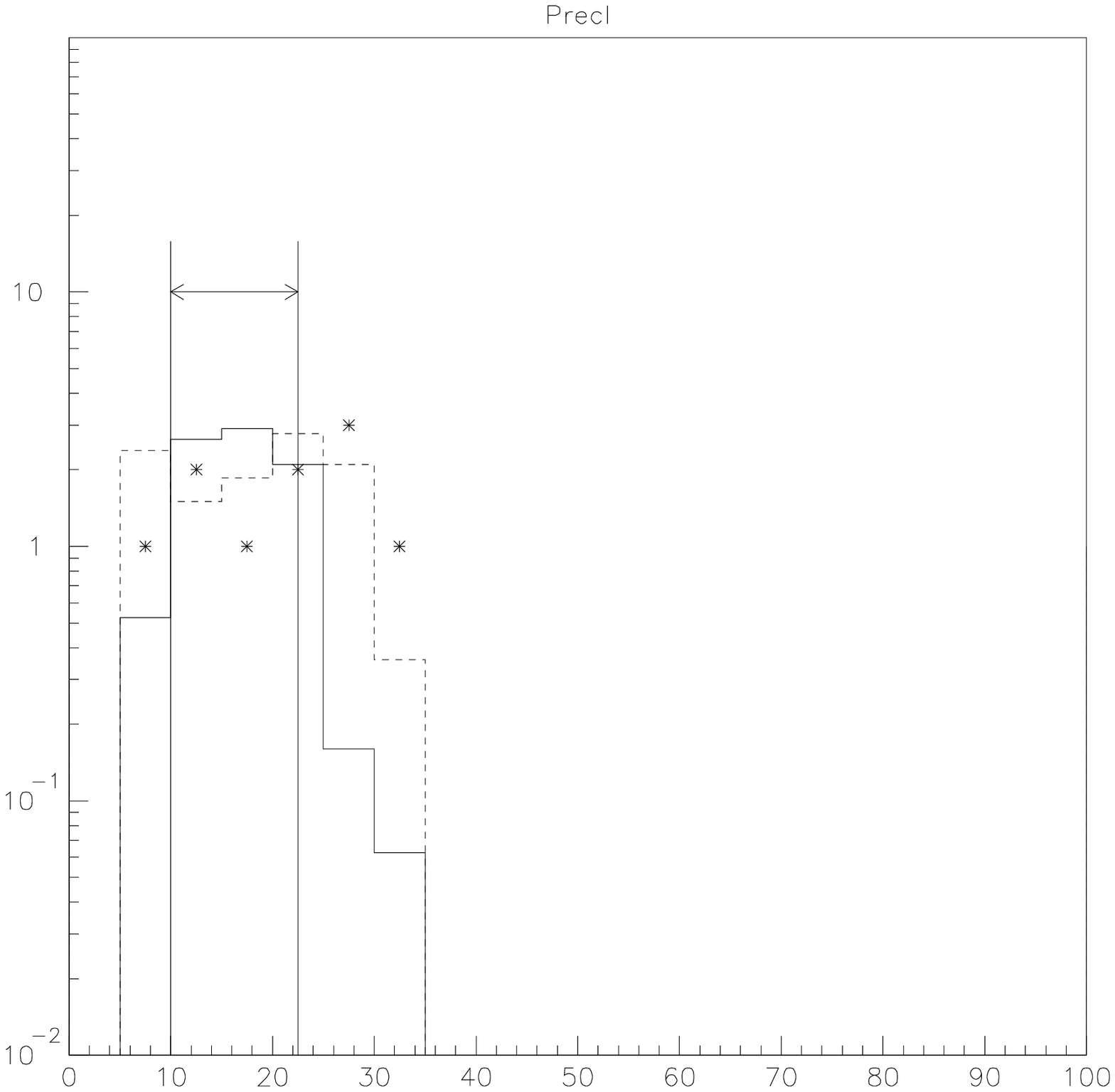}
\includegraphics[width=7.0cm]{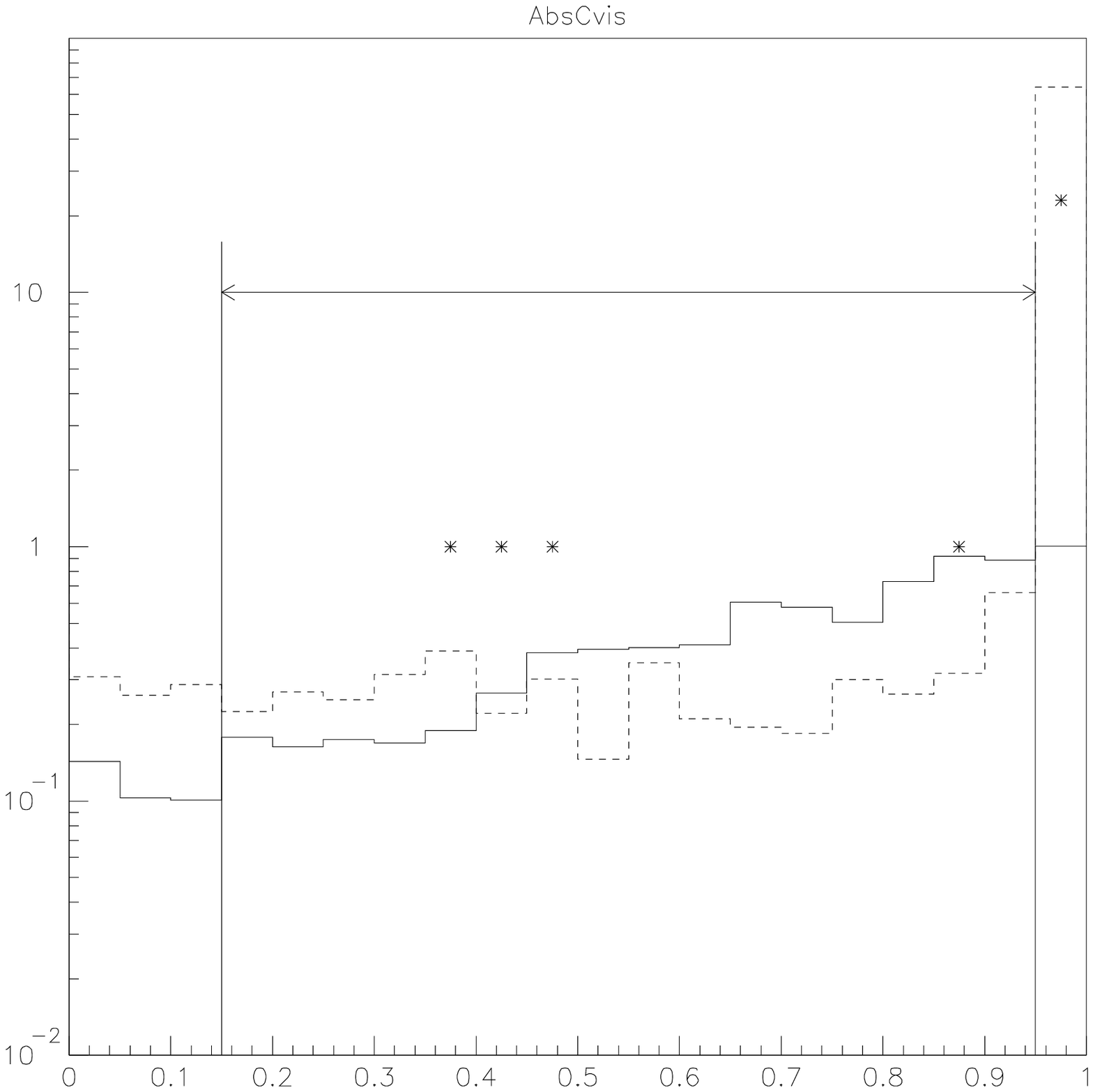}
\includegraphics[width=7.0cm]{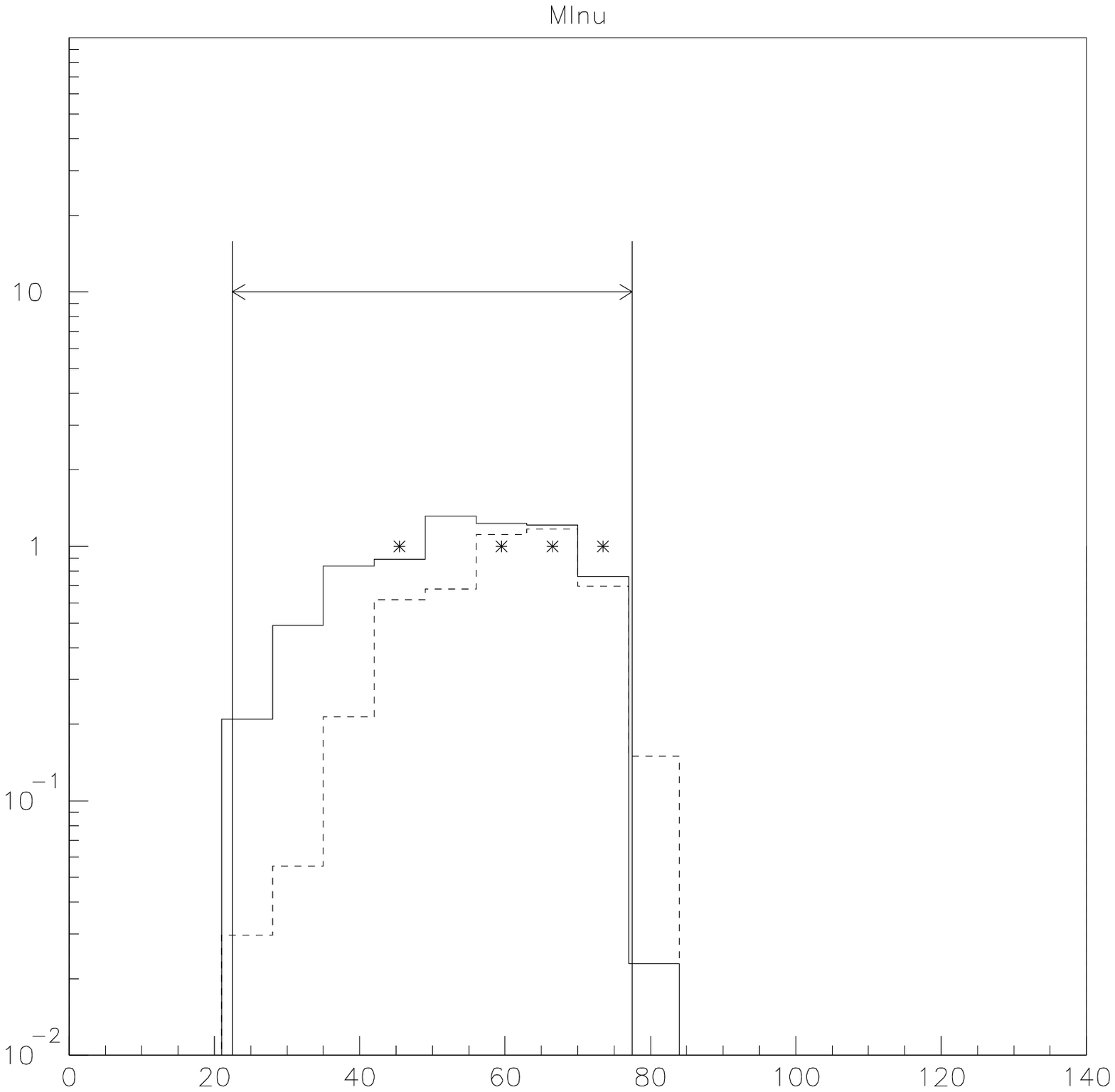}
\caption{Cuts for $M_N=$ 100 GeV in the energy region 191-200 GeV.
	The background (dotted lines) has been plotted for comparison.
	Data (stars) have been plotted for reference but have not been used
	in any way during the optimization.}
\label{fig:100PNo1} \end{center} \end{figure}

\begin{figure} \begin{center}
\includegraphics[width=7.0cm]{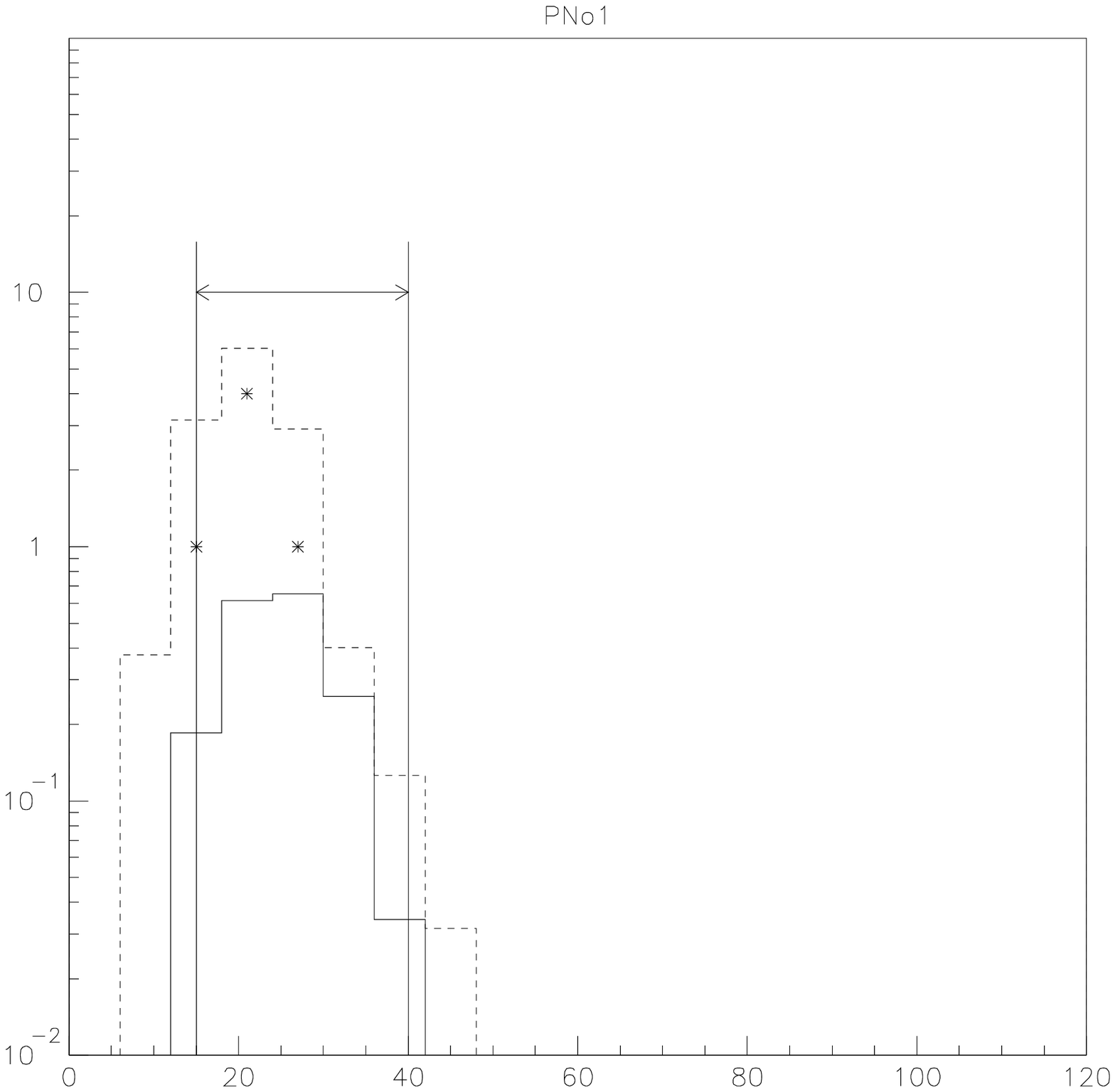}
\includegraphics[width=7.0cm]{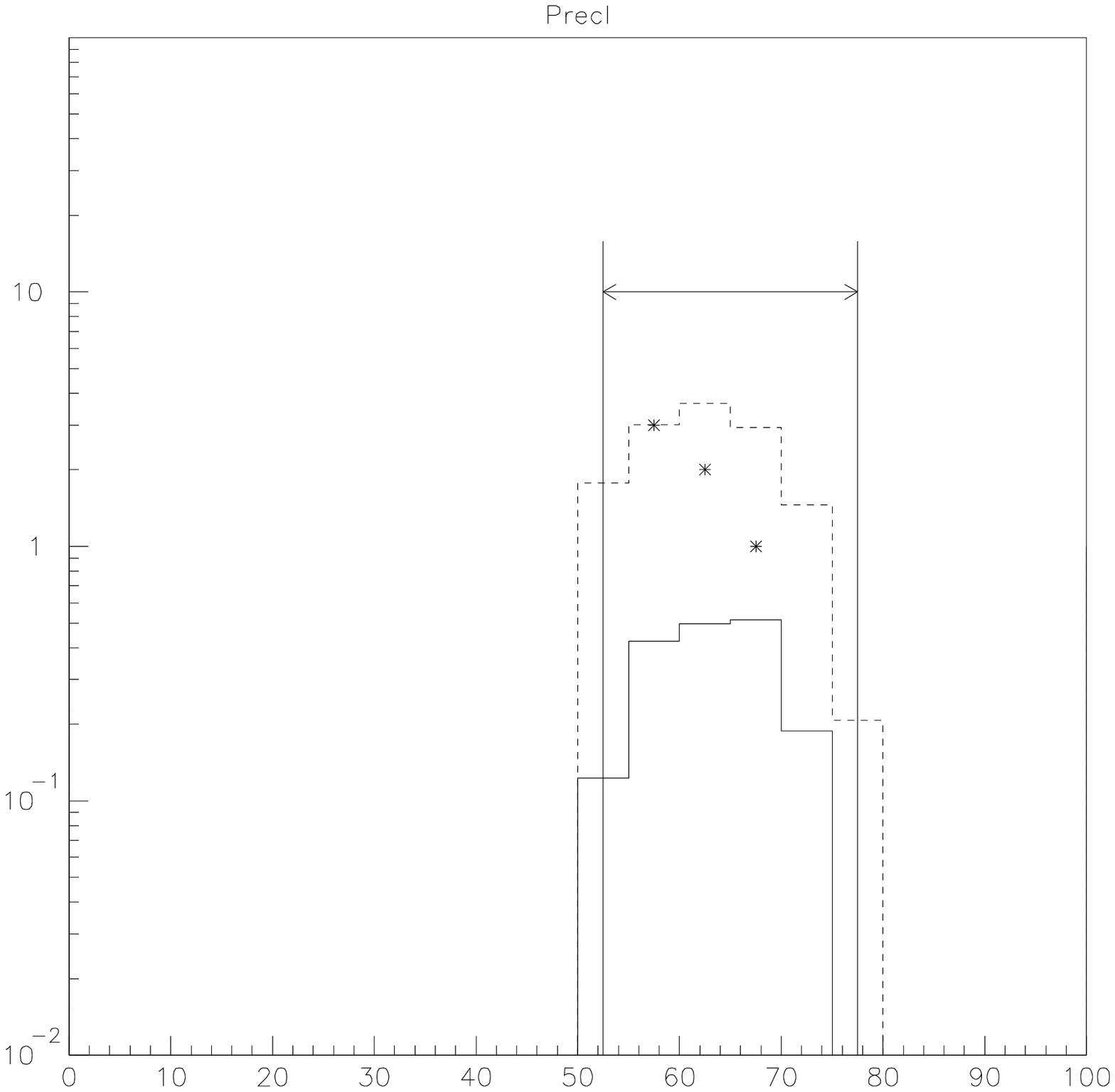}
\includegraphics[width=7.0cm]{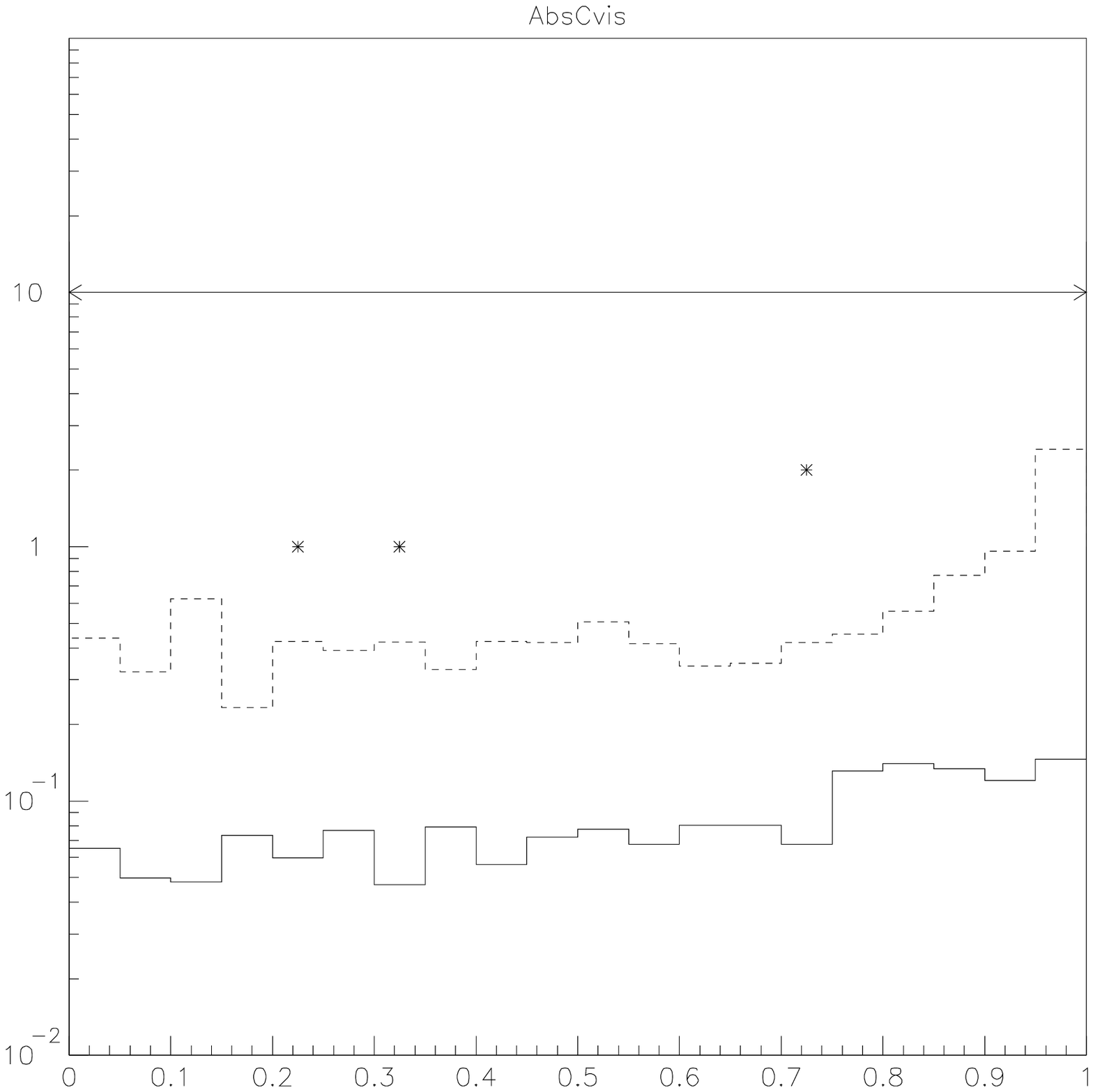}
\includegraphics[width=7.0cm]{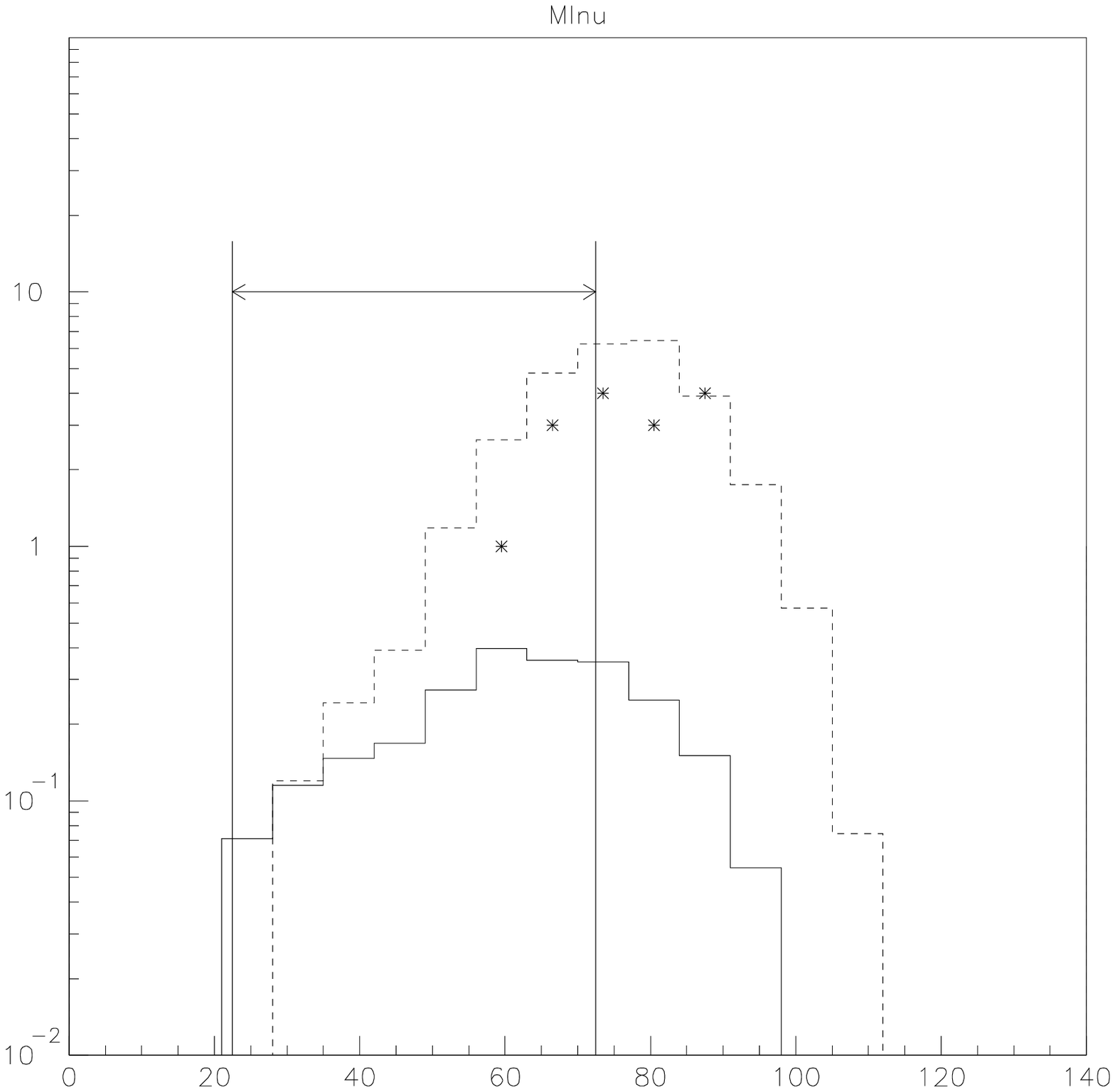}
\caption{Cuts for $M_N=$ 170 GeV in the energy region 191-200 GeV.
	The background (dotted lines) has been plotted for comparison.
	Data (stars) have been plotted for reference but have not been used
	in any way during the optimization.}
\label{fig:170PNo1} \end{center} \end{figure}

\begin{table}[here!]
\begin{tabular}{lllllll}
Cut                       &            llqq  &      qq$\gamma$  &            eeqq  &            qqqq  &    ee$\tau\tau$  &  $\gamma^*\gamma$qq  \\
$E_\nu$                   &          788.30  &          1546.9  &          124.10  &          197.50  &            1.22  &           25.23  \\
$E_\ell$                  &          126.80  &           354.5  &           32.57  &            3.02  &            0.10  &            4.93  \\
$|\cos\theta|$            &           22.52  &          126.43  &           12.58  &            1.22  &            0.03  &            0.73  \\
$M_{\ell\nu}$             &           17.54  &            6.59  &            0.25  &            0.32  &            0.01  &               0  \\
$\theta_{Wl\nu}(\hat z)$  &           14.22  &            4.44  &            0.22  &            0.19  &            0.01  &               0  \\
$\theta_{Wl}$             &           14.22  &            4.44  &            0.22  &            0.19  &            0.01  &               0  \\
$M_N$                     &            9.40  &            3.13  &            0.19  &            0.02  &               0  &               0  \\
Type=E                    &            4.08  &            0.24  &            0.04  &               0  &               0  &               0  \\

\end{tabular}
\caption{ Number of remaining events for different backgrounds before each cut
	for $M_N=100$ GeV in the energy region 191-200 GeV.}
\label{tab:VarVar100}
\end{table}

\begin{table}[here!]
\begin{tabular}{lllllll}
Cut                       &            llqq  &      qq$\gamma$  &            eeqq  &            qqqq  &    ee$\tau\tau$  &  $\gamma^*\gamma$qq  \\
$E_\nu$                   &          788.30  &          1546.9  &          124.10  &          197.50  &            1.22  &           25.23  \\
$E_\ell$                  &          321.30  &           498.7  &           34.33  &           79.34  &            0.53  &            8.09  \\
$|\cos\theta|$            &          166.50  &            7.29  &            6.32  &            0.17  &            0.08  &            1.50  \\
$M_{\ell\nu}$             &          166.50  &            7.29  &            6.32  &            0.17  &            0.08  &            1.50  \\
$\theta_{Wl\nu}(\hat z)$  &           45.22  &            3.45  &            4.12  &            0.01  &            0.07  &            0.18  \\
$\theta_{Wl}$             &           34.11  &             2.8  &            3.34  &            0.01  &            0.07  &            0.18  \\
$M_N$                     &           16.52  &            1.09  &            1.72  &               0  &            0.01  &            0.18  \\
Type=E                    &            7.15  &            1.08  &            1.72  &               0  &            0.01  &            0.18  \\

\end{tabular}
\caption{ Number of remaining events for different backgrounds before each cut
	for $M_N=170$ GeV in the energy region 191-200 GeV.}
\label{tab:VarVar170}
\end{table}

\section{Discussion of Heavy Charged Leptons}
\label{sec:HCL}
The process we are studying is shown in figure \ref{fig:Single}(a). If the
heavy lepton mass is not close to the centre of mass energy ($\sim 200$ GeV)
there will be much energy left for the recoiling lepton $\ell$. This energy
will be mostly deposited in the beam direction, due to spin conservation,
making the lepton escape the detector. Consequently, both the lepton and the
high-energetic neutrino will be missing, thus making the reconstruction very
difficult.

\begin{figure}\begin{center}
\includegraphics[width=8cm]{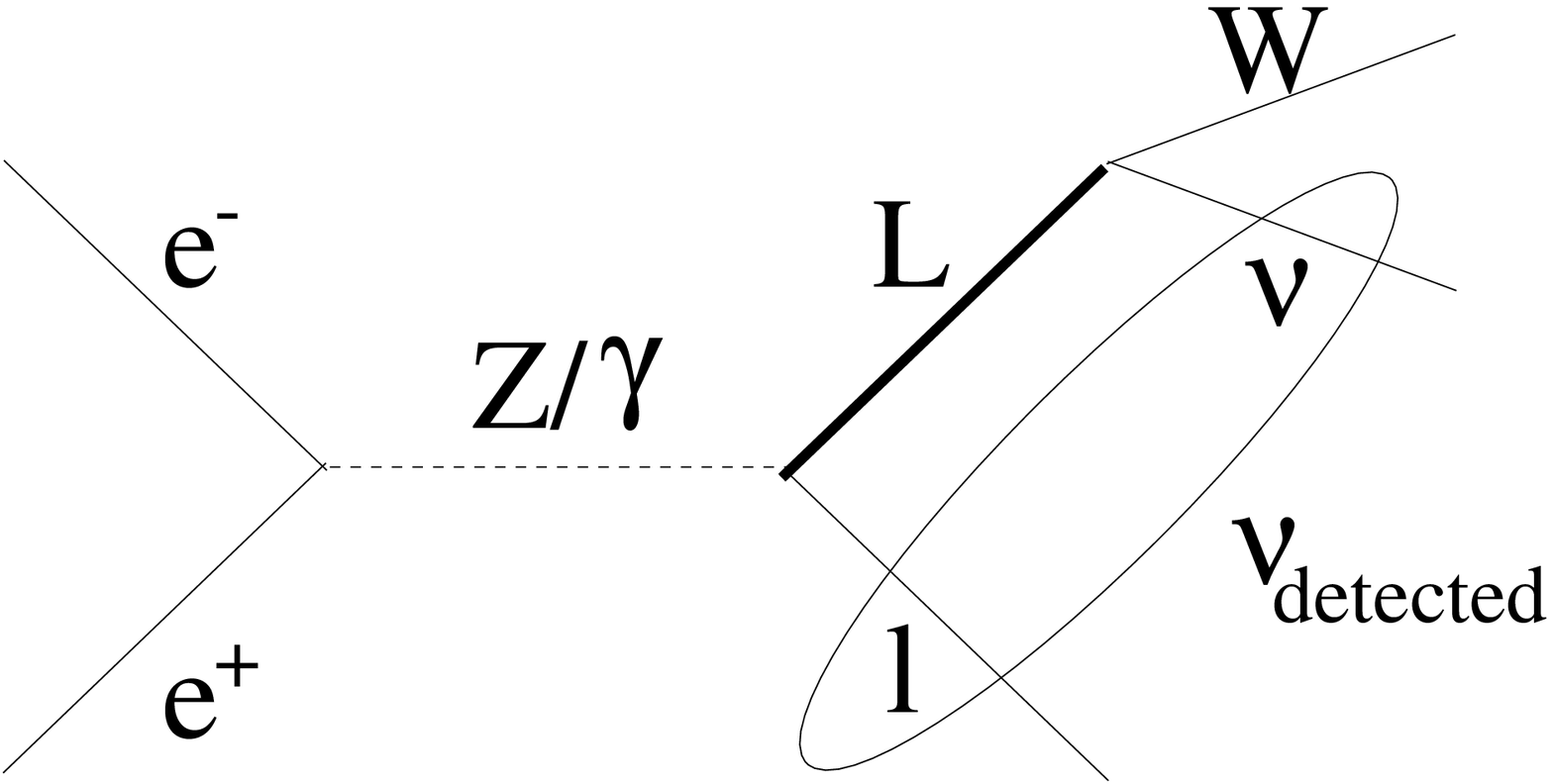}
\caption{When the primary lepton moves very close to the beam pipe, it
	escapes detection and is confounded with the neutrino.}
\label{fig:Missing}
\end{center}\end{figure}

\begin{figure}\begin{center}
\includegraphics[width=8cm]{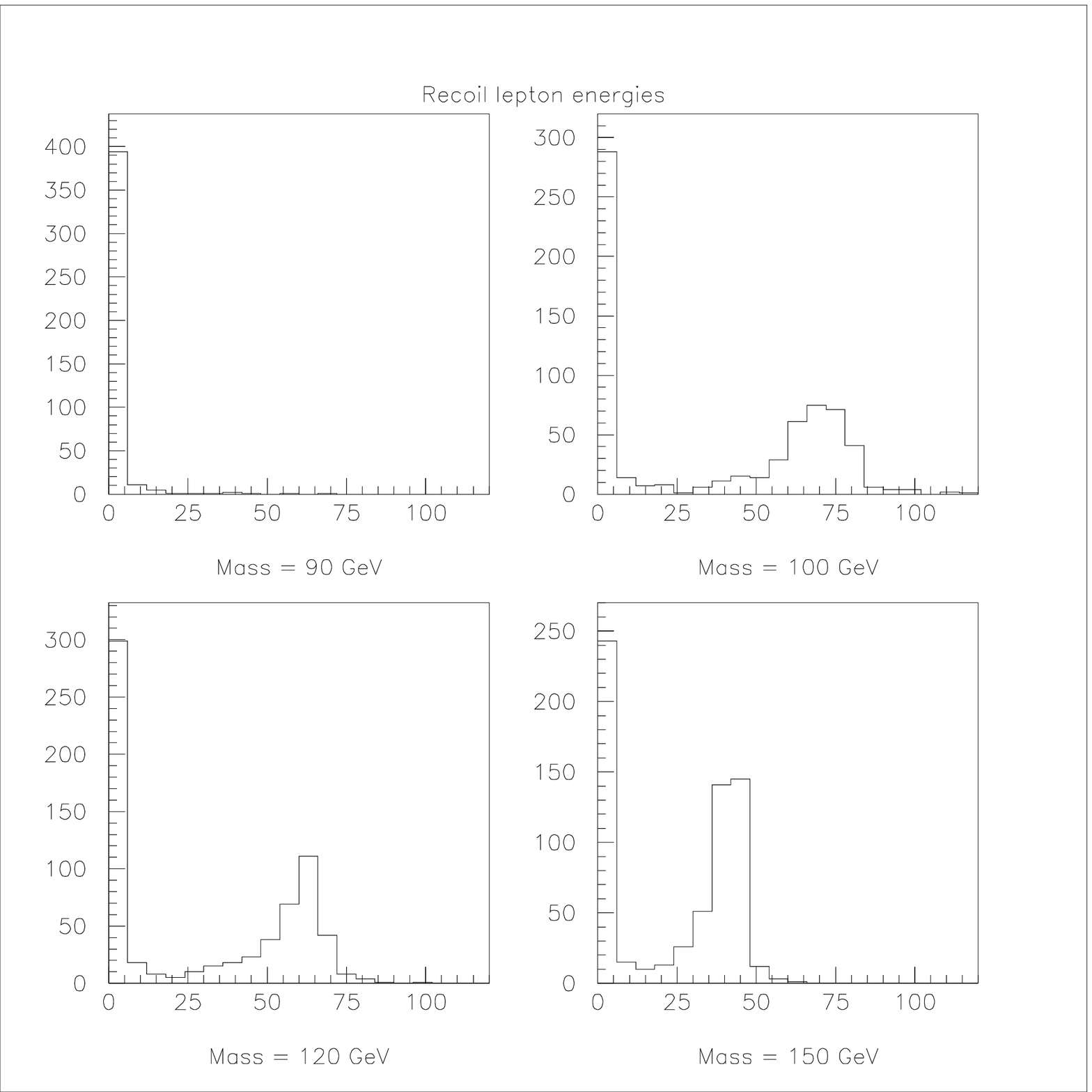} 
\caption{Energies of detected recoil leptons for $M_L=$90, 100, 120, 150 GeV.}
\label{fig:HCL_EL}
\end{center}\end{figure}

\begin{figure}\begin{center}
\includegraphics[width=8cm]{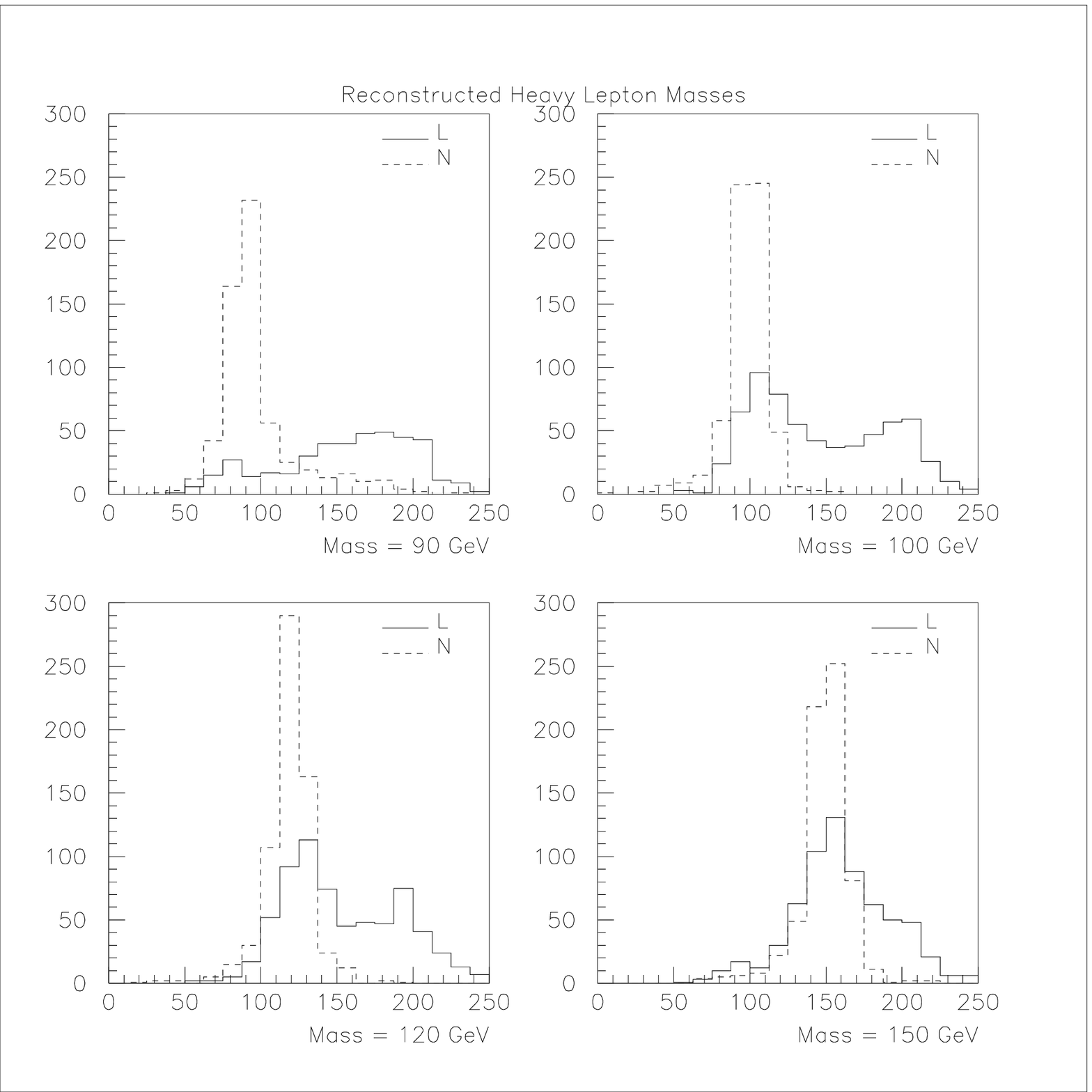}
\caption{
	Pre-detector evaluation of signals.
	Reconstructed heavy charged lepton ($L$) mass for
	$M=$90, 100, 120, 150 GeV.
	The corresponding heavy neutral lepton ($N$) reconstructed mass
	in included for reference (dotted lines).
	We note that the lower energy has a very bad resolution
	in the charged lepton case.}
\label{fig:HCL_Mass}
\end{center}\end{figure}

Hence, for heavy leptons, only really high masses are easily detectable. This is
illustrated in figure (\ref{fig:HCL_EL}). The problem with the higher masses is
that the cross sections are low (cf. figure \ref{fig:HCL_Mass}). The low
electron energy also causes problems due to the background (see figure
\ref{fig:BgVarN}).

There are two possible approximations which might help to identify and
separate the lepton and the neutrino. In the first approximation, we suppose
that the $p_T(\nu)>>p_T(\ell)\approx 0$, which could be a good estimate for
high masses. In the second approximation we assume that
$p_z(\ell)>>p_z(\nu)\approx 0$, which is a good estimate for low masses, when
the lepton has much energy and the neutrino has little. Neither of these
approximations has been tested yet.


\chapter{Data Analysis and Results}
This chapter contains analysis of the data and the results in the form of
limits on the \gloss{mixing parameter} $\zeta$. The number of measured data-events is
referred to as data, $D$, the simulated background events as $B$ and the
simulated signal $S$. Furthermore, the analysis has been grouped into three
energy regions, 183-189 GeV, 192-200 GeV and 202-209 GeV as discussed in
chapter \ref{Ch:MC}.



\section{Heavy Neutral Lepton}

Limits have been set on the mixing parameter, $\zeta$, rather than on the
production cross sections because the mixing parameter is constant for
all mass regions. The limits on the
cross sections will be different for every mass and center of mass energy.
These limits can be obtained by the use of the values in table \ref{tab:Xsec} along with the
limits on $\zeta^2$ in tables \ref{tab:Dat180} to \ref{tab:Dat200} below.
The cross section, $\sigma$ is proportional to the square of the mixing parameter:
$\sigma \propto \zeta^2$.

\begin{itemize}
\item
For more than 15 background events or more than 20 data events, the
limit is extracted with the assumption of gaussian statistics.
We look at a single-sided gaussian distribution so that a 95\% confidence
limit corresponds to $1.64\bar\sigma$ (like a 90\% double-sided limit), where
the standard deviation, $\bar\sigma = \sqrt{B+S_{95}}$, $B$ is the background, $S_{95}$ is the signal with a mixing required
	to obtain this confidence level (see figure \ref{fig:Gauss}).
\begin{figure}[here!] \begin{center}
\includegraphics[width=14.0cm]{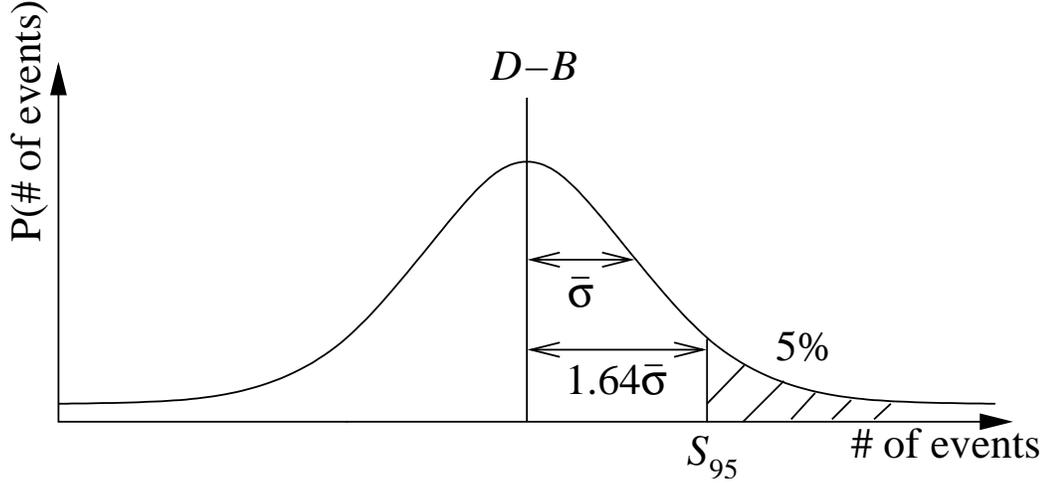}
\caption{Illustration of the statistic distribution}
\label{fig:Gauss}
\end{center} \end{figure}

\noindent
Thus, with 95\% C.L., we solve the
equation
\begin{equation}
	1.64\bar\sigma = S_{95} - (D-B),
\end{equation}
for $S_{95}$ ($D$ is the measured number of data-events).
The number of data-events is proportional to the cross section which is
proportional to the squared mixing angle, hence 
we obtain an upper limit on the mixing
angle with 95\% confidence level: $\zeta^2(95\%)=S_{95}*\zeta^2/S$ where $S$ is the simulated signal and
$\zeta^2$ is the square of the nominal mixing (see section \ref{sec:MCHL}).

\item
For small data and/or backgrounds the gaussian approximation is no longer
valid, and the method in \cite{Stat} has been used to determine the limits. In
the article, there are tables for different confidence limits, data, and
expected background.
\end{itemize}

The number of signal, background, and data events, preceeding each consecutive
cut, 
is shown in tables \ref{tab:Ev100} and \ref{tab:Ev170}. The tables show
two representative cases, $M_N=100$ GeV and $M_N=170$ GeV, both in the
191-200 GeV energy region.

In the $M_N=100$ GeV case (table \ref{tab:Ev100}), we note that the cut
on the polar angle seems to reduce the $S/\sqrt{B}$, but this is because the
result is order-dependent (as discussed in section \ref{sec:MC_Cuts}).
The final result (Type=E) has a better $S/\sqrt{B}$ with the $|\cos \theta|$ cut than without it. In this case, the cut on
$\theta_{Wl\nu}(\hat z)$ proves to be useless, which can easily be understood
in the light of table \ref{tab:Cuts190} where we see that there is no cut
on this variable for this mass. The background corresponds pretty well to the
data all along.

In the $M_N=170$ GeV case (table \ref{tab:Ev170}), we note that the $S/\sqrt{B}$ never even gets close
to that of $M_N=100$ GeV. One of the reasons for this is the lower cross section
(because of its mass), another reason is the persistence of the llqq background
(see table \ref{tab:VarVar170}) caused by the similarity of the lepton and
neutrino energy for the signal and the llqq background (see \ref{fig:170PNo1} and \ref{fig:BgVar}). In this case, the $M_{\ell\nu}$ cut turns out to be
unecessary, though it is still included for completeness.

\begin{table}[here!]
\begin{tabular}{lllllllllll}
\hline
             Cut  &    $S/\sqrt{B}$  &            Data  &              Bg  &          Signal  \\
\hline
         $E_\nu$  &            0.46  &            2674  &          2683.0  &            24.1  \\
        $E_\ell$  &            0.83  &             518  &           522.0  &            19.0  \\
  $|\cos\theta|$  &            0.76  &             146  &           163.5  &             9.7  \\
   $M_{\ell\nu}$  &            1.64  &              22  &            24.7  &             8.2  \\
  $\theta_{Wl\nu}(\hat z)$  &            1.81  &              15  &            19.1  &             7.9  \\
   $\theta_{Wl}$  &            1.81  &              15  &            19.1  &             7.9  \\
           $M_N$  &            2.08  &               9  &            12.8  &             7.4  \\
          Type=E  &            3.33  &               4  &             4.4  &             6.9  \\
\hline

\end{tabular}
\caption{
	Number of remaining events before each cut for $M_N=100$ GeV in the
	energy region 191-200 GeV.
}
\label{tab:Ev100}
\end{table}

\begin{table}[here!]
\begin{tabular}{lllllllllll}
\hline
             Cut  &    $S/\sqrt{B}$  &            Data  &              Bg  &          Signal  \\
\hline
         $E_\nu$  &            0.11  &            2674  &          2683.0  &             5.5  \\
        $E_\ell$  &            0.15  &             969  &           942.3  &             4.7  \\
  $|\cos\theta|$  &            0.26  &             150  &           181.9  &             3.5  \\
   $M_{\ell\nu}$  &            0.26  &             150  &           181.9  &             3.5  \\
  $\theta_{Wl\nu}(\hat z)$  &            0.30  &              32  &            53.1  &             2.2  \\
   $\theta_{Wl}$  &            0.32  &              24  &            40.5  &             2.0  \\
           $M_N$  &            0.38  &              13  &            19.5  &             1.7  \\
          Type=E  &            0.52  &               5  &            10.1  &             1.7  \\
\hline

\end{tabular}
\caption{
	Number of remaining events before each cut for $M_N=170$ GeV in the
	energy region 191-200 GeV.
}
\label{tab:Ev170}
\end{table}

\subsection{Lower Limits on the Mixing}
\label{Data:LowLim}
In the first attempt to analyse the data, there were considerably more data
left, after the cuts, than expected.
This
excess called for a more thorough analysis, which, however, did not confirm the
excess. There are still a couple of masses around $M_N\sim$140 GeV, which
have an excess, but considering that there are 24 results, this is not
considered significant.

Just before applying the last cut (requiring that the lepton should be an
electron) there is a noticeable excess in the data for certain points in
the $(M_N,\sqrt{\hat s})$-plane.
In these cases we can also calculate lower limits on the mixing.
In the energy region 180-190 GeV there is one lower limit with 99\% CL at $M_N=$120 GeV,
two lower limits with 90\% CL at $M_N$=100 and 130 GeV, and one lower limit with 68\% CL
at $M_N=$160 GeV. This means that the probability of {\it not} having a new particle is less
than 1\% for $(M_N,\sqrt{\hat s})=(120,180-190)$. However, the total number of events remaining
after all cuts is too small to allow any definite conclusions.
Furthermore, in the energy region 191-200 GeV, there is a lower limit with 95\%
for $M_N$=140 GeV and a lower limit with 68\% CL at $M_N$=150 GeV. Finally, in the energy region
201-208 GeV, there are two lower limit with 68\% CL at $M_N$=120 and 130 GeV.
This seems to indicate
that something special is going on at $M_N\sim$ 130 GeV. Nevertheless, it is
peculiar that this excess only appears before the final cut. After that cut, the excess
disappears and the data
corresponds well to the expected background. The signal we are
looking for requires the lepton to be of electron type. Hence, the observed excess
doesn't come from the expected process. However, if the lepton is of muon type it could
possibly give this result. This is definitely something that is
worth to investigate further.

Nevertheless, the excess is
not detected in all the energy regions, thus weakening the probability of the
excess to be a real signal.

\subsection{Upper Mixing Limits}
The upper limits on the mixing between ordinary and heavy leptons for each
energy region is presented in tables \ref{tab:Dat180} to \ref{tab:Dat200}.
We note that in most cases, the data and the expected background correspond
very well.

Furthermore, table \ref{tab:DatSum} shows the combination of the three energy
regions. The combination has been obtained like
\[
1/\zeta^2 = \sum_{i=energy regions}1/\zeta_i^2.
\]
If $\zeta_i=0$ it has been ignored, nothing can be concluded from this.
For certain masses, the upper limits are improved with respect to the
estimated 0.005 \cite{MixLim,Mixing}. The reason for the higher limits for masses
$\sim130$ GeV is a certain excess in the data (compare with section
\ref{Data:LowLim}). It is also noteworthy that the statistics are rather low
and the results should therefore be regarded with certain caution.

\begin{table}
\begin{tabular}{lllll}
\hline
$M_N$  &   Data  &     Bg  &  Signal  &  $1000*\zeta^2(95\%)$\\
\hline
  100  &      5  &    2.7  &    7.7  &    5.49\\
  110  &      1  &    0.7  &    3.5  &    6.20\\
  120  &      0  &    0.3  &    2.9  &    0.00\\
  130  &      0  &    0.5  &    2.7  &    0.00\\
  140  &      0  &    0.6  &    2.5  &    0.00\\
  150  &      1  &    1.2  &    2.0  &    9.67\\
  160  &      1  &    0.6  &    1.6  &   13.25\\
  170  &      6  &    6.6  &    1.9  &   16.78\\
\hline

\end{tabular}
\caption{Upper limits on the mixing parameter, $\zeta^2$, with 95\% CL,
	for different masses and in the energy region 180-190 GeV.
        Note that the values given for the mixing parameter have to be divided by 1000.}
\label{tab:Dat180}
\end{table}

\begin{table}
\begin{tabular}{lllll}
\hline
$M_N$  &   Data  &     Bg  &  Signal  &  $1000*\zeta^2(95\%)$\\
\hline
  100  &      4  &    4.0  &    7.2  &    3.36\\
  110  &      4  &    3.3  &    3.5  &    9.37\\
  120  &      1  &    1.3  &    2.5  &    7.92\\
  130  &      0  &    0.2  &    1.8  &    0.00\\
  140  &      2  &    2.7  &    2.3  &    8.70\\
  150  &      2  &    1.5  &    1.4  &   16.86\\
  160  &      1  &    1.2  &    1.3  &   14.88\\
  170  &      5  &    9.3  &    1.7  &    8.83\\
\hline

\end{tabular}
\caption{Upper limits on the mixing parameter, $\zeta^2$, with 95\% CL,
	for different masses and in the energy region 191-200 GeV.
        Note that the values given for the mixing parameter have to be divided by 1000.}
\label{tab:Dat190}
\end{table}

\begin{table}
\begin{tabular}{lllll}
\hline
$M_N$  &   Data  &     Bg  &  Signal  &  $1000*\zeta^2(95\%)<$\\
\hline
  100  &      5  &    7.2  &    8.9  &    2.21\\
  110  &      5  &    7.0  &    6.0  &    3.16\\
  120  &      3  &    1.7  &    2.4  &   13.44\\
  130  &      3  &    1.3  &    2.4  &   14.69\\
  140  &      0  &    0.6  &    1.7  &    0.00\\
  150  &     20  &   17.4  &    3.6  &   15.84\\
  160  &     12  &   16.7  &    3.2  &    3.88\\
  170  &     28  &   31.7  &    3.2  &   10.04\\
\hline

\end{tabular}
\caption{Upper limits on the mixing parameter, $\zeta^2$, with 95\% CL,
	for different masses and in the energy region 201-208 GeV.
        Note that the values given for the mixing parameter have to be divided by 1000.}
\label{tab:Dat200}
\end{table}

\begin{table}
\begin{tabular}{lllllllll}
\hline
$M_N$                 &    100  &    110  &    120  &    130  &    140  &    150  &    160  &    170  \\
\hline
$1000*\zeta^2 (90\% CL)<$  &   0.79  &   1.27  &   4.12  &  12.96  &   6.97  &   3.58  &   0.84  &   2.46  \\
$1000*\zeta^2 (95\% CL)<$  &   1.07  &   1.71  &   4.98  &  14.69  &   8.70  &   4.43  &   2.50  &   3.67  \\
\hline

\end{tabular}
\caption{Combined upper limits on the mixing parameter for different masses for
	the the three energy regions. The nominal value of the
        mixing parameter is $\zeta^2=0.005$.}
\label{tab:DatSum}
\end{table}

\subsection{Systematic Errors}
There are several types of systematic errors: theoretical, simulation,
normalization and reconstruction. These errors are presented below.

\subsubsection{Theoretical}
\begin{itemize}
\item	Intergenerational mixing could exist.
\item	Extra light gauge bosons (\eg $Z'$) could affect an eventual signal.
\end{itemize}
\subsubsection{Simulation}
\begin{itemize}
\item	The simulation of the backgrounds is inexact, especially the
	hadronization.
\item	More background simulations for $E_{CMS} = 189$ GeV could help as discussed
	in section \ref{MC_Bg}.
\item	The GOPAL detector simulator is not exactly correct.
\end{itemize}
\subsubsection{Normalization}
\begin{itemize}
\item	Luminosities could be somewhat different.
\end{itemize}
\subsubsection{Reconstruction}
\begin{itemize}
\item	The reconstruction of leptons is not perfect.
\item	The use of energy regions instead of optimizing the cuts individually
	for every energy.
	This can be particularly important for high masses in the center
	of mass energy region 180-190 GeV.
\item	Problems with the detector (\eg calibration).
\end{itemize}
The number of events before cuts is very similar for the background and the
data (see table \ref{tab:Ev100}), which means that the background is well
simulated. The signal would be far too small to be seen without any cuts.

In tables \ref{tab:ChCut100} and \ref{tab:ChCut170}, the sensitivity of the
cuts is shown. The first table presents $M_N=100$ GeV and the second
$M_N=170$ GeV, both in the 191-200 GeV energy region. Every cut has been
changed individually, modifying the cut $\pm 2.5$ GeV (or $\pm 0.1$ for angles).
As we can see, there is no major change in the 
results ($S/\sqrt{B}$). We note that the $|\cos\theta|$ cut in the $M_N=100$ GeV case is
very significant. This is quite normal considering the form of the background
for this variable (see figure \ref{fig:BgVar}). The low mass does not allow
for an agressive cut on the lepton energy. This results in leptons close to
the beam-pipe, \ie leptons with $|\cos\theta|\sim 1$. However, when
$|\cos\theta|$ is very close to 1, the leptons escape the detector (see section
\ref{sec:Det}).


\begin{table}
\begin{tabular}{lllllllll}
\hline
       Variable  &   $S/\sqrt{B}$  &           Data  &             Bg  &         Signal  \\
\hline
 57.5$<$$E_\nu$$<$90  &           3.45  &              7  &            5.0  &            7.7  \\
 62.5$<$$E_\nu$$<$90  &           3.60  &              3  &            3.4  &            6.6  \\
 60$<$$E_\nu$$<$92.5  &           3.58  &              4  &            4.0  &            7.2  \\
 60$<$$E_\nu$$<$87.5  &           3.58  &              4  &            4.0  &            7.2  \\
 12.5$<$$E_\ell$$<$22.5  &           3.27  &              3  &            3.5  &            6.1  \\
 7.5$<$$E_\ell$$<$22.5  &           3.30  &              4  &            5.0  &            7.4  \\
 10$<$$E_\ell$$<$20  &           3.48  &              3  &            2.9  &            5.9  \\
 10$<$$E_\ell$$<$25  &           3.38  &              5  &            5.5  &            8.0  \\
 0.1$<$$|\cos\theta|$$<$0.95  &           3.54  &              4  &            4.3  &            7.4  \\
 0.2$<$$|\cos\theta|$$<$0.95  &           3.60  &              4  &            3.8  &            7.0  \\
 0.15$<$$|\cos\theta|$$<$1  &           1.29  &             27  &           40.5  &            8.2  \\
 0.15$<$$|\cos\theta|$$<$0.9  &           3.39  &              4  &            3.5  &            6.4  \\
 25$<$$M_{\ell\nu}$$<$77.5  &           3.54  &              4  &            4.0  &            7.1  \\
 20$<$$M_{\ell\nu}$$<$77.5  &           3.58  &              4  &            4.0  &            7.2  \\
 22.5$<$$M_{\ell\nu}$$<$75  &           3.59  &              4  &            3.9  &            7.1  \\
 22.5$<$$M_{\ell\nu}$$<$80  &           3.56  &              4  &            4.1  &            7.2  \\
 0.05$<$$\theta_{Wl}$$<$0.8  &           3.31  &              4  &            3.9  &            6.6  \\
 0$<$$\theta_{Wl}$$<$0.85  &           3.50  &              4  &            4.4  &            7.3  \\
 0$<$$\theta_{Wl}$$<$0.75  &           3.63  &              4  &            3.8  &            7.0  \\
 0.05$<$$\theta_{Wl\nu}(\hat z)$$<$1.6  &           3.59  &              4  &            3.9  &            7.1  \\
 0$<$$\theta_{Wl\nu}(\hat z)$$<$1.55  &           3.57  &              4  &            4.0  &            7.2  \\
 0$<$$\theta_{Wl\nu}(\hat z)$$<$1.65  &           3.58  &              4  &            4.0  &            7.2  \\
 85$<$$M_N$$<$115  &           3.52  &              4  &            4.3  &            7.3  \\
 90$<$$M_N$$<$115  &           3.39  &              4  &            3.9  &            6.7  \\
 87.5$<$$M_N$$<$117.5  &           3.49  &              4  &            4.3  &            7.2  \\
 87.5$<$$M_N$$<$112.5  &           3.66  &              2  &            3.6  &            7.0  \\
\hline

\end{tabular}
\caption{Effect on the results for changes in cuts for $M_N=100$ GeV
	in the center of mass energy region 191-200 GeV}
\label{tab:ChCut100}
\end{table}

\begin{table}
\begin{tabular}{lllllllll}
       Variable  &   $S/\sqrt{B}$  &           Data  &             Bg  &         Signal  \\
 12.5$<$$E_\nu$$<$40  &           0.55  &              6  &           10.4  &            1.8  \\
 17.5$<$$E_\nu$$<$40  &           0.53  &              5  &            8.2  &            1.5  \\
 15$<$$E_\nu$$<$42.5  &           0.56  &              5  &            9.3  &            1.7  \\
 15$<$$E_\nu$$<$37.5  &           0.55  &              5  &            9.3  &            1.7  \\
 55$<$$E_\ell$$<$77.5  &           0.55  &              5  &            8.1  &            1.6  \\
 50$<$$E_\ell$$<$77.5  &           0.56  &              5  &           10.1  &            1.8  \\
 52.5$<$$E_\ell$$<$75  &           0.55  &              5  &            9.3  &            1.7  \\
 52.5$<$$E_\ell$$<$80  &           0.56  &              5  &            9.5  &            1.7  \\
 0.05$<$$|\cos\theta|$$<$1  &           0.54  &              4  &            9.1  &            1.6  \\
 0$<$$|\cos\theta|$$<$1.05  &           0.55  &              5  &            9.3  &            1.7  \\
 0$<$$|\cos\theta|$$<$0.95  &           0.56  &              5  &            7.7  &            1.6  \\
 25$<$$M_{\ell\nu}$$<$72.5  &           0.54  &              5  &            9.3  &            1.6  \\
 20$<$$M_{\ell\nu}$$<$72.5  &           0.56  &              5  &            9.3  &            1.7  \\
 22.5$<$$M_{\ell\nu}$$<$70  &           0.57  &              4  &            7.7  &            1.6  \\
 22.5$<$$M_{\ell\nu}$$<$75  &           0.53  &              6  &           11.3  &            1.8  \\
 1$<$$\theta_{Wl}$$<$1.4  &           0.54  &              5  &            9.9  &            1.7  \\
 1.1$<$$\theta_{Wl}$$<$1.4  &           0.57  &              4  &            8.5  &            1.7  \\
 1.05$<$$\theta_{Wl}$$<$1.45  &           0.53  &              5  &           10.3  &            1.7  \\
 1.05$<$$\theta_{Wl}$$<$1.35  &           0.53  &              5  &            8.2  &            1.5  \\
 0.05$<$$\theta_{Wl\nu}(\hat z)$$<$0.95  &           0.52  &              5  &            8.8  &            1.5  \\
 0$<$$\theta_{Wl\nu}(\hat z)$$<$0.9  &           0.56  &              5  &            8.9  &            1.7  \\
 0$<$$\theta_{Wl\nu}(\hat z)$$<$1  &           0.56  &              5  &            9.7  &            1.7  \\
 157.5$<$$M_N$$<$192.5  &           0.56  &              5  &            9.8  &            1.7  \\
 162.5$<$$M_N$$<$192.5  &           0.55  &              5  &            8.6  &            1.6  \\
 160$<$$M_N$$<$195  &           0.55  &              5  &            9.4  &            1.7  \\
 160$<$$M_N$$<$190  &           0.55  &              5  &            9.1  &            1.7  \\

\end{tabular}
\caption{Effect on the results for changes in cuts for $M_N=170$ GeV
	in the center of mass energy region 191-200 GeV.}
\label{tab:ChCut170}
\end{table}


\chapter{Conclusions and Outlook}
\section{Conclusions}
This is a search for heavy neutrinos in the single production channel with
charged current decay, which gives two jets ($e^+e^- \rightarrow N\nu
\rightarrow W\ell\nu \rightarrow 2\, $jets$+\ell\nu$). No traces of additional
neutrinos have been found, which is quantified by new limits on the mixing
parameter, established for different masses (see table
\ref{tab:DatSum}). Masses between 100 and 170 GeV have been analysed at
intervals of 10 GeV. The search has been done on OPAL data in the energy region
180-209 GeV with a total integrated luminosity of 663 pb$^{-1}$. The data has
been divided in three energy regions, 180-190 GeV, 191-200 GeV, and 201-208
GeV. Main as well as minor backgrounds have been considered and every
energy region and mass has been individually optimized.

\section{Outlook}
In order to complete the analysis, it would be interesting to investigate 
the neutral current channel as well as heavy charged leptons. For the heavy
charged leptons, an approximation has to be made (see section \ref{sec:HCL}) and
it would be important to evaluate its validity and/or look for other options.
The neutral current channel has a lower branching ratio than the charged
current channel, but for high masses it could still be a competitive
complement.

The peculiar excess in the data just before the last cut is also something that
is worth investigating. This could be a signal of a heavy muon even though
the number of measured data-events is low. Even though the
optimization would be different, there are many common points between
a heavy muon and a heavy electron and the cuts could be rather similar for the
two cases. However, for the muon it would be an $s$-channel process, which has
a lower cross section.

Furthermore, the cuts can be optimized differently. The fine-tuning described
in section \ref{sec:MC_Cuts} has not necessarily been pushed to its limit.
Some cuts could be harder for some mass cases and loser for others.
It would also be possible to do a likelihood analysis. This could be a good
method of analysis considering the many cuts and the fact that several of them are
interdependent.


	\appendix
        \chapter{Glossary}
\label{App:Glossary}
\begin{description}
\item[Antimatter]       Antimatter is constituted of \gloss{antiparticles}.
\item[Antiparticle]
        An antiparticle is defined
        as having the opposite \gloss{quantum numbers} of the corresponding particle,
        but the same mass.
        Particles with their quantum numbers zero, except for spin which should be integer, (like the photon
) are their own antiparticles.
\item[Branching ratio]
        The relative probability that a particle will decay in a specific way.
        Example: $Z^0 \rightarrow e^+e^-$ has a branching ratio of about 3 percent.
        The sum of all branching ratios should be unity.
\item[Charge conjugation($C$)]
        $C$ is an operation that reverses the charge of a particle.
\item[Charged Current (CC)]
	Charged current is an interaction via a $W$ boson.
\item[CP-violation]
	The violation of the charge (C) and parity (P) quantum numbers.
\item[CKM-matrix]
	The mixing between quarks.
\item[Decay channels]   The possible decays of a particle.\\
        Example: $$Z^0 \rightarrow
                u{\bar u},
                d{\bar d},
                s{\bar s},
                c{\bar c},
                b{\bar b},
                t{\bar t},
                e^-e^+,
                \mu^-\mu^+,
                \tau^-\tau^+,
                \nu_e{\bar \nu_e},
                \nu_\mu{\bar \nu_\mu},
                \nu_\tau{\bar \nu_\tau},$$
        where the different processes have different \gloss{branching ratios}.
\item[Exotic]
	An exotic particle is either heavy or excited.
\item[Fermion]
	A fermion is a lepton or a quark.
\item[Final state radiation]
	Final state radiation is decays of particles or emission of gluons or
	photons that occur after the event.
\item[Higgs boson]
        A hypothetical particle required in the standard model of particle physics.
        The Higgs boson explains why $W^\pm$, and $Z^0$ have a mass.
\item[Initial state radiation]
	Initial state radiation is decays of particles or emission of        
	gluons or photons that occur before the event.
\item[Preons]
	Preons are the consituents of leptons and/or quarks.
\item[Indices L and R]
	These indices refer to left-handed and right-handed
	particles (for left-handed particles the spin is antiparallell with the
	direction of the particle, the spin points "backwards").
\item[Massshell]
        On the mass shell means that the particle is real ($M^2=E^2-|\vec p|^2$).
        Off the mass shell means that it is virtual ($M^2\ne E^2-|\vec p|^2$).
\item[Mixing/Mixing parameter/Mixing Angle]
	The mixing between a heavy lepton and an ordinary lepton. For further explanation, see section \ref{sec:Mixing}.
\item[Neutral Current (NC)]
	Neutral current is an interaction via a $Z$ boson.
\item[Parity ($P$)]
	Parity is an operation that gives a particle the same properties as if
	is was observed in a ``point-like mirror''. In other words, the spin of
	the particle will be inversed. $P\left| parity=p\right> = \left|
	parity=-p\right>$. The left-parity operator is represented by $P_L =
	\frac 12(1-\gamma_5)$, where $\gamma_5$ is defined in appendix
	\ref{App:Symb}. The right-parity operator is defined as $P_R = \frac
	12(1+\gamma_5)$.
\item[Signal]
	The signal is the signal of the exotic lepton.
\item[Symmetry]
        A symmetry operation does not change the physical solution. 
        For example, the position of the origin in the coordinate system
        does not change the physical solution to problem.
\item[$s$-channel]
	The $s$-channel is the merging of two particles into one, \eg
	$e^+e^- \rightarrow \gamma^*$ (see figure \ref{fig:Single}).
\item[$t$-channel]
	The $t$-channel is the exchange of one particle between two other
	(see figure \ref{fig:tSingle}).
\item[Quantum numbers]  The numbers that can be said to best describe the state of
        a particle. Examples: electric charge ($Q$), lepton number ($L$),
        baryon number ($B$), parity ($P$), spin ($S$), isospin ($I$),
        strangeness ($S$), and charge conjugation ($C$).
\end{description}

\chapter{List of Symbols}
\label{App:Symb}

Regarding the notations used, a capital letter in fermion names signifies a
heavy fermion (e. g. $L$=heavy lepton, $E$=heavy electron) while a star
signifies an excited fermion (e. g. $u^*$=excited up-quark, $\mu^*$=excited
muon). A fermion is either a lepton or a quark, and an exotic particle is
either heavy or excited.
A lepton
$L$ can be either charged or neutral.
In both cases a $\pm$ denotes a charged fermion (e. g. $L^\pm$=charged heavy
lepton).
A bar denotes an anti-particle and unless otherwise specified, $c=\hbar=1$.

\section{Exotic Particles}
\begin{tabular}{ll}
	$F^*$	&	Exotic fermion\\
	$L^*$	&	Exotic lepton\\
	$F$	&	Heavy fermion\\
	$L$	&	Heavy lepton\\
	$L^\pm$	&	Heavy charged lepton\\
	$N$	&	Heavy neutral lepton (neutrino)\\
	$E$	&	Heavy electron\\	
	$N_E$	&	Heavy electron neutrino\\	
	$Q$	&	Heavy quark\\
	$U$	&	Heavy up-quark\\
	$D$	&	Heavy down-quark\\
	$l^*,\ell^*$&	Excited lepton\\
	$l^{*\pm},\ell^{*\pm}$&	Excited charged lepton\\
	$\nu^*$	&	Excited neutrino\\
	$e^*$	&	Excited electron\\
	$\nu^*_e$&	Excited electron neutrino\\
	$\mu^*$	&	Excited muon\\
	$\nu^*_\mu$&	Excited muon neutrino\\
	$\tau^*$&	Excited tau\\
	$\nu^*_\tau$&	Excited tau neutrino\\
\end{tabular}

\section{Standard Model Particles}
\begin{tabular}{ll}
        $\gamma$&       Photon\\
        $f$     &       Fermion (quark or lepton, origin: Enrico Fermi)\\
        $l,\ell$&       Lepton (origin: greek leptos=small)\\
	$e$	&	Electron (origin: greek elektron = amber,\\
		&	the rubbing of which causes electrostatic phenomena)\\
	$\nu_e$	&	Electron neutrino (origin: latin "small neutron")\\
	$\mu$	&	Muon\\
	$\nu_\mu$&	Muon neutrino\\
	$\tau$	&	Tauon\\
	$\nu_\tau$&	Tau neutrino\\
        $q$     &       Quark (origin: "Three quarks for Muster Mark" in\\
		&	James Joyce's Finnegans Wake (1939))\\
	$u$	&	Up quark \\
	$d$	&	Down quark \\
	$s$	&	Strange quark \\
	$c$	&	Charm quark\\
        $t$     &       Top quark\\
        $b$     &       Bottom quark\\
        $jet$   &       Jet of particles, produced by a quark or a gluon\\

\end{tabular}
\section{Variables}
\begin{tabular}{ll}
        $p$     &       Momentum\\
        $p_T$   &       Transverse momentum, $p_T \equiv \sqrt{p_x^2+p_y^2}$ if $z$ is the beam direction\\
        $\cal L$&       Luminosity\\
        $\sigma$&       Cross-section\\
	$\theta$&	Polar angle\\
	$\phi$	&	Azimuthal angle\\
        $t$     &       Time\\
	$\bar\sigma$&	Standard deviation\\
        $i$     &       Imaginary number, defined as $i^2=-1$ \\
        $\gamma$ matrices&      
                        $\gamma_0 = \left[\begin{smallmatrix} 1 & 0 & 0 & 0 \\ 0 & 1 & 0 & 0 \\ 0 & 0 & -1 & 0 \\ 0 & 0 & 0 & -1 \end{smallmatrix}\right]$
                        $\gamma_5 = \left[\begin{smallmatrix} 0 & 0 & 1 & 0 \\ 0 & 0 & 0 & 1 \\ 1 & 0 & 0 & 0 \\ 0 & 1 & 0 & 0 \end{smallmatrix}\right]$\\
                &       $\gamma_1 = \left[\begin{smallmatrix} 0 & 0 & 0 & 1 \\ 0 & 0 & 1 & 0 \\ 0 & -1 & 0 & 0 \\ -1 & 0 & 0 & 0 \end{smallmatrix}\right]$
                        $\gamma_2 = \left[\begin{smallmatrix} 0 & 0 & 0 & -i \\ 0 & 0 & i & 0 \\ 0 & i & 0 & 0 \\ -i & 0 & 0 & 0 \end{smallmatrix}\right]$
                        $\gamma_3 = \left[\begin{smallmatrix} 0 & 0 & 1 & 0 \\ 0 & 0 & 0 & -1 \\ -1 & 0 & 0 & 0 \\ 0 & 1 & 0 & 0 \end{smallmatrix}\right]$
\end{tabular}

\section{List of Constants}
\begin{tabular}{lll}
        $\theta_W$&     28.7 degrees&   The weak mixing angle $\sin^2 \theta_W  \approx 0.23 $\\
        $\hbar$ &       1.054571597$\times 10^{-34}$ Js&        $\hbar=h/2\pi$, where $h$ is Planck's constant\footnotemark\\
        $c$     &       299792458 ms$^{-1}$&    Speed of light in vacuum\footnotemark[\value{footnote}]\\
        $k_B$   &       1.3806503$\times 10^{-23}$ JK$^{-1}$&   Boltzmann's constant\footnotemark[\value{footnote}]\\
        eV      &       1.6022$\times 10^{-19}$ J&              Electron Volt
\end{tabular}
\footnotetext{
        Unless otherwise specified, we set $\hbar=c=k_B=1$}

\section{Abbreviations}
\begin{tabular}{ll}
        OPAL    &       Omni-Purpose Apparatus at LEP\\
        CERN    &       European Laboratory For Particle Physics\\
                &       (earlier: Conseil Europ\'een pour la Recherche Nucl\'eaire)\\
        LEP     &       Large Electron Positron collider\\
        MSSM    &       Minimal Supersymmetric Standard Model\\
	GUT     &       Grand Unified Theory\\
        PYTHIA  &       A FORTRAN program to simulate collisions between particles\\
	grc4f	&	GRC (Grace) four fermions\\
	kk2f	&	KK two fermions\\
        QCD     &       Quantum ChromoDynamics, the field theory for strong interactions\\
        QED     &       Quantum ElectroDynamics, the field theory for electromagnetic interactions\\
        SUSY    &       SUper SYmmetry\\
\end{tabular}

        \pagebreak

%
%
\newcommand{\ZP}[3]    {Z. Phys. {\bf C#1} (#2) #3.}
\newcommand{\PL}[3]    {Phys. Lett. {\bf B#1} (#2) #3.}
\newcommand{\PhysLett}  {Phys.~Lett.}
\newcommand{\PRL} {Phys.~Rev.\ Lett.}
\newcommand{\PhysRep}   {Phys.~Rep.}
\newcommand{\PhysRev}   {Phys.~Rev.}
\newcommand{\NPhys}  {Nucl.~Phys.}
\newcommand{\NIM} {Nucl.~Instr.\ Meth.}
\newcommand{\CPC} {Comp.~Phys.\ Commun.}
\newcommand{\ZPhys}  {Z.~Phys.}
\newcommand{\IEEENS} {IEEE Trans.\ Nucl.~Sci.}

\end{document}